\documentclass[12pt]{article}
\usepackage{amsmath,amssymb,amsfonts,color,graphicx,cite,feynarts,soul}
\usepackage[usenames,dvipsnames,svgnames,table]{xcolor}
\input paperdef

\graphicspath{{figs/}}

\evensidemargin \oddsidemargin
\allowdisplaybreaks

\begin{document}
\thispagestyle{empty}

\def\thefootnote{\fnsymbol{footnote}}

\begin{flushright}
%arXiv:yymm.nnnn [hep-ph]
\end{flushright}

\vspace{0.5cm}

\begin{center}

{\large\sc {\bf Neutralino Decays in the Complex MSSM at One-Loop:}}

\vspace{0.4cm}

{\large\sc {\bf a Comparison of On-Shell Renormalization Schemes}}

\vspace{1cm}

{\sc
A.~Bharucha$^{1}$%
\footnote{email: Aoife.Bharucha@desy.de}%
, S.~Heinemeyer$^{2}$%
\footnote{email: Sven.Heinemeyer@cern.ch}%
, F.~von der Pahlen$^{2}$%
\footnote{email: pahlen@ifca.unican.es}%
\footnote{MultiDark Fellow}%
,~and~C.~Schappacher$^{3}$%
\footnote{email: cs@particle.uni-karlsruhe.de}%
}

\vspace*{.7cm}

{\sl
$^1$II.~Institut f\"ur Theoretische Physik, Universit\"at Hamburg, Lupurer Chaussee 149, D--22761 Hamburg, Germany

\vspace{0.1cm}

$^2$Instituto de F\'isica de Cantabria (CSIC-UC), E-39005 Santander, Spain

\vspace*{0.1cm}

$^3$Institut f\"ur Theoretische Physik, Karlsruhe Institute of Technology, \\
D--76128 Karlsruhe, Germany%
\footnote{Former address}

}

\end{center}

\vspace*{0.1cm}

\begin{abstract}
\noindent
We evaluate two-body decay modes of neutralinos in 
the Minimal Supersymmetric Standard Model with complex parameters
(cMSSM). Assuming heavy scalar quarks we take into account all two-body decay channels
involving charginos, neutralinos, (scalar) leptons, 
Higgs bosons and Standard Model gauge bosons.
The evaluation of the decay widths is based on a full one-loop calculation 
including hard and soft QED radiation. 
Of particular phenomenological interest are
 decays involving the 
Lightest Supersymmetric Particle (LSP), 
i.e.\ the lightest neutralino, or a neutral or charged Higgs boson.
For the chargino/neutralino sector we employ two different
renormalization schemes, which differ in the treatment of the complex
phases. In the numerical analysis we concentrate 
on the decay of the
heaviest neutralino and show the results in the two different schemes. 
The higher-order corrections of the heaviest neutralino decay widths
involving the LSP can easily reach 
a level of about $10{\rm -}15\%$, 
while the corrections to the
decays to Higgs bosons are up to $20{\rm -}30\%$,
translating into
corrections of similar size in the respective branching ratios.
The difference between the two schemes, indicating the size of unknown two-loop corrections, 
is less than ${\cal O}(0.1\%)$.
These corrections are important for the correct interpretation of 
LSP and Higgs production at the LHC and at a future linear $e^+e^-$
collider.
The results will be implemented into the Fortran code {\fh}.
\end{abstract}
%\pacs{}

\def\thefootnote{\arabic{footnote}}
\setcounter{page}{0}
\setcounter{footnote}{0}
\newpage

\newcommand{\DecayCCh[1]}{\cha{2} \to \cha{1} h_{#1}}
\newcommand{\DecayCCZ}{\cha{2} \to \cha{1} Z}
\newcommand{\DecayCNH}[2]{\cha{#1} \to \neu{#2} H^\mp}
\newcommand{\DecayCNW}[2]{\cha{#1} \to \neu{#2} W^\mp}
\newcommand{\DecayCnSl}[3]{\cham{#1} \to \nu_{#2}\, \tilde{#2}_{#3}^{-}}
\newcommand{\DecayClSn}[2]{\cham{#1} \to {#2}^{-}\, \tilde{\nu_{#2}}}
\newcommand{\DecayCxy}[1]{\cha{#1} \to {\rm xy}}

\newcommand{\DecayNCW}[2]{\neu{#1} \to \champ{#2} W^\pm}
\newcommand{\DecayNCH}[2]{\neu{#1} \to \champ{#2} H^\pm}
\newcommand{\DecayNlpmSl}[3]{\neu{#1} \to {#2}^\mp\, \tilde{#2}_{#3}^\pm}
\newcommand{\DecayNnSnnbarSn}[2]{\neu{#1} \to \bar{\nu}_{#2}\, \tilde{\nu}_{#2}  /  {\nu}_{#2}\, \tilde{\nu}_{#2}^\dagger  }

\newcommand{\decayCCh}{\DecayCCh{k}}
\newcommand{\decayCCZ}{\DecayCCZ}
\newcommand{\decayCNH}{\DecayCNH{i}{j}}
\newcommand{\decayCNW}{\DecayCNW{i}{j}}
\newcommand{\decayCnSl}{\DecayCnSl{i}{l}{k}}
\newcommand{\decayClSn}{\DecayClSn{i}{l}}
\newcommand{\decayCxy}{\DecayCxy{i}}

\newcommand{\DecayNxy}[1]{\neu{#1} \to {\rm xy}}
\newcommand{\DecayNCmH}[2]{\neu{#1} \to \cham{#2} H^+}
\newcommand{\DecayNCmW}[2]{\neu{#1} \to \cham{#2} W^+}
\newcommand{\DecayNCpH}[2]{\neu{#1} \to \chap{#2} H^-}
\newcommand{\DecayNCpW}[2]{\neu{#1} \to \chap{#2} W^-}
\newcommand{\DecayNNh}[3]{\neu{#1} \to \neu{#2} h_{#3}}
\newcommand{\DecayNNZ}[2]{\neu{#1} \to \neu{#2} Z}
\newcommand{\DecayNnSn}[2]{\neu{#1} \to {\nu}_{#2}\, \tilde{\nu}_{#2}^\dagger}
\newcommand{\DecayNlSl}[3]{\neu{#1} \to {#2}^-\, \tilde{#2}_{#3}^+}
\newcommand{\DecayNbarnSn}[2]{\neu{#1} \to \bar{\nu}_{#2}\, \tilde{\nu}_{#2}}
\newcommand{\DecayNlpSl}[3]{\neu{#1} \to {#2}^+\, \tilde{#2}_{#3}^-}

\newcommand{\decayNCH}{\DecayNCH{i}{j}}
\newcommand{\decayNCW}{\DecayNCW{i}{j}}
\newcommand{\decayNlpmSl}{\DecayNlpmSl{i}{\ell}{k}}
\newcommand{\decayNnSnnbarSn}{\DecayNnSnnbarSn{i}{\ell}}

\newcommand{\decayNCmH}{\DecayNCmH{i}{j}}
\newcommand{\decayNCmW}{\DecayNCmW{i}{j}}
\newcommand{\decayNNh}{\DecayNNh{i}{j}{k}}
\newcommand{\decayNNZ}{\DecayNNZ{i}{j}}
\newcommand{\decayNnSn}{\DecayNnSn{i}{\ell}}
\newcommand{\decayNlSl}{\DecayNlSl{i}{\ell}{k}}
\newcommand{\decayNbarnSn}{\DecayNbarnSn{i}{\ell}}
\newcommand{\decayNlpSl}{\DecayNlpSl{i}{\ell}{k}}

\newcommand{\decayNxy}{\DecayNxy{i}}

\newcommand{\DecayCmCh}[1]{\cham{2} \to \cham{1} h_{#1}}
\newcommand{\DecayCmCZ}{\cham{2} \to \cham{1} Z}
\newcommand{\DecayCmNH}[2]{\cham{#1} \to \neu{#2} H^-}
\newcommand{\DecayCmNW}[2]{\cham{#1} \to \neu{#2} W^-}
\newcommand{\DecayCmnSl}[3]{\cham{#1} \to \bar{\nu}_{#2}\, \tilde{#2}_{#3}^-}
\newcommand{\DecayCmlSn}[2]{\cham{#1} \to {#2}^-\, \tilde{\nu}_{#2}^\dagger}
\newcommand{\DecayCmxy}[1]{\cham{#1} \to {\rm xy}}

\newcommand{\decayCmCh}{\DecayCmCh{k}}
\newcommand{\decayCmCZ}{\DecayCmCZ}
\newcommand{\decayCmNH}{\DecayCmNH{i}{j}}
\newcommand{\decayCmNW}{\DecayCmNW{i}{j}}
\newcommand{\decayCmnSl}{\DecayCmnSl{i}{l}{k}}
\newcommand{\decayCmlSn}{\DecayCmlSn{i}{l}}
\newcommand{\decayCmxy}{\DecayCmxy{i}}

%%%%%%%%%%%%%%%%%%%%%%%%%%%%%%%%%%%%%%%%%%%%%%%%%%%%%%%%%%%%%%%%%%%%%%%%%%%%%%%
%%%%%%%%%%%%%%%%%%%%%%%%%%%%%%%%%%%%%%%%%%%%%%%%%%%%%%%%%%%%%%%%%%%%%%%%%%%%%%%
\section{Introduction}

One of the important tasks at the LHC is to search for 
physics beyond the Standard Model (SM), where the 
Minimal Supersymmetric Standard Model (MSSM)~\cite{mssm} is one of the 
leading candidates.

Two related important tasks are investigating the mechanism of electroweak
symmetry breaking, as well as the production and measurement of
the properties of Cold Dark Matter (CDM).
The Higgs searches currently ongoing at the LHC (and previously 
carried out at the {Tevatron~\cite{TevatronHiggs} and LEP~\cite{LEPHiggs}})
address both those goals. The spectacular 
discovery of a Higgs-like particle 
with a mass around $M_H\simeq 125\gev$, which has just been announced by ATLAS
and CMS~\cite{discovery}, marks a milestone of an effort that has been 
ongoing for almost half a century and opens a new era of particle physics. 
Both ATLAS and CMS reported a clear excess around $\sim 125 \gev$
in the two photon channel, as well as in the $ZZ^{(*)}$ channel, whereas
the analyses in other channels have a lower mass resolution 
and at present are largely less mature. The combined sensitivity
in each of the experiments reaches about $\sim 5 \si$.
The discovery is consistent with the predictions for the Higgs boson in the
{SM~\cite{LHCHXSWS}}, 
as well as with the predictions for the lightest Higgs boson in the 
{MSSM~\cite{{LHCHXSWS,thHiggsMSSM}}}.
The latter model also offers a natural candidate for CDM, the
Lightest Supersymmetric Particle (LSP), i.e.\ the lightest 
neutralino,~$\neu{1}$~\cite{EHNOS}.
Supersymmetry (SUSY) predicts two scalar partners for all SM fermions as well
as fermionic partners to all SM bosons.
Contrary to the case of the SM, in the MSSM 
two Higgs doublets are required.
This results in five physical Higgs bosons instead of the single Higgs
boson in the SM. These are the light and heavy $\cp$-even Higgs bosons, $h$
and $H$, the $\cp$-odd Higgs boson, $A$, and the charged Higgs bosons,
$H^\pm$.
In the MSSM with complex parameters (cMSSM) the three neutral Higgs
bosons mix~\cite{mhiggsCPXgen,Demir,mhiggsCPXRG1,mhiggsCPXFD1}, 
giving rise to the $\cp$-mixed states $\He, \Hz, \Hd$.

If SUSY is realized in nature and the scalar quarks and/or the gluino
are in the kinematic reach of the LHC, it is expected that these
strongly interacting particles are copiously produced. 
The primarily produced strongly interacting particles subsequently
decay via cascades to 
SM particles and (if $R$-parity conservation is assumed, as we do) the
LSP. One step in these decay chains is often the decay of a neutralino, 
$\neu{2,3,4}$,  
to a SM particle and the LSP, or as a competing process the
neutralino decay to another SUSY particle accompanied by a
SM particle. Also neutral and charged Higgs bosons are expected to be produced this way.
Via these decays some characteristics of the LSP and/or Higgs bosons can be
measured, see, e.g., \citeres{atlas,cms} and references therein. 
At any future $e^+e^-$ collider (such as ILC or CLIC)
a precision determination of the properties of the observed particles is
expected~\cite{teslatdr,ilc}. 
(For combined LHC/ILC analyses and further
prospects see \citere{lhcilc}.) 
Thus, if kinematically accessible, the pair production of neutralinos 
with a subsequent decay to the LSP and/or Higgs bosons 
can yield important information about the lightest neutralino and 
the Higgs sector of the model.

In order to yield a sufficient accuracy, at least one-loop corrections to
the various neutralino decay modes have to be considered. 
In this paper we evaluate full one-loop corrections to neutralino decays
in the cMSSM.
If scalar quarks are sufficiently heavy (as in many GUT based models
such as CMSSM, GMSB or AMSB, see for instance
\citere{newbenchmark}) 
a neutralino decay to a quark and a scalar 
quark is kinematically forbidden. Assuming heavy squarks 
we calculate the full one-loop correction to
all two body decay modes (which are non-zero at the tree-level),
\begin{align}
\label{NCH}
&\Ga(\decayNCH) \qquad (i = 2,3,4;\; j = 1,2)~, \\
\label{NCW}
&\Ga(\decayNCW) \qquad (i = 2,3,4;\; j = 1,2)~, \\
\label{NNh}
&\Ga(\decayNNh) \qquad (i = 2,3,4;\; j < i;\; k = 1,2,3)~, \\
\label{NNZ}
&\Ga(\decayNNZ) \qquad (i = 2,3,4;\; j < i)~, \\
\label{NSll}
&\Ga(\decayNlpmSl) \qquad (i = 2,3,4;\; {\ell} = e, \mu, \tau;\; k = 1,2)~, \\
\label{NSnn}
&\Ga(\decayNnSnnbarSn) \qquad (i = 2,3,4;\; {\ell} = e, \mu, \tau)~.
\end{align}
The total width is defined as the sum of the channels (\ref{NCH}) to
(\ref{NSnn}), where we neglect the decays to colored particles as these will not 
be kinematically allowed for the scenarios considered in the numerical analysis.
It should be noted that several modes are closed for nearly the whole
MSSM parameter space due to the structure of the chargino and neutralino
mass matrices (see below). 
Therefore, while we have evaluated analytically {\em all} neutralino
decays, in our numerical analysis we will concentrate on the decays of
the heaviest neutralino, $\neu{4}$.

As explained above, 
we are especially interested in the branching ratios (BR) of the
decays involving a Higgs boson, \refeqs{NCH}, (\ref{NNh})
as part of an evaluation of a Higgs production cross section, and/or
involving the LSP, \refeqs{NNh}, (\ref{NNZ}) as part of the measurement
of CDM properties at the LHC, the ILC or CLIC.
Consequently, it is not necessary to investigate three- or four-body decay
modes. These only play a significant role once the two-body modes
are kinematically forbidden, and thus the relevant BR's are zero. 
The same applies to two-body decay modes that exist only at the
one-loop level, such as $\neu{i} \to \neu{j} \ga$ (see, for instance,
\citere{chachaga}). While this channel is 
of \order{\al^2}, the size of the one-loop corrections to \refeqs{NCH}
to (\ref{NSnn}) is of \order{\al}. We have numerically verified that the
contribution of $\Ga(\neu{i} \to \neu{j} \ga)$ to the total width is
completely negligible. 

Tree-level results for the neutralinos decays in the MSSM were presented in \citeres{Baer:neudec,chachaga,haber3}. 
The code {\tt SDECAY}~\cite{SDECAY} includes all two-body decays of neutralinos at tree level. 
Tree-level studies of neutralino decays have shown that they could be invaluable in distinguishing between different patterns of supersymmetry breaking~\cite{Huitu:2010me}, 
as well as in detecting $\cp$ violating effects 
at a linear collider~\cite{Choi:2003pq,Choi:2003fs,Bartl:2004jj,Choi:2005gt,Choi:2006vh,Bartl:2007qy}
or a muon collider~\cite{mucollNCP}.
Higher-order corrections to neutralino decays have been evaluated in
various analyses over the last decade in the real MSSM (rMSSM), for which the on-shell renormalization of
the chargino-neutralino sector was developed in Refs.~\cite{Lahanas:1993ib,Pierce:1993gj,Pierce:1994ew,Eberl:2001eu,Fritzsche:2002bi,Oller:2003ge,Drees:2006um,dissAF,Fowler:2009ay,onshellCNmasses}. In \citere{ChaDecCPC} three-body decays into the LSP and quarks were calculated 
including corrections to the masses of third generation fermions and SUSY
particles. In \citere{Drees:2006um}, decays of the next-to-lightest neutralino to the lightest neutralino and two leptons were calculated at one-loop. The one-loop electroweak corrections to all two-body decay channels 
of neutralinos, evaluated in  an on-shell renormalization scheme, 
have been implemented in the code {\tt SloopS}~\cite{BaroII}. 
Radiative corrections to a number of neutralino decay channels were also recently studied in \citere{dissBS} for the case of real parameters.
A full one-loop  calculation of the electroweak corrections to the partial width of the decay of a neutralino into a chargino
and a $W$~boson in the MSSM and NMSSM is presented in \citere{liebler},  
and made available with the code {\tt CNNDecays}.
In the cMSSM, the on-shell renormalization of the chargino--neutralino sector was first studied in \citere{dissAF}, and subsequently in \citere{bfmw}, and decays of type $\Ga(\decayNCH)$ were studied in \citeres{dissAF,Fowler:2009ay}.
The approach to the renormalization of the complex parameters differs slightly from that 
used in \citere{Stop2decay,LHCxC}, where chargino decays in the cMSSM at the
one-loop level were analyzed. 
One important part of this work consists of a comparison of these two schemes.

In this paper we present
for the first time a full one-loop calculation for all non-hadronic 
two-body decay channels of a neutralino, taking into
account soft and hard QED radiation, simultaneously and
consistently evaluated in the cMSSM.
The calculation is based on two independent set-ups that differ slightly
in the inclusion of higher-order corrections to quantities used in
one-loop corrections, i.e.\ in effects beyond the one-loop level. 
The two set-ups furthermore employ renormalization schemes in the 
chargino/neutralino sector that differ in their treatment 
of complex phases~\cite{LHCxC,Stop2decay,dissAF,Fowler:2009ay}. 
The numerical results are shown for both set-ups and the (small)
differences indicate the size of theoretical uncertainties beyond the one-loop level.

The paper is organized as follows. 
In \refse{sec:cMSSM} we review the relevant sectors of the cMSSM
and give all the details about the two different renormalization schemes
in the chargino/neutralino sector. 
Details about the calculation can be
found in \refse{sec:calc}. The numerical results for all decay
channels are presented in \refse{sec:numeval}. The conclusions can be
found in \refse{sec:conclusions}.
The evaluation of the branching ratios of the neutralinos will be 
implemented into the Fortran code 
{\fh}~\cite{feynhiggs,mhiggslong,mhiggsAEC,mhcMSSMlong}.

%%%%%%%%%%%%%%%%%%%%%%%%%%%%%%%%%%%%%%%%%%%%%%%%%%%%%%%%%%%%%%%%%%%%%%%%%%%%%%%
%%%%%%%%%%%%%%%%%%%%%%%%%%%%%%%%%%%%%%%%%%%%%%%%%%%%%%%%%%%%%%%%%%%%%%%%%%%%%%%
\section{The relevant sectors of the complex MSSM}
\label{sec:cMSSM}

All channels (\ref{NCH}) -- (\ref{NSnn}) are calculated at the
one-loop level, including real QED radiation. This requires the
simultaneous renormalization of several sectors of the cMSSM. In
the following subsections we introduce our notation for these sectors.
Details about the two renormalization schemes used in the
chargino/neutralino sector are given. The renormalization of the other
sectors can be found in \citeres{Stop2decay,LHCxC,Stau2decay}.

%%%%%%%%%%%%%%%%%%%%%%%%%%%%%%%%%%%%%%%%%%%%%%%%%%%%%%%%%%%%%%%%%%%%%%%%%%%%%%%
\subsection{The chargino/neutralino sector of the cMSSM}
\label{sec:chaneu}
While many details about the renormalization of the cMSSM can already 
be found in \citeres{LHCxC,Stop2decay,dissAF,Fowler:2009ay},
we repeat here the most relevant aspects in order to give a complete picture and 
to facilitate the comparison between the two employed renormalization schemes. 
The chargino/neutralino sector contains two soft SUSY-breaking gaugino mass
parameters $M_1$ and $M_2$ corresponding to the bino and the wino fields,
respectively, as well as the Higgs superfield mixing parameter $\mu$,
which, in general, can be complex. 
Since not all the possible phases of the cMSSM Lagrangian are physical,
it is possible (without loss of generality) to choose some parameters
real. This applies in particular to one out of the three parameters
$M_1$, $M_2$, and $M_3$, the gluino mass parameter. For the numerical analysis in \refse{sec:numeval} we
choose $M_2$ to be real, however for the renormalization scheme~I introduced below~\cite{LHCxC,Stop2decay}, we do not make such an assumption, and the analytical derivation of the renormalization constants is 
performed for a complex $M_2$, discussed further in \refse{sec:numeval}.

The starting point for the renormalization procedure of the
chargino/neutralino sector is the part of the Fourier transformed MSSM
Lagrangian which is bilinear in the chargino and neutralino fields,
\begin{align}
\cL^{\text{bil.}}_{\cham{},\tilde{\chi}^0} &= 
  \overline{\cham{i}}\, \pslash\, \OM \cham{i} 
+ \overline{\cham{i}}\, \pslash\, \OP \cham{i} 
- \overline{\cham{i}}\, [\matr{V}^* \matr{X}^\top \matr{U}^\dagger]_{ij} \,
  \OM \cham{j} 
- \overline{\cham{i}}\, [\matr{U} \matr{X}^* \matr{V}^{\top}]_{ij} \,
  \OP \cham{j} \non \\
&\quad + \frac{1}{2} \KL
  \overline{\neu{k}}\, \pslash\, \OM \neu{k}, 
+ \overline{\neu{k}}\, \pslash\, \OP \neu{k} 
- \overline{\neu{k}}\, [\matr{N}^*\matr{Y} \matr{N}^\dagger]_{kl} \,
  \OM \neu{l} 
- \overline{\neu{k}}\, [\matr{N} \matr{Y}^* \matr{N}^{\top}]_{kl} \,
  \OP \neu{l} \KR~, 
\end{align}
already expressed in terms of the chargino and neutralino mass eigenstates
$\cham{i}$ and $\neu{k}$, respectively, 
and $i,j = 1,2$ and $k,l = 1,2,3,4$.
The mass eigenstates can be determined via unitary 
transformations where the corresponding matrices diagonalize the chargino and
neutralino mass matrix, $\matr{X}$ and $\matr{Y}$, respectively. 

In the chargino case, two $2 \times 2$ matrices $\matr{U}$ and
$\matr{V}$ are necessary for the diagonalization of the chargino mass
matrix~$\matr{X}$, 
\begin{align}
\matr{M}_{\cham{}} = \matr{V}^* \, \matr{X}^\top \, \matr{U}^{\dagger} =
  \begin{pmatrix} m_{\tilde{\chi}^\pm_1} & 0 \\ 
                  0 & m_{\tilde{\chi}^\pm_2} \end{pmatrix}  \quad
\text{with} \quad
  \matr{X} =
  \begin{pmatrix}
    \MTwo & \sqrt{2} \sinb \MW \\
    \sqrt{2} \cosb \MW & \mu
\label{eq:X}
  \end{pmatrix}~,
\end{align}
where $\matr{M}_{\cham{}}$ is the diagonal mass matrix with the chargino
masses $\mcha{1}, \mcha{2}$ as entries, which are determined as the
(real and positive) singular values of $\matr{X}$. 
The singular value decomposition of $\matr{X}$ also yields results for 
$\matr{U}$ and~$\matr{V}$. 
Using the transformation matrices $\matr{U}$ and $\matr{V}$, the interaction
Higgsino and wino spinors $\tilde{H}^-_1$, $\tilde{H}^+_2$ and
$\tilde{W}^\pm$, which are two component Weyl spinors, can be transformed into
the mass eigenstates 
\begin{align}
\cham{i} = 
\begin{pmatrix} \psi^L_i
   \\ \overline{\psi^R_i} \end{pmatrix}
\quad \text{with} \quad \psi^L_i = U_{ij} \begin{pmatrix} \tilde{W}^-
  \\ \tilde{H}^-_1 \end{pmatrix}_j \quad \text{and} \quad
 \psi^R_i = V_{ij} \begin{pmatrix} \tilde{W}^+
  \\ \tilde{H}^+_2 \end{pmatrix}_j
\end{align}
where the $i$th mass eigenstate can be expressed in terms of either 
the Weyl spinors $\psi^L_i$ and $\psi^R_i$  or the Dirac spinor $\cham{i}$.

In the neutralino case, as the neutralino mass matrix $\matr{Y}$ is
symmetric, one $4 \times 4$~matrix is sufficient for the diagonalization
\begin{align}
\matr{M}_{\neu{}} = \matr{N}^* \, \matr{Y} \, \matr{N}^{\dagger} =
\text{\bf diag}(m_{\neu{1}}, m_{\neu{2}}, m_{\neu{3}}, m_{\neu{4}})
\end{align}
with
\begin{align}
\matr{Y} &=
  \begin{pmatrix}
    \MOne                  & 0                & -\MZ \, \sw \cosb
    & \MZ \, \sw \sinb \\ 
    0                      & \MTwo            & \quad \MZ \, \cw \cosb
    & -\MZ \, \cw \sinb \\ 
    -\MZ \, \sw \cosb      & \MZ \, \cw \cosb & 0
    & -\mu             \\ 
    \quad \MZ \, \sw \sinb & -\MZ \, \cw \sinb & -\mu              & 0
  \end{pmatrix}~.
\label{eq:Y}
\end{align}
$\MZ$ and $\MW$ are the masses of the $Z$~and $W$~boson, 
$\cw = \MW/\MZ$ and $\sw = \sqrt{1 - \cw^2}$. 
The unitary 4$\times$4 matrix $\matr{N}$ and the physical neutralino
(tree-level) masses $\mneu{k}$ ($k = 1,2,3,4$) result from a numerical Takagi 
factorization \cite{Takagi} of $\matr{Y}$. 
Starting from the original bino/wi\-no/higg\-si\-no basis, the mass
eigenstates can be determined with the help of the transformation matrix~$\matr{N}$, 
\begin{align}
\neu{k} = \begin{pmatrix} \psi^0_k \\[.2em] \overline{\psi^0_k} 
\end{pmatrix} 
\qquad \text{with} \qquad 
\psi^0_k = N_{kl}
{\KL \tilde{B}^0, \tilde{W}^0, \tilde{H}^0_1, \tilde{H}^0_2 \KR^{\top}_l}
\end{align}
where $\psi^0_k$ denotes the two component Weyl spinor and $\neu{k}$
the four component Majorana spinor of the $k$th neutralino field.

\newcommand{\chapm}{\cham}
Concerning the renormalization of this sector, we implement two prescriptions that
differ in the treatment of the complex phases. The first prescription is
based on \citeres{dissTF,Stop2decay}, while the second one is based on
\citere{Fowler:2009ay,dissAF,bfmw,Bharucha:2012qr}. We will emphasize the points where the two schemes
deviate from each other. 

The following replacements of the parameters and the
fields are performed according to the multiplicative renormalization
procedure, which is formally identical for the two set-ups:
\begin{align}
M_1 \; &\to \; M_1 + \de M_1 ~, \\
M_2 \; &\to \; M_2 + \de M_2 ~, \\
\mu \; &\to \; \mu + \de \mu ~, \\
\OM \chapm{i} \; &\to \; \KKL \id + \edz \dZm{\chapm{}}^L \KKR_{ij}
                         \OM \chapm{j} \qquad (i,j = 1,2)~, \\
\OP \chapm{i} \; &\to \; \KKL \id + \edz \dZm{\chapm{}}^R \KKR_{ij}
                         \OP \chapm{j} \qquad (i,j = 1,2)~, \\
\OM \neu{k} \; &\to \; \KKL \id + \edz \dZm{\neu{}}^{} \KKR_{kl}
                       \OM \neu{l} \qquad (k,l = 1,2,3,4)~, \\
\OP \neu{k} \; &\to \; \KKL \id + \edz \dZm{\neu{}}^* \KKR_{kl}
                       \OP \neu{l} \qquad (k,l = 1,2,3,4)~.
\label{dZNeuR}
\end{align}
It should be noted that the parameter counterterms are complex
counterterms which each need two renormalization conditions to be fixed, (except for $\de M_2$, which in scheme~II is real).
The transformation matrices are not renormalized, so that, using the notation 
of replacing a matrix by its renormalized matrix and a counterterm matrix 
\begin{align}
\label{deX}
\matr{X} &\to \matr{X} + \de\matr{X} ~, \\
\matr{Y} &\to \matr{Y} + \de\matr{Y} ~
\end{align}
with
\begin{align}\label{deltaX}
\de\matr{X} &= 
  \begin{pmatrix} \de M_2 & \sqrt{2}\, \de(\MW \Sbe) \\
                  \sqrt{2}\, \de(\MW \Cb) & \de \mu
  \end{pmatrix}~, \\[.4em]
\de\matr{Y} &= 
  \begin{pmatrix} 
      \de M_1 & 0 & -\de(\MZ\sw\Cb) & \de(\MZ\sw\Sbe) \\
      0 & \de M_2 & \de(\MZ\cw\Cb) & -\de(\MZ\cw\Sbe) \\
      -\de(\MZ\sw\Cb) & \de(\MZ\cw\Cb) & 0 & -\de\mu  \\
      \de(\MZ\sw\Sbe) & -\de(\MZ\cw\Sbe) & -\de\mu & 0
  \end{pmatrix}~,
\end{align}
the replacements of the matrices $\matr{M}_{\cham{}}$ and $\matr{M}_{\neu{}}$
can be expressed as
\begin{align}
\matr{M}_{\cham{}} &\to \matr{M}_{\cham{}} + \de\matr{M}_{\cham{}}
   = \matr{M}_{\cham{}} + \matr{V}^* \de\matr{X}^\top \matr{U}^\dagger \\
\label{Mneu}
\matr{M}_{\neu{}} &\to \matr{M}_{\neu{}} + \de\matr{M}_{\neu{}}
   = \matr{M}_{\neu{}} + \matr{N}^* \de\matr{Y} \matr{N}^\dagger~. 
\end{align}

\noindent
For convenience, we decompose the self-energies into left- and right-handed
vector and scalar coefficients via
\begin{equation}
\KKL \Si_{\tilde \chi}^{}(p^2)\KKR_{nm} = \displaystyle{\not}p\, {\omega_-}
\KKL \Si_{\tilde \chi}^L(p^2) \KKR_{nm} 
+ \displaystyle{\not}p\, {\omega_+} \KKL \Si_{\tilde \chi}^R(p^2) \KKR_{nm}
+ {\omega_-} \KKL \Si_{\tilde \chi}^{SL}(p^2) \KKR_{nm} 
+ {\omega_+} \KKL \Si_{\tilde \chi}^{SR}(p^2) \KKR_{nm}~.
\label{decomposition}
\end{equation}
Now the coefficients of the renormalized self-energies are given by $(i,j = 1,2; k,l = 1,2,3,4)$
\begin{align}\label{renSEcha_L}
\KKL \hSi_{\chapm{}}^L(p^2) \KKR_{ij} &=
  \KKL \Si_{\chapm{}}^L(p^2) \KKR_{ij} 
+ \edz \KKL \dZm{\chapm{}}^L + \dZm{\chapm{}}^{L\dagger} \KKR_{ij}~, \\
\KKL \hSi_{\chapm{}}^R(p^2) \KKR_{ij} &=
  \KKL \Si_{\chapm{}}^R(p^2) \KKR_{ij} 
+ \edz \KKL \dZm{\chapm{}}^R + \dZm{\chapm{}}^{R\dagger} \KKR_{ij}~, \\
\KKL \hSi_{\chapm{}}^{SL}(p^2) \KKR_{ij} &=
  \KKL \Si_{\chapm{}}^{SL}(p^2) \KKR_{ij}
  - \KKL \edz \dZm{\chapm{}}^{R\dagger} \matr{M}_{\cham{}}
        + \edz \matr{M}_{\cham{}} \dZm{\chapm{}}^L
        + \de\matr{M}_{\cham{}} \KKR_{ij}~, \\
\KKL \hSi_{\chapm{}}^{SR}(p^2) \KKR_{ij} &=
  \KKL \Si_{\chapm{}}^{SR}(p^2) \KKR_{ij}
  - \KKL \edz \dZm{\chapm{}}^{L\dagger} \matr{M}_{\cham{}}^\dagger
       + \edz \matr{M}_{\cham{}}^\dagger \dZm{\chapm{}}^R
       + \de\matr{M}_{\cham{}}^\dagger \KKR_{ij}~, \\[.4em]
\KKL \hSi_{\neu{}}^L(p^2) \KKR_{kl} &=
  \KKL \Si_{\neu{}}^L(p^2) \KKR_{kl} 
+ \edz \KKL \dZm{\neu{}} + \dZm{\neu{}}^{\dagger} \KKR_{kl}~, \\
\KKL \hSi_{\neu{}}^R(p^2) \KKR_{kl} &=
  \KKL \Si_{\neu{}}^R(p^2) \KKR_{kl} 
+ \edz \KKL \dZm{\neu{}}^* + \dZm{\neu{}}^{\top} \KKR_{kl}~, \\
\KKL \hSi_{\neu{}}^{SL}(p^2) \KKR_{kl} &=
  \KKL \Si_{\neu{}}^{SL}(p^2) \KKR_{kl}
  - \KKL \edz \dZm{\neu{}}^{\top} \matr{M}_{\neu{}}
       + \edz \matr{M}_{\neu{}} \dZm{\neu{}}
        + \de\matr{M}_{\neu{}} \KKR_{kl}~, \\
\label{renSEneu_SR}
\KKL \hSi_{\neu{}}^{SR}(p^2) \KKR_{kl} &=
  \KKL \Si_{\neu{}}^{SR}(p^2) \KKR_{kl}
  - \KKL \edz \dZm{\neu{}}^{\dagger} \matr{M}_{\neu{}}^{\dagger} 
       + \edz \matr{M}_{\neu{}}^{\dagger} \dZm{\neu{}}^*
       + \de\matr{M}_{\neu{}}^\dagger \KKR_{kl}~.
\end{align}

Instead of choosing the three complex
parameters $M_1$, $M_2$ and $\mu$ to be independent parameters we impose on-shell
conditions for the two chargino masses and the mass of the lightest
neutralino (however, slightly differently in the two schemes; see below)  
and extract the expressions for the counterterms of  $M_1$,
$M_2$ and $\mu$, accordingly. It was shown in \citere{dissAF,Bharucha:2012qr,bfmw} 
that for numerically stable results, one bino-,  wino-, Higgsino-like
particle should be chosen on-shell.
Further, in a recent analysis~\cite{onshellCNmasses} it was emphasized that 
in the case of the renormalization of two chargino and one neutralino
mass always the most bino-like neutralino has to be renormalized in order
to ensure numerical stability.
In \citere{BaroII} the problem of large unphysical
contributions due to a non-bino like lightest neutralino is  
also discussed. In our numerical set-up, see
\refse{sec:numeval}, the lightest neutralino is always rather bino-like.  
On the other hand, it would be trivial to change
our prescription from the lightest neutralino being on-shell to any other
neutralino, ensuring that a neutralino with a large bino component is renormalized on-shell. 
In \citere{onshellCNmasses} it was also suggested that the
numerically most stable result is obtained via the renormalization of
one chargino and two neutralinos. 
However, in our approach, this choice would lead to IR divergences, 
since the chargino mass changes (from the tree-level mass to the 
one-loop pole mass) by a finite shift due to the renormalization procedure. 
Using the shifted mass for the external particles 
and the tree-level mass for internal particles results in IR divergences.
On the other hand, in general, inserting the shifted chargino mass 
internally yields UV divergences.
Consequently, we stick to our choice of imposing on-shell conditions 
for the two charginos and one neutralino.

As stated before, the numerical analysis is carried out
using two different renormalization schemes. 
It should be noted that the differences arise
in the renormalization of the parameters, but the expressions for the field
renormalization constants are identical. This means that only the renormalization
of the phases of the complex parameters  differ. Therefore the schemes
are identical in the real 
MSSM. 
We will briefly describe the two schemes in the following, making an attempt to highlight the differences.

%%%%%%%%%%%%%%%%%%%%%%%%%%%%%%%%% Scheme I  %%%%%%%%%%%%%%%%%%%%%%%%%%%%%%%%% 

\begin{itemize}
\item{{\bf Scheme~I}~\cite{LHCxC,Stop2decay}{\bf :}}
The on-shell conditions in this scheme read
\begin{align}
\label{mcha-OS_org}
\Bigl(\KKL \wtre \hSi_{\cham{}} (p)\KKR_{ii} 
     \cham{i}(p)\Bigr)\Big|_{p^2 = \mcha{i}^2} &= 0 \qquad  (i = 1,2)~, \\
\label{mneu-OS_org}
\quad \Bigl(\KKL\wtre \hSi_{\neu{}} (p)\KKR_{11} 
      \neu{1}(p)\Bigr)\Big|_{p^2 = \mneu{1}^2} &= 0~.
\end{align}
These conditions can be rewritten in terms of six equations defining six
real parameters or three complex ones,
\begin{align}
\label{mcha-OS}
\wtre \KKL \mcha{i} \KL \hSi_{\cham{}}^{L}(\mcha{i}^2)
                       +\hSi_{\cham{}}^{R}(\mcha{i}^2) \KR 
                       +\hSi_{\cham{}}^{SL}(\mcha{i}^2)
                       +\hSi_{\cham{}}^{SR}(\mcha{i}^2) 
      \KKR_{ii} &= 0~, \\
\label{mcha-OS-2}
\wtre \KKL \mcha{i} \KL \hSi_{\cham{}}^{L}(\mcha{i}^2)
                       -\hSi_{\cham{}}^{R}(\mcha{i}^2) \KR 
                       -\hSi_{\cham{}}^{SL}(\mcha{i}^2)
                       +\hSi_{\cham{}}^{SR}(\mcha{i}^2) 
      \KKR_{ii} &= 0~, \\
\label{mneu-OS}
\wtre \KKL \mneu{1} \KL \hSi_{\neu{}}^{L}(\mneu{1}^2)
                       +\hSi_{\neu{}}^{R}(\mneu{1}^2) \KR
                       +\hSi_{\neu{}}^{SL}(\mneu{1}^2)
                       +\hSi_{\neu{}}^{SR}(\mneu{1}^2) 
      \KKR_{11} &= 0~, \\
\label{mneu-OS-2}
\wtre \KKL \mneu{1} \KL \hSi_{\neu{}}^{L}(\mneu{1}^2)
                       -\hSi_{\neu{}}^{R}(\mneu{1}^2) \KR
                       -\hSi_{\neu{}}^{SL}(\mneu{1}^2)
                       +\hSi_{\neu{}}^{SR}(\mneu{1}^2) 
      \KKR_{11} &= 0~.
\end{align}
Equations~(\ref{mcha-OS-2}) and (\ref{mneu-OS-2}) are related to the
axial and axial-vector component of the renormalized self energy and therefore
the l.h.s.\ vanishes in the case of real couplings.
Therefore, in the rMSSM only Eqs.~(\ref{mcha-OS}) and (\ref{mneu-OS}) remain.
It should be noted that since the lightest neutralino is stable there are no absorptive
contributions from its self energy and $\wtre$ can be dropped from 
Eqs.~(\ref{mneu-OS_org},\ref{mneu-OS},\ref{mneu-OS-2}). 
We retain it here in order to allow for these on-shell conditions to be
generalized to other neutralinos.

For the further determination of the field renormalization constants,
applicable to both schemes, we also impose
\begin{align}
\lim_{p^2 \to \mcha{i}^2} 
\Bigl(\frac{(\pslash\, + \mcha{i}) \bigl[ \wtre \hSi_{\cham{}}(p)\bigr]_{ii}}
           {p^2 - \mcha{i}^2} \cham{i}(p)\Bigr) &= 0 \qquad (i = 1,2)~, 
\label{fieldRCcha}\\ 
 \lim_{p^2 \to {\mneu{k}^2}} 
\Bigl(\frac{(\pslash\, + {\mneu{k}})\bigl[ \wtre \hSi_{\neu{}}(p)\bigr]_{{kk}}}
           {p^2 - {\mneu{k}^2}}{\neu{k}}(p)\Bigr) &= 0 \qquad (k = 1,2,3,4)~,
\label{fieldRCneu}
\end{align} 
{
which, together with \refeqs{mcha-OS-2} and (\ref{mneu-OS-2}),
lead to the following set of equations (for $i = 1,2$;~$k = 1$)
}
\begin{align}
\non
\wtre\Bigl[\edz \KL \hSi_{\cham{}}^{L}(\mcha{i}^2)
                  + \hSi_{\cham{}}^{R}(\mcha{i}^2) \KR
                  + \mcha{i}^2 \KL \hSi_{\cham{}}^{L'}(\mcha{i}^2)
                  + \hSi_{\cham{}}^{R'}(\mcha{i}^2) \KR \quad\ &\\
                  + \mcha{i} \KL \hSi_{\cham{}}^{SL'}(\mcha{i}^2)
                  + \hSi_{\cham{}}^{SR'}(\mcha{i}^2) \KR 
      \Bigr]_{ii} &= 0~, \label{Zchadiag-OS}\\
\wtre \KKL \hSi_{\cham{}}^{L}(\mcha{i}^2)
                       -\hSi_{\cham{}}^{R}(\mcha{i}^2) 
      \KKR_{ii} &= 0~, \label{Zchadiag-OS-2}\\
\non
\wtre \Bigl[\edz \KL \hSi_{\neu{}}^{L}(\mneu{{1}}^2)
                   + \hSi_{\neu{}}^{R}(\mneu{{1}}^2) \KR 
                   + \mneu{{1}}^2 \KL \hSi_{\neu{}}^{L'}(\mneu{{1}}^2) 
                   + \hSi_{\neu{}}^{R'}(\mneu{{1}}^2) \KR \quad\ &\\
                   + \mneu{{1}} \KL \hSi_{\neu{}}^{SL'}(\mneu{{1}}^2)
                   + \hSi_{\neu{}}^{SR'}(\mneu{{1}}^2) \KR
      \Bigr]_{{11}} 
                   &= 0~, \label{Zneudiag-OS}\\
\wtre \KKL \hSi_{\neu{}}^{L}(\mneu{{1}}^2)
                       -\hSi_{\neu{}}^{R}(\mneu{{1}}^2)
      \KKR_{{11}} 
                 &= 0~,
\label{Zneudiag-OS-2}
\end{align}
where we have used the short-hand 
$\Si'(m^2) \equiv (\partial \Si/\partial p^2)|_{p^2 = m^2}$. 
It should be noted that \refeq{Zneudiag-OS-2} is already fulfilled due to the 
Majorana nature of the neutralinos. 

\smallskip
Inserting \refeqs{renSEcha_L}~--~(\ref{renSEneu_SR}) for the
renormalized self-energies in \refeqs{mcha-OS}~--~(\ref{mneu-OS-2})
and solving for 
$\KKL \de\matr{M}_{\cham{}}\KKR_{ii}$ and $\KKL \de\matr{M}_{\neu{}} \KKR_{11}$
results in
\begin{align}
\label{eq:redZcha}
\re\KKL \de\matr{M}_{\cham{}} \KKR_{ii} &= \frac{1}{2}
 \wtre \KKL \mcha{i} \KL \Si_{\chapm{}}^L(\mcha{i}^2) 
                        + \Si_{\chapm{}}^R(\mcha{i}^2) \KR
                        + \Si_{\chapm{}}^{SL}(\mcha{i}^2)
                        + \Si_{\chapm{}}^{SR}(\mcha{i}^2) \KKR_{ii}~, \\
\im\KKL\de\matr{M}_{\cham{}}\KKR_{ii} &= \frac{i}{2}
   \wtre \KKL \Si_{\chapm{}}^{SR}(\mcha{i}^2) 
            - \Si_{\chapm{}}^{SL}(\mcha{i}^2) \KKR_{ii}
 - \frac{1}{2} \mcha{i} \im \KKL \dZm{\cham{}}^L
           - \dZm{\cham{}}^R \KKR_{ii}~, \label{eq:imdZcha}\\[.4em]
\re\KKL\de\matr{M}_{\neu{}}\KKR_{11} &= \frac{1}{2}
   \wtre \KKL \mneu{1} \KL \Si_{\neu{}}^L(\mneu{1}^2)
                         + \Si_{\neu{}}^R(\mneu{1}^2) \KR
                         + \Si_{\neu{}}^{SL}(\mneu{1}^2)
                         + \Si_{\neu{}}^{SR}(\mneu{1}^2) \KKR_{11}~, \label{eq:redZneu}\\
\im\KKL\de\matr{M}_{\neu{}}\KKR_{11} &= \frac{i}{2}
   \wtre \KKL \Si_{\neu{}}^{SR}(\mneu{1}^2)
             -\Si_{\neu{}}^{SL}(\mneu{1}^2) \KKR_{11}
   -\mneu{1} \im \dZZm{\neu{}}_{11}~,\label{eq:imdZneu}
\end{align}
where we have used the 
relations~(\ref{Zchadiag-OS-2})~and~(\ref{Zneudiag-OS-2}).
Using \refeqs{deltaX}~--~(\ref{Mneu}), these conditions lead
to~\cite{dissTF,diplTF} 
\begin{align}
\de M_1 &= \frac{1}{(N_{11}^*)^2} \Big(
         \de\tmneu{1}
- N_{12}^{*2}\, \de M_2  +  2 N_{13}^* N_{14}^*\, \de\mu
   \non \\
&\qquad\qquad
   + 2 N_{11}^* \KKL N_{13}^*\, \de(\MZ\sw\Cb) - N_{14}^*\, \de(\MZ\sw\Sbe) \KKR
  \non \\
&\qquad\qquad
  - 2 N_{12}^* \KKL  N_{13}^*\, \de(\MZ\cw\Cb)
                      - N_{14}^*\, \de(\MZ\cw\Sbe) 
                  \KKR
        \Big)~, 
\label{deltaM1}\\[.2em]
% -------------------------------------------------------------------
\de M_2 &= \frac{1}{ 2 \KL U_{11}^* U_{22}^* V_{11}^* V_{22}^*
                         -U_{12}^* U_{21}^* V_{12}^* V_{21}^* \KR } \times
  \non \\
&\qquad\qquad\Big( 
        2 U_{22}^* V_{22}^*\,
       \de\tmcha{1}
       -2 U_{12}^* V_{12}^*\, 
       \de\tmcha{2}
       \non \\
&\qquad\qquad
+  (U_{12}^* U_{21}^* - U_{11}^* U_{22}^*) V_{12}^* V_{22}^*
          \,\de(\sqrt{2}\MW\Sbe) \non \\
&\qquad\qquad
+  U_{12}^* U_{22}^* (V_{12}^* V_{21}^* - V_{11}^* V_{22}^*)
          \,\de(\sqrt{2}\MW\Cb) \Big)~, 
\label{deltaM2}\\[.2em]
% -------------------------------------------------------------------
\de \mu &= \frac{1}{2 \KL U_{11}^* U_{22}^* V_{11}^* V_{22}^*
                         -U_{12}^* U_{21}^* V_{12}^* V_{21}^* \KR } \times
  \non \\
&\qquad\qquad \Big( 
        2 U_{11}^* V_{11}^*\, 
       \de\tmcha{2}
       -2 U_{21}^* V_{21}^*\, 
       \de\tmcha{1}
       \non \\
&\qquad\qquad +  (U_{12}^* U_{21}^* - U_{11}^* U_{22}^*) V_{11}^* V_{21}^*
          \,\de(\sqrt{2}\MW\Cb) \non \\
&\qquad\qquad +  U_{11}^* U_{21}^* (V_{12}^* V_{21}^* - V_{11}^* V_{22}^*)
          \,\de(\sqrt{2}\MW\Sbe) \Big)~,
\label{deltamu}
\end{align}
where, combining \refeqs{eq:redZcha} -- (\ref{eq:imdZcha})
and \refeqs{eq:redZneu} -- (\ref{eq:imdZneu}), 
we introduced the short-hand notation
\begin{align}
\de\tmneu{1} &= \frac{1}{2}\wtre \KKL {\mneu{1}}
                \KL \Si_{\neu{}}^L(\mneu{1}^2) + \Si_{\neu{}}^R(\mneu{1}^2) \KR
                + 2 \Si_{\neu{}}^{SL}(\mneu{1}^2) \KKR_{11}~,
\label{eq:dmneu}
\\
\de\tmcha{i} &=  \frac{1}{2}\wtre \KKL {\mcha{i}}
                \KL \Si_{\cha{}}^L(\mcha{i}^2) + \Si_{\cha{}}^R(\mcha{i}^2) \KR
                + 2 \Si_{\cha{}}^{SL}(\mcha{i}^2) \KKR_{ii}~.
\label{eq:dmcha}
\end{align}
Here we have used already \refeqs{eq:dZcha} and (\ref{eq:dZneu}), see below,
which fix the expressions for the imaginary parts of the chargino and neutralino 
field renormalization constants $\dZm{\chapm{}}^{L/R}$ and $\dZm{\neu{}}$, respectively.

%%%%%%%%%%%%%%%%%%%%%%%%%%%%%%%%% Scheme II %%%%%%%%%%%%%%%%%%%%%%%%%%%%%%%%%

\item{{\bf Scheme~II}~\cite{dissAF,Fowler:2009ay}{\bf :}}
Here on the other hand, it is the real part of the corrections to the self-energies
of the on-shell particles that are required to vanish,

\begin{align}
\label{mcha-OS_orgII}
\Bigl(\KKL \re \hSi_{\cham{}} (p)\KKR_{ii} 
     \cham{i}(p)\Bigr)\Big|_{p^2 = \mcha{i}^2} &= 0 \qquad  (i = 1,2)~, \\
\label{mneu-OS_orgII}
\quad \Bigl(\KKL\re \hSi_{\neu{}} (p)\KKR_{11} 
      \neu{1}(p)\Bigr)\Big|_{p^2 = \mneu{1}^2} &= 0~.
\end{align}
It should be noted that for the derivation of the field renormalization constants, in order to ensure 
that the on-shell propagator has only a scalar and vector part, 
we impose the additional conditions
\begin{align}
 \KKL\hSi_{\tilde \chi}^{L}(m_{\tilde{\chi}_{j}}^2)\KKR_{jj} &= 
 \KKL\hSi_{\tilde \chi}^{R}(m_{\tilde{\chi}_{j}}^2)\KKR_{jj}~, \\ 
 \KKL\hSi_{\tilde \chi}^{SL}(m_{\tilde{\chi}_{j}}^2)\KKR_{jj} &= 
 \KKL\hSi_{\tilde \chi}^{SR}(m_{\tilde{\chi}_{j}}^2)\KKR_{jj}~, 
\end{align}
with $\tilde{\chi}_j$ 
denoting either a chargino (with $j=1,2$) or a neutralino  (with $j=1,\ldots,4$).
The first equation, relating the vector coefficients,
is automatically satisfied in the cMSSM for both charginos and neutralinos. 
This means that the conditions (\ref{Zchadiag-OS-2}) and (\ref{Zneudiag-OS-2}) applied in scheme I, 
describing the axial and axial-vector components of the renormalized self-energy, are also satisfied in scheme II. 
It should be noted that here, however, we have dropped the $\wtre$\footnote{See the discussion on absorptive contributions at the end of this subsection.}. 
On expanding \refeqs{mcha-OS_orgII} and (\ref{mneu-OS_orgII}), in analogy to \refeqs{mcha-OS} and (\ref{mneu-OS}), we find
\begin{align}
\label{mcha-OSII}
\re \KKL \mcha{i} \KL \hSi_{\cham{}}^{L}(\mcha{i}^2)
                       +\hSi_{\cham{}}^{R}(\mcha{i}^2) \KR 
                       +\hSi_{\cham{}}^{SL}(\mcha{i}^2)
                       +\hSi_{\cham{}}^{SR}(\mcha{i}^2) 
      \KKR_{ii} &= 0~, \\
%%%
\label{mneu-OSII}
\re \KKL \mneu{1} \KL \hSi_{\neu{}}^{L}(\mneu{1}^2)
                       +\hSi_{\neu{}}^{R}(\mneu{1}^2) \KR
                       +\hSi_{\neu{}}^{SL}(\mneu{1}^2)
                       +\hSi_{\neu{}}^{SR}(\mneu{1}^2) 
      \KKR_{11} &= 0~.
\end{align}

We are left with three on-shell conditions for scheme~II, \refeqs{mcha-OSII} and (\ref{mneu-OSII}). 
Therefore one can only fix the renormalization constants%
\footnote{Here we adopt the notation $\deII$ 
for scheme~II 
to distinguish the renormalization constants from those in scheme~I.} 
$\deII |M_1|$, $\deII M_2$ (note that $M_2$ is chosen to be real) and $\deII |\mu|$, but not $\deII\phiMe$ or $\deII\phi_{\mu}$, 
where $\deII M_1=\deII |M_1|e^{i\phiMe}$ and $\deII \mu=\deII |\mu|e^{i\phi_{\mu}}$.
This, however, is not a problem, as it turns out that the phases of $M_1$ and $\mu$ are UV finite at one-loop (see e.g.~\citere{dissAF}), 
and need not be renormalized at all, 
i.e.\ they can be set to zero,
 $\deII\phiMe=\deII\phi_{\mu}=0$. 
It should be noted that $\deII |\mu|$, $\deII |M_1|$, $\deII M_2$ are related to $\deII\matr{M}_{\cham{}}$ and $\deII\matr{M}_{\neu{}}$ via \refeqs{deltaX}~--~(\ref{Mneu}). 
Expressions for these renormalization constants can then easily be obtained by inserting \refeqs{renSEcha_L}~--~(\ref{renSEneu_SR}) for the
renormalized self-energies in \refeqs{mcha-OSII}~--~(\ref{mneu-OSII}), resulting in (for details see \citere{Fowler:2009ay,dissAF,Bharucha:2012qr,bfmw})
\\
\begin{align}
\non\deII |M_1| &= -\frac{1}{\re\,(e^{-i\phiMe}\,N_{11}^2) S} \times\non\\
\non &\quad \Big( 
            \KKL \re\,(U_{11} V_{11})\re\,(e^{-i\phi_{\mu}}U_{22} V_{22})-\re\,(e^{-i\phi_{\mu}}U_{12} V_{12})\re\,(U_{21} V_{21}) \KKR N_{1}\qquad\\
\non &\quad +\KKL 2\re\,(e^{-i\phi_{\mu}}N_{13} N_{14}) \re\,(U_{11} V_{11})+ \re\,(N_{12}^2)\re\,(e^{-i\phi_{\mu}}U_{12} V_{12})\KKR C_{2}\\
&\quad -\KKL\re\,( N_{12}^2)\re\,(e^{-i\phi_{\mu}}U_{22} V_{22}) +2\re\,(e^{-i\phi_{\mu}}N_{13} N_{14})\re\,(U_{21} V_{21})\KKR C_{1} \Big) 
\label{deltam1II}~,\\[.2em]
\deII M_2 &= \frac{1}{S}\KKL\re\,(e^{-i\phi_{\mu}}U_{12} V_{12})C_{2}-\re\,(e^{-i\phi_{\mu}}U_{22} V_{22}) C_{1}\KKR \label{deltam2II}~,\\[.2em]
\deII |\mu| &= -\frac{1}{S}\KKL\re\,(U_{11} V_{11})C_{2}-\re\,(U_{21} V_{21})C_{1}\KKR\label{deltamuII}~,
\end{align}
where we use the abbreviations
\begin{align}
C_{i} &\equiv \mathrm{Re} \KKL\mcha{i}[\Si^L_{\chapm{}}(\mcha{i}^2)+ \Si^R_{\chapm{}}(\mcha{i}^2)]+
      \Sigma^{SL}_{\chapm{}}(\mcha{i}^2)+\Sigma^{SR}_{\chapm{}}(\mcha{i}^2)\KKR_{ii}\non\\
&\quad -2\de(\MW \Sbe) \re\,(U_{i2}V_{i1})- 2\de(\MW \Sbe)\re\,(U_{i1}V_{i2}),\\[.2em]
\non N_{i} &\equiv \mathrm{Re} \KKL\mneu{i}[\Si^L_{\neu{}}(\mneu{i}^2)+\Si^R_{\neu{}}(\mneu{i}^2)] +\Si^{SL}_{\neu{}}(\mneu{i}^2)+ \Si^{SR}_{\neu{}}(\mneu{i}^2)\KKR_{ii} \non\\
\non &\quad + 4\de(\MZ\sw\Cb)\re\,(N_{i1}N_{i3})- 4\de(\MZ\cw\Cb)\re\,(N_{i2}N_{i3})\\
&\quad - 4\de(\MZ\sw\Sbe)\re\,(N_{i1} N_{i4}) + 4\de(\MZ\cw\Sbe)\re\,(N_{i2}N_{i4})~,\\[.2em]
\label{eqn:CiNishorthand}S &\equiv 2\BL\re\,(U_{21} V_{21}) \re\,(e^{-i\phi_{\mu}}U_{12} V_{12})-\re\,(U_{11} V_{11})\re\,(e^{-i\phi_{\mu}}U_{22} V_{22})\BR~.
\end{align}
$\deII\matr{M}_{\cham{}}$ and $\deII\matr{M}_{\neu{}}$ are simply obtained by the replacements $\de M_1$, $\de M_2$, and $\de \mu$ by, 
respectively, $\deII M_1$, $\deII M_2$, and $\deII \mu$ in \refeqs{deX}~--~(\ref{Mneu}).
\end{itemize}
\noindent The following discussion of the 
neutralino mass shifts and the 
field renormalization is applicable to both schemes.
\medskip 

Since the chargino masses $\mcha{1}, \mcha{2}$ and the lightest
neutralino mass $\mneu{1}$ have been chosen as independent parameters,
the one-loop masses of the heavier neutralinos $\neu{i}$
($i$ = 2,3,4) 
are obtained from the tree-level ones via the shifts
\begin{align}
\De \mneu{i} = -\frac{1}{2}\re \KKKL \mneu{i} \KL \hSi_{\neu{i}}^{L}(\mneu{i}^2)
                       +\hSi_{\neu{i}}^{R}(\mneu{i}^2) \KR
                       +\hSi_{\neu{i}}^{SL}(\mneu{i}^2)
                       +\hSi_{\neu{i}}^{SR}(\mneu{i}^2) 
     \KKKR 
           ~.
\label{Deltamneu}
\end{align}

Where necessary we distinguish the tree-level mass  $\mneu{i}$ from the on-shell mass,
\begin{align}
\mneuOS{i} &= \mneu{i} + \De\mneu{i}~.
\label{mneuOS}
\end{align}

We use $\mneuOS{i}$ for all externally appearing neutralino masses, 
which includes the (on-shell) momentum in the employed neutralino
self-energies. 
In order to yield UV-finite results we use the tree-level values $\mneu{i}$ 
for all internally appearing neutralino masses in loop calculations.

\smallskip
\refeqs{Zchadiag-OS} and (\ref{Zneudiag-OS}) define the real part of
the diagonal field renormalization constants of the chargino fields and of
the lightest neutralino field. 
By extending \refeq{Zneudiag-OS}
to apply also for $k = 2,3,4$, we can generalize the result for the
diagonal field renormalization constants of the lightest neutralino to the other  
neutralino fields.

The imaginary parts of the diagonal field renormalization 
constants are still undefined. 
However, these can be obtained using
\refeqs{eq:imdZcha} and (\ref{eq:imdZneu}), 
where the latter is generalized to include $k = 2,3,4$.
Now in scheme~I (or scheme~II), for the
charginos and the lightest neutralino, \refeqs{eq:imdZcha} and
(\ref{eq:imdZneu}) define
the imaginary parts of $\KKL\de\matr{M}_{\cha{}}\KKR_{ii}$ (or 
$\KKL\deII\matr{M}_{\cha{}}\KKR_{ii}$)
($i = 1,2$), and $\KKL\de\matr{M}_{\neu{}}\KKR_{11}$ (or 
$\KKL\deII\matr{M}_{\neu{}}\KKR_{11}$) in terms of the imaginary part of the field
renormalization constants. Therefore these are simply set to zero (see below \refeqs{eq:dZcha} 
and \eqref{eq:dZneu}), which is possible as all divergences are absorbed by other counterterms.

\smallskip
The off-diagonal field renormalization constants are fixed by the
condition that 
\begin{align}
\Bigl(\KKL \wtre \hSi_{\cham{}} (p)\KKR_{ij} 
    \cham{j}(p)\Bigr)\Big|_{p^2 = \mcha{j}^2} &= 0 \qquad (i,j = 1,2)~,
\\
\Bigl(\KKL\wtre \hSi_{\neu{}} (p)\KKR_{kl} 
    \neu{l}(p)\Bigr)\Big|_{p^2 = \mneu{l}^2} &= 0 \qquad (k,l = 1,2,3,4)~.
\end{align}
Finally, this yields for the field renormalization
constants~\cite{dissTF} (where we now make the correct dependence on
tree-level and on-shell masses explicit), 
\begin{align}
\re \KKL \dZm{\chapm{}}^{L/R} \KKR_{ii} &=
        - \wtre \Big[ \Si_{\chapm{}}^{L/R}(\mcha{i}^2) 
\label{dZcha_iiRe}\\
&\qquad + \mcha{i}^2 \KL \Si_{\chapm{}}^{L'}(\mcha{i}^2)
                       + \Si_{\chapm{}}^{R'}(\mcha{i}^2) \KR
        + \mcha{i} \KL \Si_{\chapm{}}^{SL'}(\mcha{i}^2)
                    +  \Si_{\chapm{}}^{SR'}(\mcha{i}^2) \KR
      \Big]_{ii}~, \non \\
\im \KKL \dZm{\chapm{}}^{L/R} \KKR_{ii} &= 
       \pm \frac{1}{\mcha{i}} \KKL \frac{i}{2} 
    \wtre \KKKL \Si_{\chapm{}}^{SR}(\mcha{i}^2) 
              - \Si_{\chapm{}}^{SL}(\mcha{i}^2) \KKKR
    - \im \de\matr{M}_{\cham{}} \KKR_{ii} \stackrel{\mathrm{S_I}}{:=} 0~,\label{eq:dZcha} 
\\
\KKL \dZm{\chapm{}}^{L/R} \KKR_{ij} &= \frac{2}{\mcha{i}^2 - \mcha{j}^2} 
  \wtre \Big[ \mcha{j}^2 \Si_{\chapm{}}^{L/R}(\mcha{j}^2) 
             +\mcha{i} \mcha{j} \Si_{\chapm{}}^{R/L}(\mcha{j}^2)
\label{dZcha_ij} \\
&\qquad + \mcha{i} \Si_{\chapm{}}^{SL/SR}(\mcha{j}^2)
        + \mcha{j} \Si_{\chapm{}}^{SR/SL}(\mcha{j}^2)
        - \mcha{i/j} \de\matr{M}_{\cham{}}
        - \mcha{j/i} \de\matr{M}_{\cham{}}^{\dagger} \Big]_{ij}~, \non \\
\re \KKL \dZm{\neu{}}^{} \KKR_{kk} &= 
  -\wtre \Big[  \Si_{\neu{}}^L(\mneuOS{k}^2) \\
&\qquad  + \mneu{k}^2 \KL \Si_{\neu{}}^{L'}(\mneuOS{k}^2)
                         +\Si_{\neu{}}^{R'}(\mneuOS{k}^2) \KR
         + \mneu{k}   \KL \Si_{\neu{}}^{SL'}(\mneuOS{k}^2)
                         +\Si_{\neu{}}^{SR'}(\mneuOS{k}^2) \KR
       \Big]_{kk}~, \non \\
\im \KKL \dZm{\neu{}}^{} \KKR_{kk} &= 
         \frac{1}{\mneu{k}} \KKL \frac{i}{2} 
    \wtre \KKKL \Si_{\neu{}}^{SR}(\mneuOS{k}^2) \label{eq:dZneu}
              - \Si_{\neu{}}^{SL}(\mneuOS{k}^2) \KKKR
    - \im \de\matr{M}_{\neu{}} \KKR_{kk} \stackrel{k=1,\,\,\mathrm{S_I}}{:=} 0~, \\
\KKL \dZm{\neu{}}^{} \KKR_{kl} &= \frac{2}{\mneu{k}^2 - \mneu{l}^2}
 \wtre \Big[ \mneu{l}^2 \Si_{\neu{}}^L(\mneuOS{l}^2) 
            +\mneu{k}\mneu{l} \Si_{\neu{}}^R(\mneuOS{l}^2) \non \\
&\qquad + \mneu{k} \Si_{\neu{}}^{SL}(\mneuOS{l}^2)
        + \mneu{l} \Si_{\neu{}}^{SR}(\mneuOS{l}^2)
        - \mneu{k} \de\matr{M}_{\neu{}} 
        - \mneu{l} \de\matr{M}_{\neu{}}^\dagger \Big]_{kl}~, 
\label{eq:dZneu_ij}
\end{align}
within scheme~I ($\mathrm{S_I}$). 
Making the replacements $\de\matr{M}_{\neu{}} \to \deII\matr{M}_{\neu{}}$
and $\de\matr{M}_{\cham{}} \to \deII\matr{M}_{\cham{}}$ for scheme~II, \refeqs{eq:dZcha} and (\ref{eq:dZneu}) no longer vanish.

\medskip
Contributions to the partial
decay widths can arise from the product of the imaginary parts of the
loop-functions (absorptive contributions) of the self-energy type
contributions in the external legs and the imaginary parts of
complex couplings entering the decay vertex or the self-energies. 
It is possible to combine these additional contributions 
with the field renormalization constants in a single ``$Z$~factor'',
$\mathcal Z$, see e.g.~\citere{LHCxC,Stop2decay} and references therein. 
In our notation they read (unbarred for an incoming neutralino or a
negative chargino, barred for an outgoing neutralino or negative
chargino, and not making the difference between scheme~I and~II explicit), 
\begin{align}
\KKL \dcZ{\chapm{}}^{L/R} \KKR_{ii} &=
        - \Big[ \Si_{\chapm{}}^{L/R}(\mcha{i}^2) \\
&\qquad + \mcha{i}^2 \KL \Si_{\chapm{}}^{L'}(\mcha{i}^2)
                       + \Si_{\chapm{}}^{R'}(\mcha{i}^2) \KR
        + \mcha{i} \KL \Si_{\chapm{}}^{SL'}(\mcha{i}^2)
                    +  \Si_{\chapm{}}^{SR'}(\mcha{i}^2) \KR \Big]_{ii}~\non \\
&\qquad \pm \ed{2 \mcha{i}} \KKL 
                      \Si_{\chapm{}}^{SL}(\mcha{i}^2)
                    - \Si_{\chapm{}}^{SR}(\mcha{i}^2)
                    - \de\matr{M}_{\cham{}}
                    + \de\matr{M}_{\cham{}}^{*} \KKR_{ii}~, \non \\
\KKL \dcZ{\chapm{}}^{L/R} \KKR_{ij} &= \frac{2}{\mcha{i}^2 - \mcha{j}^2} 
        \Big[ \mcha{j}^2 \Si_{\chapm{}}^{L/R}(\mcha{j}^2) 
             +\mcha{i} \mcha{j} \Si_{\chapm{}}^{R/L}(\mcha{j}^2) \\
&\qquad + \mcha{i} \Si_{\chapm{}}^{SL/SR}(\mcha{j}^2)
        + \mcha{j} \Si_{\chapm{}}^{SR/SL}(\mcha{j}^2)
        - \mcha{i/j} \de\matr{M}_{\cham{}}
        - \mcha{j/i} \de\matr{M}_{\cham{}}^\dagger \Big]_{ij}~, \non \\
\KKL \dcZ{\neu{}}^{L/R} \KKR_{kk} &= 
  - \Big[ \Si_{\neu{}}^{L/R}(\mneuOS{k}^2) \\
&\qquad     + \mneu{k}^2 \KL \Si_{\neu{}}^{L'}(\mneuOS{k}^2)
                            +\Si_{\neu{}}^{R'}(\mneuOS{k}^2) \KR
            + \mneu{k} \KL \Si_{\neu{}}^{SL'}(\mneuOS{k}^2)
                           +\Si_{\neu{}}^{SR'}(\mneuOS{k}^2) \KR
       \Big]_{kk} \non \\
&\qquad \pm \ed{2 \mneu{k}} \KKL \Si_{\neu{}}^{SL}(\mneuOS{k}^2)
            - \Si_{\neu{}}^{SR}(\mneuOS{k}^2)
            - \de\matr{M}_{\neu{}} 
            + \de\matr{M}_{\neu{}}^{*} \KKR_{kk}~, \non \\
\KKL \dcZ{\neu{}}^{L/R} \KKR_{kl} &= \frac{2}{\mneu{k}^2 - \mneu{l}^2}
  \Big[ \mneu{l}^2 \Si_{\neu{}}^{L/R}(\mneuOS{l}^2) 
            + \mneu{k}\mneu{l} \Si_{\neu{}}^{R/L}(\mneuOS{l}^2) \\
&\qquad + \mneu{k} \Si_{\neu{}}^{SL/SR}(\mneuOS{l}^2)
        + \mneu{l} \Si_{\neu{}}^{SR/SL}(\mneuOS{l}^2)
        - \mneu{k/l} \de\matr{M}_{\neu{}} 
        - \mneu{l/k} \de\matr{M}_{\neu{}}^{\dagger} \Big]_{kl}~, \non
\end{align}
\begin{align}
\KKL \dbcZ{\chapm{}}^{L/R} \KKR_{ii} &=
        - \Big[ \Si_{\chapm{}}^{L/R}(\mcha{i}^2) \\
&\qquad + \mcha{i}^2 \KL \Si_{\chapm{}}^{L'}(\mcha{i}^2)
                       + \Si_{\chapm{}}^{R'}(\mcha{i}^2) \KR
        + \mcha{i} \KL \Si_{\chapm{}}^{SL'}(\mcha{i}^2)
                    +  \Si_{\chapm{}}^{SR'}(\mcha{i}^2) \KR \Big]_{ii}~\non \\
&\qquad \mp \ed{2 \mcha{i}} \KKL 
                      \Si_{\chapm{}}^{SL}(\mcha{i}^2)
                    - \Si_{\chapm{}}^{SR}(\mcha{i}^2)
                      - \de\matr{M}_{\cham{}}
                      + \de\matr{M}_{\cham{}}^{*} \KKR_{ii}~, \non \\
\KKL \dbcZ{\chapm{}}^{L/R} \KKR_{ij} &= \frac{2}{\mcha{j}^2 - \mcha{i}^2} 
        \Big[ \mcha{i}^2 \Si_{\chapm{}}^{L/R}(\mcha{i}^2) 
             +\mcha{i} \mcha{j} \Si_{\chapm{}}^{R/L}(\mcha{i}^2) \\
&\qquad + \mcha{i} \Si_{\chapm{}}^{SL/SR}(\mcha{i}^2)
       + \mcha{j} \Si_{\chapm{}}^{SR/SL}(\mcha{i}^2)
       - \mcha{i/j} \de\matr{M}_{\cham{}}
       - \mcha{j/i} \de\matr{M}_{\cham{}}^\dagger \Big]_{ij}~, \non \\
\label{diagneu}
\KKL \dbcZ{\neu{}}^{L/R} \KKR_{kk} &= 
 \KKL \dcZ{\neu{}}^{R/L} \KKR_{kk}~,  \\
\label{offdiagneu}
\KKL \dbcZ{\neu{}}^{L/R} \KKR_{kl} &= 
 \KKL \dcZ{\neu{}}^{R/L} \KKR_{lk}~, 
\end{align}
within scheme~I, and with $\de\matr{M}_{\neu{}} \to \deII\matr{M}_{\neu{}}$
and $\de\matr{M}_{\cham{}} \to \deII\matr{M}_{\cham{}}$ for scheme~II.
The char\-gi\-no/neu\-tra\-li\-no $\cZ$ factors obey
$\wtre\,\dbcZ{\tilde{\chi}}^{L/R} = 
[\wtre\,\dcZ{\tilde{\chi}}^{L/R}]^\dagger =
[\dZm{\tilde{\chi}}^{L/R}]^\dagger$, 
which is exactly the case without absorptive contributions.
The Eqs.~\eqref{diagneu} and \eqref{offdiagneu} hold due to the Majorana 
character of the neutralinos. We will use these $\cZ$ factors rather than 
the field renormalization constants defined in \refeqs{dZcha_iiRe} to (\ref{eq:dZneu_ij}) in
the following numerical analysis. 

\medskip
Special care has to be taken in the regions of the cMSSM parameter space where 
the gaugino-Higgsino mixing in the chargino sector is maximal,
i.e.\ where {$|\mu| \approx M_2$}. 
Here 
     $\de M_2$ (see \refeq{deltaM2}) and $\de \mu$ (see \refeq{deltamu}) 
diverge as $(U^*_{11}U^*_{22}V^*_{11}V^*_{22} - U^*_{12}U^*_{21}V^*_{12}V^*_{21})^{-1}$ for scheme I 
or $2\BL\re\,(U_{21} V_{21}) \re\,(e^{-i\phi_{\mu}}U_{12} V_{12})-\re\,(U_{11} V_{11})\re\,(e^{-i\phi_{\mu}}U_{22} V_{22})\BR$ for scheme II,
and the loop calculation does not yield a reliable result%
\footnote{Similar divergences appearing in the on-shell
renormalization in the sbottom sector, occurring for ``maximal sbottom
mixing'', have been observed and discussed in 
\citeres{SbotRen,Stop2decay,LHCxC}.}.%
~It should be noted that the {singularity} arises in both schemes  for $|\mu|=M_2$.
~These kind of divergences were also discussed in
\citere{dissAF,onshellCNmasses,bfmw}.
\medskip

Our two renormalization schemes differ in the treatment of complex contributions in the chargino and neutralino sector.
The first difference is that in scheme~I, in the derivation of the renormalization conditions $\de M_2$ is  
allowed to be complex. Therefore six real conditions must be imposed in order to
renormalize the mass matrices $\matr{X}$ and $\matr{Y}$, \refeqs{eq:X} and (\ref{eq:Y}),
as opposed to the five in scheme~II. 
In addition, in scheme~I, the renormalization of the phases is obtained 
by imposing that the imaginary parts of the relevant diagonal
renormalization constants \refeqs{eq:dZcha} and (\ref{eq:dZneu}) vanish.
In scheme~II the phases are not renormalized as they are found to be UV finite.

Scheme~I is based on the idea that the absorptive contributions should not enter the renormalization procedure.
Therefore it requires that the absorptive contributions are not included in the on-shell mass renormalization 
conditions \refeqs{mcha-OS} and (\ref{mneu-OS}). 
While the phases of the complex counterterms of the mass matrix parameters are 
also found to be UV finite, no condition is imposed on them.
Notice that it is always possible to rephase the  parameters of the cha\-gi\-no/neu\-tra\-li\-no sector since only the relative phases are physically relevant.
Scheme~II is supported by the argument that the phases are not renormalized, i.e.~they are the same at tree level and loop level. 
Therefore it is clear which of the fundamental parameters are chosen to be real and which complex, 
i.e. in scheme~II $M_2$ is fixed to be real while in scheme~I its counterterm is allowed to be complex.
As stated earlier, the results in both schemes should agree in the case of real parameters. In the complex case, 
we expect the differences to be very small, as they are of higher order.

%%%%%%%%%%%%%%%%%%%%%%%%%%%%%%%%%%%%%%%%%%%%%%%%%%%%%%%%%%%%%%%%%%%%%%%%%%%%%%%

\subsection{The lepton/slepton sector of the cMSSM}
\label{sec:slepton}
For the discussion of the one-loop contributions to the decay channels 
in \refeqs{NSll} and (\ref{NSnn}) a description of the scalar lepton 
($\Sl$) and neutrino ($\Sn$) sector as well as their fermionic SM
partners is needed (we assume no generation 
mixing and discuss the case for one generation only).
The bilinear part of the $\Sl$ and $\Sn$ Lagrangian,
\begin{align}
\cL_{\Sl/\Sn}^{\text{mass}} &= - \begin{pmatrix}
\sll^{\dagger}, \slr^{\dagger} \end{pmatrix}
\matr{M}_{\Sl} \begin{pmatrix} \sll \\ \slr
\end{pmatrix} 
- \begin{pmatrix} \Sn^{\dagger} \end{pmatrix}
\matr{M}_{\Sn}\begin{pmatrix} \Sn \end{pmatrix}~,
\end{align}
contains the slepton and sneutrino mass matrices
$\matr{M}_{\Sl}$ and $\matr{M}_{\Sn}$,
given by 
\begin{align}\label{Sfermionmassenmatrix}
\matr{M}_{\Sl} &= \begin{pmatrix} 
\MslL^2 + \ml^2 + M_Z^2 c_{2 \beta} (I_\ell^3 - Q_\ell \sw^2) & 
 \ml \Xl^* \\[.2em]
 \ml \Xl &
\MslR^2 + \ml^2 +M_Z^2 c_{2 \beta} Q_\ell \sw^2
\end{pmatrix}~, \\[.5em]
\matr{M}_{\Sn} &= \MslL^2 
+ I_\nu^3 c_{2\be} M_Z^2
\end{align}
with
\begin{align}
\Xl &= \Al - \mu^* \tb~.
\end{align}
$\MslL$ and $\MslR$ are the soft SUSY-breaking mass
parameters, where $\MslL$ is equal for all members of an
$SU(2)_L$ doublet.
$\ml$ and $Q_{\ell}$ are, respectively, the mass and the charge of the
corresponding lepton, $I_{\ell/\nu}^3$ denotes the isospin of $\ell/\nu$,  
and $A_\ell$ is the trilinear soft-breaking parameter.
We use the
short-hand notations $c_{x} = \cos(x)$, $s_x = \sin(x)$.
The mass matrix~$\matr{M}_{\Sl}$ can be diagonalized with the help of a unitary
transformation ${\matr{U}}_{\Sl}$,
\begin{align}\label{transformationkompl}
\matr{D}_{\Sl} &= 
\matr{U}_{\Sl}\, \matr{M}_{\Sl} \, {\matr{U}}_{\Sl}^\dagger = 
\begin{pmatrix} \msle^2 & 0 \\ 0 & \mslz^2 \end{pmatrix}~, \qquad
{\matr{U}}_{\Sl}= 
\begin{pmatrix} U_{\Sl_{11}} & U_{\Sl_{12}} \\  
                U_{\Sl_{21}} & U_{\Sl_{22}} \end{pmatrix}
~.
\end{align}
The mass eigenvalues depend only on $|\Xl|$. 
The scalar lepton masses will always be mass ordered, i.e.\
$m_{\sle} \le m_{\slz}$:
\begin{align}
\label{MSlep}
m_{\Sl_{1,2}}^2 &= \edz \KL M_{\Sl_L}^2 + M_{\Sl_R}^2 \KR
       + \ml^2 + \edz I_\ell^3 c_{2\be} \MZ^2 \\
&\quad \mp \edz \sqrt{\KKL \MslL^2 - \MslR^2
       + \MZ^2 c_{2\be} (I_\ell^3 - 2 Q_\ell \sw^2) \KKR^2 + 4 \ml^2 |\Xl|^2}~, 
\non\\[.5em]
m_{\Sn}^2 &= \MslL^2 + I_\nu^3 c_{2\be} M_Z^2~.
\end{align}

A detailed description of the renormalization of this sector can be found in
\citeres{Stop2decay,LHCxC,Stau2decay}. Here we just briefly review the restoration of the $SU(2)_L$ relation 
for the renormalized slepton mass parameters.
Since our on-shell approach results in an independent renormalization of the 
charged sleptons and of the scalar neutrino, 
we need to restore the $SU(2)_L$ relation at one-loop to avoid problems concerning UV- and IR-finiteness as
discussed in detail in \citere{SbotRen}.
This is achieved via a shift in the $M_{\Sl_L}$ parameter entering 
the $\Sl$~mass matrix (see also \citeres{stopsbot_phi_als,dr2lA,Stop2decay,LHCxC}).
Requiring the $SU(2)_L$ relation
to be valid at the loop level induces the following shift in 
$M^2_{\Sl_L}(\Sl)$ 
\begin{align}
M_{\Sl_L}^2(\Sl) = M_{\Sl_L}^2(\Sn) 
   + \de M_{\Sl_L}^2(\Sn) - \de M_{\Sl_L}^2(\Sl)
\label{MSnushift}
\end{align}
with
\begin{align}
\de M_{\Sl_L}^2(\Sl) &= |U_{\Sl_{11}}|^2 \de\msle^2
   + |U_{\Sl_{12}}|^2 \de\mslz^2
   - U_{\Sl_{22}} U_{\Sl_{12}}^* \de Y_\ell
   - U_{\Sl_{12}} U_{\Sl_{22}}^* \de Y_\ell^* - 2 \ml \de\ml \non \\
&\quad  + \MZ^2\, c_{2\be}\, Q_\ell\, \de \sw^2 
        - (I_\ell^3 - Q_\ell \sw^2) (c_{2\be}\, \de \MZ^2 + \MZ^2\, \de c_{2\be})~, 
\\[.5em]
\de M_{\Sl_L}^2(\Sn) &= \de\msn^2 
   - I_\nu^3(c_{2\be}\, \de \MZ^2 + \MZ^2\, \de c_{2\be})~.
\label{MSnushift-detail}
\end{align}
Such shifts however mean that both slepton masses are no longer on-shell.
An additional shift in $M_{\Sl_R}$ restores at least one slepton mass to be on-shell. 
\begin{align}
M_{\Sl_R}^2(\Sl_i) = \frac{\ml^2\, |\Al^* - \mu \tb|^2}
  {M_{\Sl_L}^2(\tilde{\ell}) + \ml^2 
   + \MZ^2\, c_{2\be} (I_\ell^3 - Q_\ell \sw^2) - \msli^2} 
  - \ml^2 - \MZ^2\, c_{2\be}\, Q_\ell\, \sw^2+ \msli^2~.
\label{backshift}
\end{align}
A ``natural'' choice is to preserve the character of the sleptons in the
renormalization process, and this additional shift relates the mass of the chosen slepton to
the slepton parameter $M_{\Sl_R}$.
As $\msle \le \mslz$ (see above) and $M_{\Sl_L}^2 < M_{\Sl_R}^2$ for both scenarios considered later 
(see \refta{tab:para}),
we choose to insert $\mslz$ into \refeq{backshift} and recover its original value from the
re-diagonalization after applying this shift.

The renormalization of the quark/squark sector is described in detail in \citere{Stop2decay}.
As for the slepton sector, the restoration of the  $SU(2)_L$ relation in the squark sector
leads to shifts analogous to \refeqs{MSnushift} and (\ref{backshift}) 
in the left and right squark mass parameters, respectively, $\MsqL$ and $\MsqR$.
In our subsequent numerical analysis these shifts have been included for the results calculated in scheme I,
but their effect is found to be negligible for our choice of parameters.
%%%%%%%%%%%%%%%%%%%%%%%%%%%%%%%%%%%%%%%%%%%%%%%%%%%%%%%%%%%%%%%%%%%%%%%%%%%%%%%
\subsection{The Higgs boson sector of the cMSSM}
\label{sec:higgs}
The two Higgs doublets of the cMSSM are decomposed in the following way,
\begin{align}
\label{eq:higgsdoublets}
\cHe = \begin{pmatrix} H_{11} \\ H_{12} \end{pmatrix} &=
\begin{pmatrix} v_1 + \tfrac{1}{\sqrt{2}} (\phi_1-i \chi_1) \\
  -\phi^-_1 \end{pmatrix}, \notag \\ 
\cHz = \begin{pmatrix} H_{21} \\ H_{22} \end{pmatrix} &= e^{i \xi}
\begin{pmatrix} \phi^+_2 \\ v_2 + \tfrac{1}{\sqrt{2}} (\phi_2+i
  \chi_2) \end{pmatrix}. 
\end{align}
Besides the vacuum expectation values $v_1$ and $v_2$, in 
\refeq{eq:higgsdoublets} a possible new phase $\xi$ between the two Higgs doublets is introduced. 
After a rotation to the physical fields 
one obtains for the terms linear and bilinear in the fields,
\begin{align}
\label{VHiggs}
\VHiggs &=  \ldots + T_{h}\, h + T_{H}\, H + T_{A}\, A \non \\ 
&\quad  - \edz \begin{pmatrix} h, H, A, G 
        \end{pmatrix} 
\matr{M}_{hHAG}^{\rm diag}
\begin{pmatrix} h \\ H \\ A \\ G  \end{pmatrix} + 
\begin{pmatrix} H^{+}, G^{+}  \end{pmatrix}
\matr{M}_{H^\pm G^\pm}^{\rm diag}
\begin{pmatrix} H^{-} \\ G^{-} \end{pmatrix} + \ldots~,
\end{align}
where the tree-level masses are denoted as
$\mh$, $\mH$, $\mA$, $\mG$, $\MHp$, $\mGp$.
With the help of a Peccei-Quinn
transformation~\cite{Peccei} $\mu$ and the complex soft SUSY-breaking
parameters in the Higgs sector can be 
redefined~\cite{MSSMcomplphasen} such that the complex phases
vanish at tree-level.

Including higher-order corrections the three neutral Higgs bosons can
mix~\cite{mhiggsCPXgen,mhiggsCPXRG1,mhiggsCPXFD1,mhcMSSMlong}, 
\begin{align}
\KL h, H, A \KR \quad \longrightarrow \quad \KL \He, \Hz, \Hd \KR~,
\end{align} 
where we define the loop corrected masses according to
\begin{align}
\MHe \le \MHz \le \MHd~.
\end{align}
Details about the renormalization and the $\matr{Z}$~factors, ensuring the on-shell
properties of external Higgs bosons can be found in \citeres{mhcMSSMlong,Stop2decay}.
For the renormalization of $\tb$ and the Higgs field
renormalization the \DRbar\ scheme is
chosen~\cite{mhcMSSMlong,Stop2decay}. This leads to the introduction
of the scale $\mu_R$, which will be fixed later to the
mass of the decaying particle.

%%%%%%%%%%%%%%%%%%%%%%%%%%%%%%%%%%%%%%%%%%%%%%%%%%%%%%%%%%%%%%%%%%%%%%%%%%%%%%%
%%%%%%%%%%%%%%%%%%%%%%%%%%%%%%%%%%%%%%%%%%%%%%%%%%%%%%%%%%%%%%%%%%%%%%%%%%%%%%%
\section{Calculation of loop diagrams}
\label{sec:calc}

\begin{figure}[t!]
\begin{center}
\includegraphics[width=0.90\textwidth]{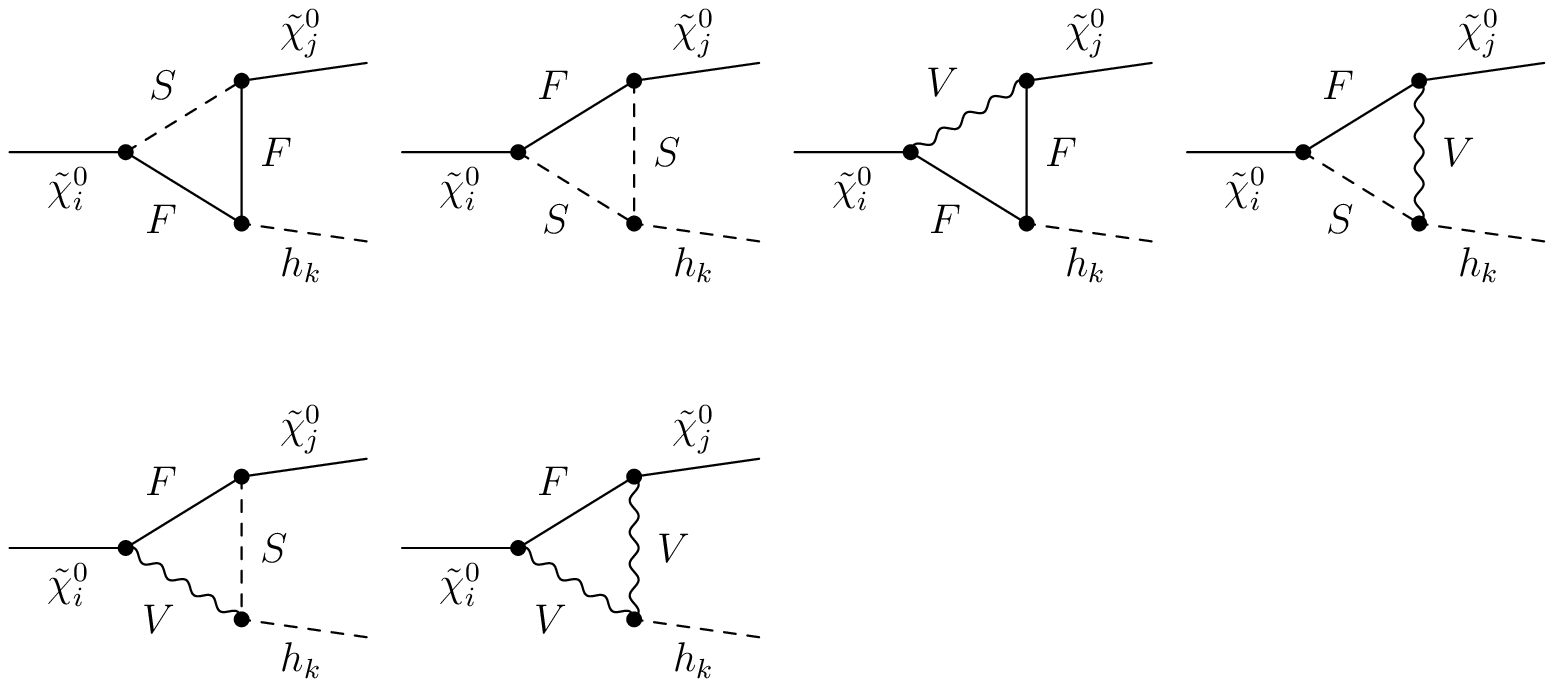}
\caption{
  Generic Feynman diagrams for the decay 
  $\decayNNh$ 
  ($i=2,3,4;\; j < i ;\; k = 1,2,3$).
  $F$ can be a SM fermion, chargino, or neutralino; 
  $S$ can be a sfermion or a Higgs boson; 
  $V$ can be a $\ga$, $Z$, or $W^\pm$. 
  Not shown are the diagrams with a $Z$--$h_{k}$ or $G$--$h_{k}$ 
  transition contribution on the external Higgs boson leg. 
}
\label{fig:fdNNh}
\end{center}
\vspace{-2em}
%\vspace{-1em}
\end{figure}

\begin{figure}[ht!]
\vspace{2em}
\begin{center}
\includegraphics[width=0.90\textwidth]{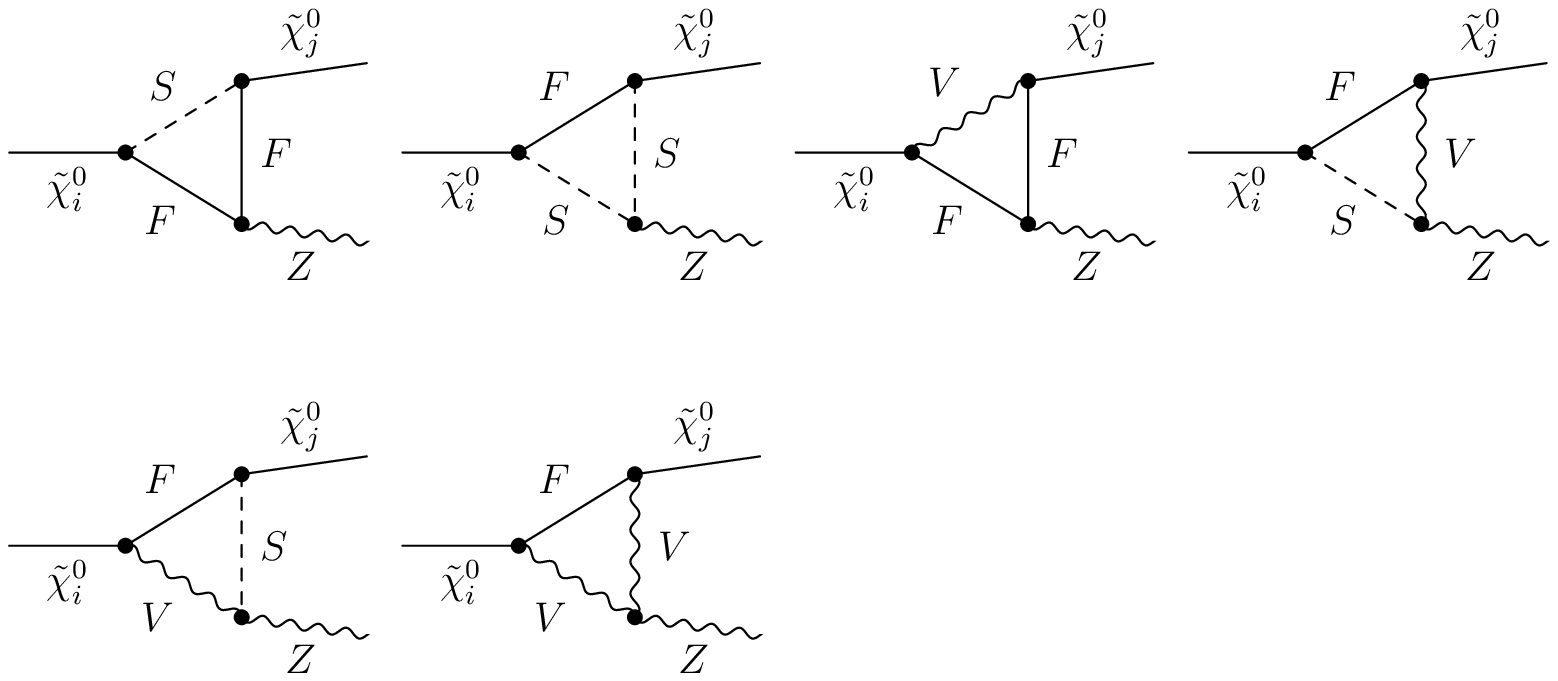}
\caption{
  Generic Feynman diagrams for the decay 
  $\decayNNZ$ 
  ($i=2,3,4;\; j < i$).
  $F$ can be a SM fermion, chargino, or neutralino;
  $S$ can be a sfermion or a Higgs boson;
  $V$ can be a $\ga$, $Z$, or $W^\pm$. 
}
\label{fig:fdNNZ}
\end{center}
\vspace{-1em}
\end{figure}

\begin{figure}[ht!]
\begin{center}
\includegraphics[width=0.90\textwidth]{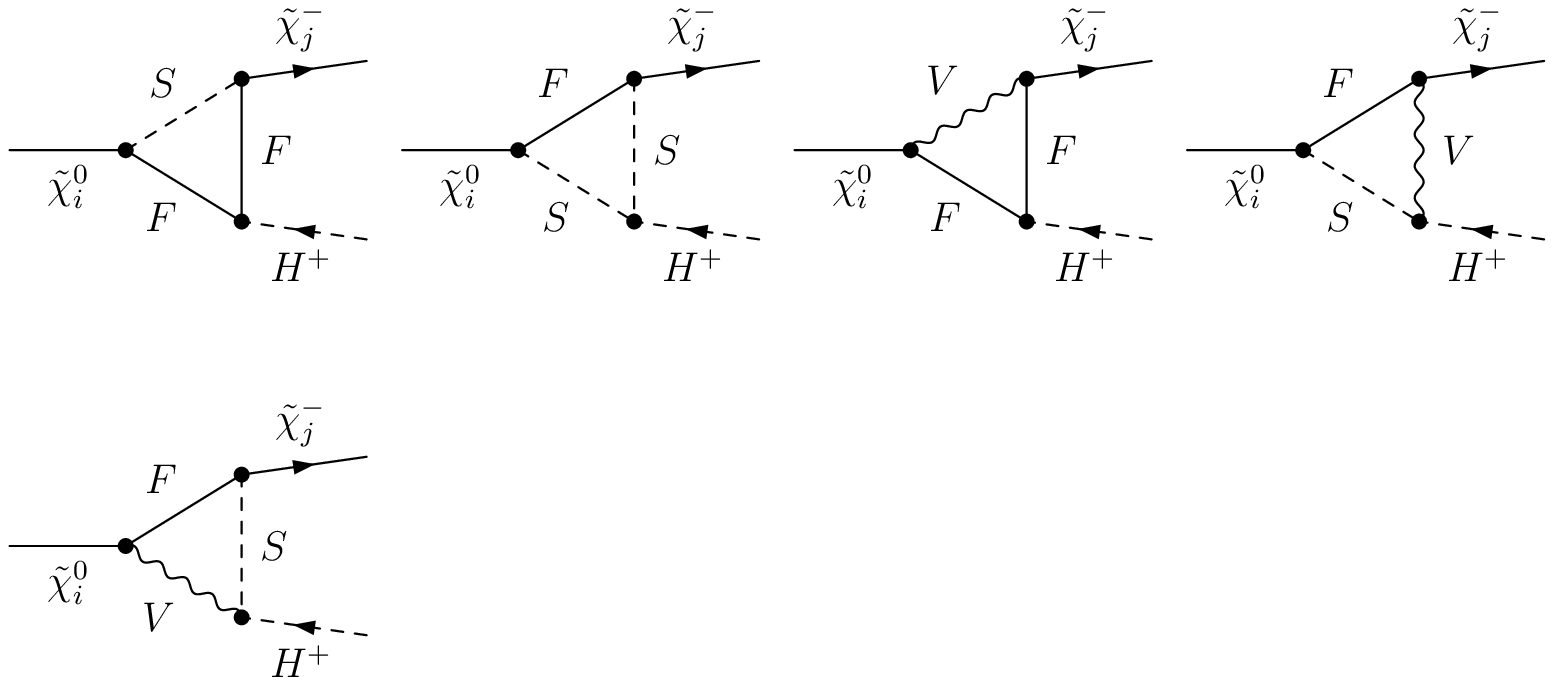}
\caption{
  Generic Feynman diagrams for the decay 
  $\decayNCmH$ ($i = 2,3,4;\; j = 1,2$).
  $F$ can be a SM fermion, chargino, or neutralino; 
  $S$ can be a sfermion or a Higgs boson; 
  $V$ can be a $\ga$, $Z$, or $W^\pm$. 
  Not shown are the diagrams with a $W^{+}$--$H^{+}$ or $G^{+}$--$H^{+}$ transition 
  contribution on the external Higgs boson leg. 
}
\label{fig:fdNCH}
\end{center}
\vspace{-1em}
\end{figure}

\begin{figure}[ht!]
\begin{center}
\includegraphics[width=0.90\textwidth]{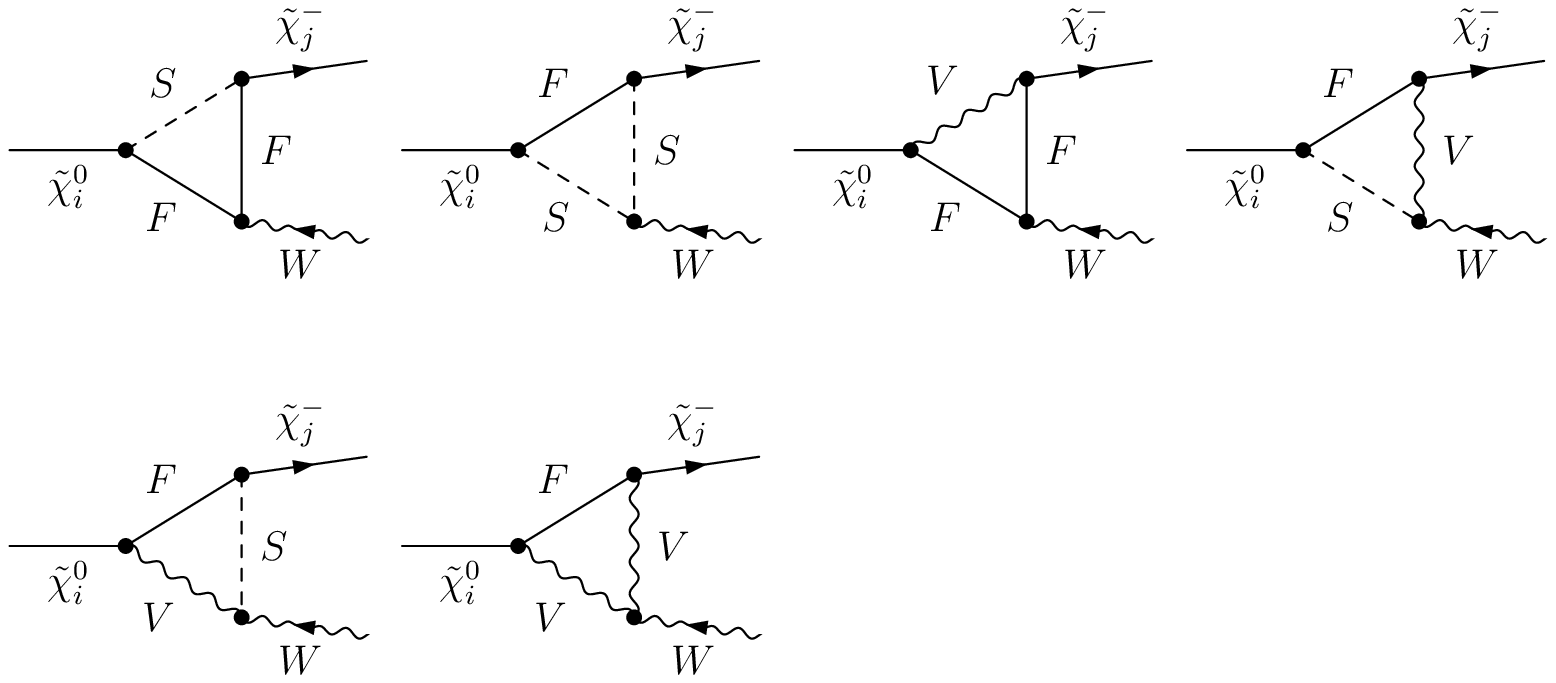}
\caption{
  Generic Feynman diagrams for the decay 
  $\decayNCmW$ ($i = 2,3,4;\; j = 1,2$).
  $F$ can be a SM fermion, chargino, or neutralino; 
  $S$ can be a sfermion or a Higgs boson; 
  $V$ can be a $\ga$, $Z$, or $W^\pm$. 
}
\label{fig:fdNCW}
\end{center}
\vspace{-1em}
\end{figure}

\begin{figure}[ht!]
\begin{center}
\includegraphics[width=0.90\textwidth]{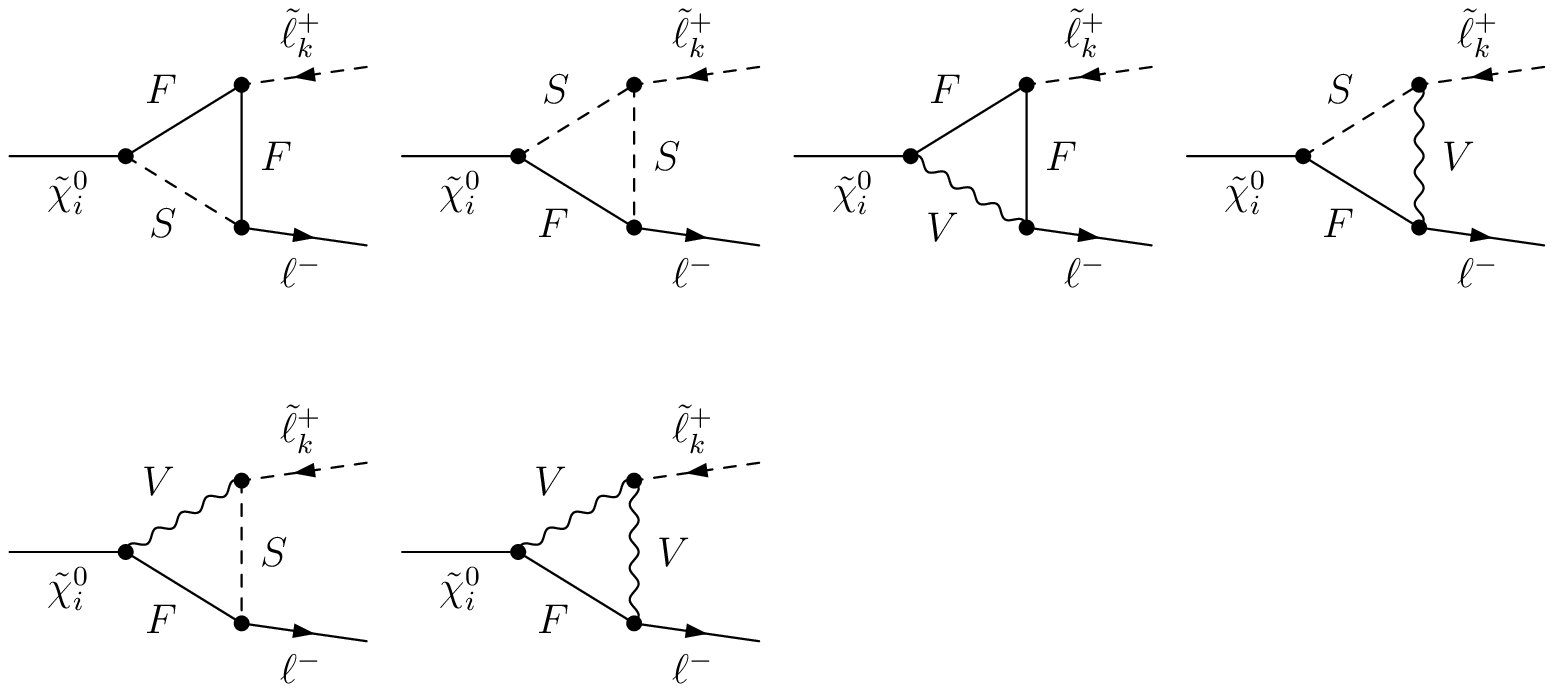}
\caption{
  Generic Feynman diagrams for the decay 
  $\decayNlSl$ ($i = 2, 3, 4;\; \ell = e, \mu, \tau;\; k = 1,2$).
  $F$ can be a SM fermion, chargino, or neutralino; 
  $S$ can be a sfermion or a Higgs boson;
  $V$ can be a $\ga$, $Z$, or $W^\pm$. 
}
\label{fig:fdNSln}
\end{center}
\vspace{-1em}
\end{figure}

\begin{figure}[ht!]
\begin{center}
\includegraphics[width=0.90\textwidth]{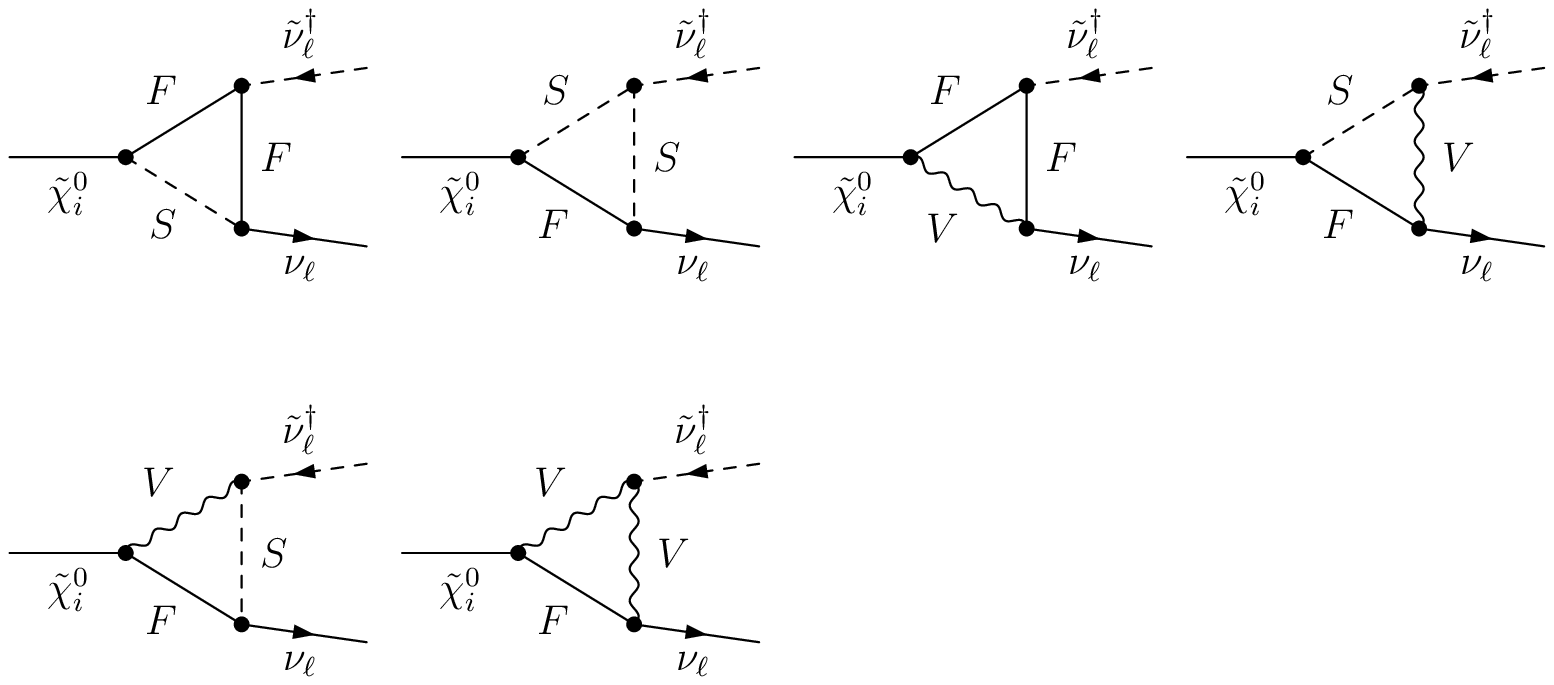}
\caption{
  Generic Feynman diagrams for the decay 
  $\decayNnSn$ ($i = 2, 3, 4;\; \ell = e, \mu, \tau$).
  $F$ can be a SM fermion, chargino, or neutralino;
  $S$ can be a sfermion or a Higgs boson;
  $V$ can be a $\ga$, $Z$, or $W^\pm$. 
}
\label{fig:fdNSnl}
\end{center}
\vspace{-1em}
\end{figure}

In this section we give some details about the calculation of the
higher-order corrections to the neutralino decays. Sample diagrams are
shown in \reffis{fig:fdNNh} -- \ref{fig:fdNSnl}. 
Here the generic internal particles are labeled as follows:
$F$ denotes a SM fermion, chargino, or neutralino,
$S$ denotes a sfermion or a Higgs and $V$ denotes a $\ga$, 
$Z$, or $W^\pm$. Concerning the diagrams for decays into charged particles, 
although we only show diagrams for decays into~$\cham{i}$ or leptons, 
we also include the corresponding diagrams for the decays into~$\chap{i}$ or antileptons.
Not shown are the diagrams for real (hard or soft) photon 
radiation. These are obtained from the corresponding tree-level diagrams
by attaching a photon to the electrically charged external
particles. It should be noted that the expressions for the tree-level diagrams are given explicitly in \refapp{sec:treeresults}.
Counterterm diagrams are also not shown, but and these can be obtained
from the corresponding tree-level diagrams, replacing the tree-level vertex by
the counterterm vertex.

Internally appearing Higgs bosons do not receive higher-order
corrections in their masses or couplings, which would correspond to
effects beyond one-loop. Furthermore, we found that using loop corrected
Higgs boson masses and couplings for the internal Higgs bosons leads to
a divergent result.
For external Higgs bosons, as mentioned in \refse{sec:higgs},
the appropriate $\matr{Z}$~factors 
are applied, following the prescription of \citere{mhcMSSMlong}.
For the numerical analysis, these factors, as well as the loop corrected masses for external Higgs bosons, are obtained from \fh -2.9.0.
Diagrams with a gauge boson/Goldstone--Higgs
self-energy contribution on the external Higgs boson leg, 
absent from the Higgs $\matr{Z}$~factors~\cite{mhcMSSMlong}, are also required for 
the decays $\decayNNh$, 
($i = 2,3,4;\; j < i; \; {k} = 1,2,3$), \reffi{fig:fdNNh}, 
with a $Z/G$--$h_{k}$ transition, and similarly for 
the decay 
$\decayNCmH$
($i = 2,3,4; \; j = 1,2$),
\reffi{fig:fdNCH}, with a $W^{+}$/$G^{+}$--$H^{+}$ transition. On the other hand, 
Goldstone--Higgs/gauge boson
self-energy corrections for the neutralino decay to
a chargino/neutralino and a gauge boson 
$\decayNNZ$ ($i=2,3,4;\; j < i$), 
or $\decayNCmW$, ($i=2,3,4; \; j=1,2$), can be neglected, as these
vanish on mass shell, 
i.e.\ for $p^2 = \MZ^2$ ($p^2 = \MW^2$) due to $\eps \cdot p = 0$, 
where $p$ denotes the external momentum and $\eps$ the polarization
vector of the gauge boson.

%%%%%%%%%%%%%%%%%%%%%%%%% F I G U R E %%%%%%%%%%%%%%%%%%%%%%%%%%%%%%%%%%%%%%%%%
%%%%%%%%%%%%%%%%%%%%%%%%% F I G U R E %%%%%%%%%%%%%%%%%%%%%%%%%%%%%%%%%%%%%%%%%

%%%%%%%%%%%%%%%%%%%%%%%%% F I G U R E %%%%%%%%%%%%%%%%%%%%%%%%%%%%%%%%%%%%%%%%%
%%%%%%%%%%%%%%%%%%%%%%%%% F I G U R E %%%%%%%%%%%%%%%%%%%%%%%%%%%%%%%%%%%%%%%%%

%%%%%%%%%%%%%%%%%%%%%%%%% F I G U R E %%%%%%%%%%%%%%%%%%%%%%%%%%%%%%%%%%%%%%%%%
%%%%%%%%%%%%%%%%%%%%%%%%% F I G U R E %%%%%%%%%%%%%%%%%%%%%%%%%%%%%%%%%%%%%%%%%

%%%%%%%%%%%%%%%%%%%%%%%%% F I G U R E %%%%%%%%%%%%%%%%%%%%%%%%%%%%%%%%%%%%%%%%%
%%%%%%%%%%%%%%%%%%%%%%%%% F I G U R E %%%%%%%%%%%%%%%%%%%%%%%%%%%%%%%%%%%%%%%%%

%%%%%%%%%%%%%%%%%%%%%%%%% F I G U R E %%%%%%%%%%%%%%%%%%%%%%%%%%%%%%%%%%%%%%%%%
%%%%%%%%%%%%%%%%%%%%%%%%% F I G U R E %%%%%%%%%%%%%%%%%%%%%%%%%%%%%%%%%%%%%%%%%

%%%%%%%%%%%%%%%%%%%%%%%%% F I G U R E %%%%%%%%%%%%%%%%%%%%%%%%%%%%%%%%%%%%%%%%%
%%%%%%%%%%%%%%%%%%%%%%%%% F I G U R E %%%%%%%%%%%%%%%%%%%%%%%%%%%%%%%%%%%%%%%%%

The diagrams and corresponding amplitudes have been obtained with 
\fa~\cite{feynarts}. 
The model file for calculations in scheme I, including the MSSM counterterms, 
is  described in more detail in \citere{Stop2decay}, 
and the model file used for calculations in scheme II 
is based on that discussed in \citeres{dissAF,bfmw}.
~The further evaluation has been performed with 
\fc\ (and \looptools)~\cite{formcalc}. 
As regularization scheme for the UV-divergences we
have used constrained differential renormalization~\cite{cdr}, 
which has been shown to be equivalent to 
dimensional reduction~\cite{dred} at the \onel\ level~\cite{formcalc}. 
Thus the employed regularization preserves SUSY~\cite{dredDS,dredDS2}. 
All UV-divergences cancel in the final result.

The IR-divergences from diagrams with an internal photon have
to cancel with the ones from the corresponding real soft radiation, 
where we have included the soft photon contribution
following the description given in \citere{denner}. 
The IR-divergences arising from the diagrams involving a $\ga$ 
are regularized by introducing a finite photon mass,
$\lambda$. 
All IR-divergences, i.e.\ all divergences in the limit
$\lambda \to 0$, cancel to all orders
once virtual and real diagrams for one decay
channel are added. 
The only exception are the decays $\neu{2,3,4} \to \cha{1} W^\mp$. 
The shift to the neutralino on-shell masses via \refeq{Deltamneu}
results in an IR divergence at the two-loop level, i.e.\ here we find a
cancellation of the divergences ``only'' at the one-loop level, as
required for our one-loop calculation. 
The remaining two-loop IR divergence could be eliminated by a symmetry restoring
counterterm in the $\neu{2,3,4} \to \cha{1} W^-$
vertex, similar to the evaluation of
the decay $\Stopz \to \Sbot_{1,2} W^+$ in \citere{Stop2decay}.
~We have furthermore checked that our result does not depend on $\De E$
defining the energy cut that separates the soft from the hard
radiation. Our numerical results have been obtained for 
$\De E = 10^{-5} \times \mneu{i}$ 
for all channels.%
\footnote{
The larger cut is necessary to obtain a better convergence of the 
integration over the three body phase space.
The contribution from nearly collinear photons 
(along the direction of the electron) leads to numerical instabilities 
in the integration.
}

%%%%%%%%%%%%%%%%%%%%%%%%%%%%%%%%%%%%%%%%%%%%%%%%%%%%%%%%%%%%%%%%%%%%%%%%%%%%%%
\section{Numerical analysis}
\label{sec:numeval}

In this section we will first introduce and motivate the scenarios studied, discussing the current
experimental constraints considered, then introduce the observables calculated, 
and finally present our results for each of the decay channels of the heavier neutralino ($\DecayNxy{4}$)
as a function of $\phiMe$.

%%%%%%%%%%%%%%%%%%%%%%%%%%%%%%%%%%%%%%%%%%%%%%%%%%%%%%%%%%%%%%%%%%%%%%%%%%%%%

%%%%%%%%%%%%%%%%%%%%% T A B L E %%%%%%%%%%%%%%%%%%%%%%%%%%%%%%%%%%%%%%%%%%%%%%
\begin{table}[ht!]
\renewcommand{\arraystretch}{1.5}
\BC
\begin{tabular}{|c|c|c|c|c|c|c|c|c|c|c|}
\hline
$\tb$ & $\MHp$ & $\mcha{2}$ & $\mcha{1}$ 
& $\MslL$ & $\MslR$ & $\Al$ 
& $\MsqL$ & $\MsqR$ & $\Aq$
\\ \hline\hline
$ 20$ & $ 160$ & $ 600$ & $ 350$ & $ 300$ & $ 310$ & $ 400$ 
& $ 1300$ & $ 1100$ & $ 2000$
\\ \hline
\end{tabular}
\caption{MSSM parameters for the initial numerical
  investigation; all 
  mass parameters are in$\gev$. $\MOne$, $\MTwo$ and $\mu$ are chosen such that the
  values for $\mcha{1}$ and $\mcha{2}$ and \refeq{M1M2} are fulfilled
(see text).
}
\label{tab:para}
\EC
\renewcommand{\arraystretch}{1.0}
\end{table}
%%%%%%%%%%%%%%%%%%%%% T A B L E %%%%%%%%%%%%%%%%%%%%%%%%%%%%%%%%%%%%%%%%%%%%%%

\noindent
As stated earlier, we present our results in two scenarios. In both,
the absolute value of $\MOne$ (see above) is fixed via the GUT
relation (with $|\MTwo| \equiv \MTwo$)
\begin{align}
|\MOne| &= \frac{5}{3} \tan^2 \thw \MTwo \approx \edz \MTwo~.
\label{M1M2}
\end{align}
For the numerical analysis we obtain $\MTwo$ and $\mu$ 
from the fixed chargino masses $\mcha{1,2}$, and $|\MOne|$ via \refeq{M1M2}, 
leaving $\phiMe$ as a free parameter.
Our two scenarios arise due to the ambiguity in calculating $\mu$ and $M_2$ from $\mcha{1,2}$.
This ambiguity can be resolved by choosing an addition condition, $\mu>M_2$ or $\mu<M_2$.
The first choice, denoted by $\Sh$, results in a Higgsino-like $\neu{4}$
while the second choice, denoted  by $\Sg$, results in a gaugino-like $\neu{4}$.

The values of the parameters for these scenarios are given in \refta{tab:para}, 
where, in analogy to the slepton parameters
$\MslL$, $\MslR$ and $\Al$ defined in \refse{sec:slepton} for the sleptons, 
$\MsqL$ and $\MsqR$ are the left- and right-handed soft SUSY-breaking mass parameters and $\Aq$ is the trilinear soft-breaking parameter for the squarks. 
These are chosen such that most decay modes are open
simultaneously to permit an analysis of as many channels as possible.
Only decays into the heavier chargino $\cha{2}$, in general degenerate with the heavier neutralino,
and the decay channels
$\neu{4} \to \neu{3} h_k/Z,\, (k=1,2,3)$, in $\Sh$
are kinematically closed.
We also ensure that the scenarios are consistent with the 
MSSM Higgs boson searches at 
LEP~\cite{LEPHiggs}, 
Tevatron~\cite{TevatronMSSM} and LHC~\cite{LHCMSSM}.
The light Higgs mass scale
  together with the value of $\tb = 20$ are in potential conflict with the
  recent MSSM Higgs search results, 
which, however, have
  only been obtained in the {\mhmax}~scenario~\cite{mhiggsmax}.
 We stick to our 
  parameter combination to facilitate the numerical analysis with all
  decay channels involving Higgs bosons being open simultaneously. 
On the other hand the 
recent discovery of the (lightest) Higgs boson at the LHC~\cite{discovery} allows for 
$\tb \gsim {9}$ as given in \refta{tab:para}.%
\footnote{The Higgs mass in the allowed range can be obtained varying the squark trilinear coupling $\Aq$.}
%
%%%%%%%%%%%%%%%%%%%%%%%%%%%%%%%%%%%%%%%%%%%%%%%%%%%%%%%%%%%%%%
%
Furthermore, 
the following exclusion limits for neutralinos~\cite{pdg} hold in
our numerical scenarios: 
\begin{align}
\mneu{1} &> 46 \gev, \;
\mneu{2} > 62 \gev, \;
\mneu{3} > 100 \gev, \;
\mneu{4} > 116 \gev~.
\end{align}
It should be noted that the limit for $\mneu{1}$ arises solely when
\refeq{M1M2} is assumed to hold. In the absence of this condition, no limit on a light
neutralino mass exists, see \citere{masslessx} and references therein.

The most restrictive experimental constraints on the phase of $\phiMe$
arise due to the bounds on the electric dipole moments (EDM's) of the 
neutron~$d_n$, mercury~$d_{\rm Hg}$, 
and Thallium~$d_{\rm Tl}$~\cite{Baker:2006ts,Regan:2002ta,Griffith:2009zz}.%
\footnote{In
    addition, the heavy quarks~\cite{EDMDoink}, the
    electron~\cite{EDMrev2,EDMPilaftsis} and the deuteron~\cite{EDMRitz} EDM's
    should be taken into account.} 
~Using
  \texttt{CPsuperH2.2}~\cite{Lee:2003nta}%
, we have calculated these EDM's, and
  find that in the scenarios studied the bounds due to the Thallium EDM are
  the most constraining, $\phiMe\lesssim \pi/100$. This is mainly because, in
  order to keep all channels open the selectrons in our scenarios are light. 
On increasing the masses of the lower generation of sleptons to $1.2\tev$
$\phiMe$ is unconstrained. 
It should be noted that such a change would only affect our results such that the decays to
the lower generation slepton channels are no longer open: 
the loop corrections to the other decays are largely independent of these masses.

% -------------------------------------------- ILC1000 -------------------------------------------
The chargino and neutralino masses for $\phiMe=0$
are shown in \refta{tab:chaneu},
while the Higgs and slepton masses are shown in \refta{tab:higgsslep}. 
Here $\Hz$ corresponds to the pure $\cp$-odd Higgs boson. 
 For $\phiMe = 90^\circ$ (i.e.\ the maximal 
  $\cp$-violation possible in our numerical analysis) we find 
  the same Higgs boson masses within the precision of \refta{tab:higgsslep}. 
In this case $\Hz$ receives a 
  very small $\cp$-even admixture of $\lsim 0.003\%$ in both scenarios,  
  while $\He$ and $\Hd$ remain
  correspondingly a nearly pure $\cp$-even state.
The masses $\mneu{i}$ are  chosen such that the neutralinos would be copiously produced in SUSY cascades at the LHC.
Furthermore, the production of $\neu{i}\neu{j}$, for $i=1, 2$ and $j = 2,3,4$,
at the ILC(1000), i.e.\ with $\sqrt{s} = 1000 \gev$, via 
$e^+e^- \to \neu{i}\neu{j}$ will be possible, where unpolarized tree-level cross sections in the scenarios $\Sg$ and $\Sh$ 
are shown in \refta{tab:ILCXS}. 
All the subsequent decay modes (\ref{NNh}) -- (\ref{NSnn})
would be (in principle) open, and the clean environment would permit a detailed study of neutralino decays~\cite{ilc,lhcilc}.
Higher-order corrections to the production cross sections would change the values in \refta{tab:ILCXS}
by up to \order{10\%}~\cite{Oller:2005xg}, and
choosing appropriate polarized beams, could enhance the cross-sections 
by a factor~$2$ to $3$.
% %
The accuracy of the relative branching ratio \refeq{brrel} at the ILC would be close to the statistical
uncertainty, and
from the high-luminosity running of the
ILC(1000), a determination of the branching ratios at the percent
level might be achievable. 
%
%%%%%%%%%%%%%%%%%%%%% T A B L E %%%%%%%%%%%%%%%%%%%%%%%%%%%%%%%%%%%%%%%%%%%%%%
\begin{table}[t!]
\renewcommand{\arraystretch}{1.6}
\BC
\begin{tabular}{|c||c|r|r|r|r|r|c||c|c|c|}
\hline
Scenario
& 
$\mcha{2}$ & $\mcha{1}$
      & $\mneu{4}$ & $\mneu{3}$ & $\mneu{2}$ & $\mneu{1}$ 
  & $\mu$  & $\MTwo$    & $\MOne$  

\\ \hline\hline
{$\Sg$} 
           & $ 600.0$ & $ 350.0$ & $ 600.0$ & $ 364.2$ & $ 359.6$ & $ 267.2$ &  
            $ 362.1$ & $ 581.8$ & $ 277.7$ 
 \\ \hline
{$\Sh$} 
           & $ 600.0$ &  $350.0 $ & $ 600.1$ & $ 586.2$ & $ 349.9$ & $ 171.4$ &  
             $581.8$ & $ 362.1$ & $ 172.8$ 
 \\ \hline
\end{tabular}
\caption{The chargino and neutralino masses in the scenarios {$\Sg$}\ and {$\Sh$}.
We also show the values for the ``derived'' parameters $\MOne$, 
$\MTwo$ and $\mu$. 
All mass parameters are in$\gev$, rounded to $0.1 \gev$ to show the 
  size of small mass
  differences, which can determine whether a certain decay channel is
  kinematically closed or open.
}
\label{tab:chaneu}
\EC
\renewcommand{\arraystretch}{1.0}
\end{table}
%%%%%%%%%%%%%%%%%%%%% T A B L E %%%%%%%%%%%%%%%%%%%%%%%%%%%%%%%%%%%%%%%%%%%%%%
%
%%%%%%%%%%%%%%%%%%%%% T A B L E %%%%%%%%%%%%%%%%%%%%%%%%%%%%%%%%%%%%%%%%%%%%%%
\begin{table}[htb!]
\renewcommand{\arraystretch}{1.6}
\BC
\begin{tabular}{|c||c|r|r|r|r|r||r|r|r|r|}
\hline

Scenario
& 
$\ms{\mu}{1} $ & 
$\ms{\mu}{2} $ & 
$\ms{\tau}{1}$ & 
$\ms{\tau}{2}$ & 
${m_{\tilde{\nu}_\mu}} $ & 
${m_{\tilde{\nu}_\tau}} $ & 
$\MHp$ & $\mHe$  & $\mHz$ & $\mHd$ 
\\ \hline\hline
{$\Sg$} & $ 303.2 $ & $ 313.1 $ & $ 287.3 $ & $ 328.0 $ & $ 293.0 $ & $ 293.0 $ & $ 160.0 $ & $ 125.8 $ & 
        $137.2 $ & $ 140.3$
 \\ \hline
{$\Sh$} & $ 302.9 $ & $ 313.3 $ & $ 273.7 $ & $ 339.5 $ & $ 293.0 $ & $ 293.0 $ & $ 160.0 $ & $ 125.8 $ & 
        $ 137.4$ &$ 140.3$
 \\ \hline
\end{tabular}
\caption{The slepton and Higgs masses in the scenarios {$\Sg$}\ and {$\Sh$}.
The selectron and electron sneutrino masses 
are equal to those of the corresponding smuon and muon sneutrino
up to a few tenths of$\gev$. 
  All masses are in$\gev$, rounded to $0.1 \gev$.
}
\label{tab:higgsslep}
\EC
\renewcommand{\arraystretch}{1.0}
\end{table}
%%%%%%%%%%%%%%%%%%%%% T A B L E %%%%%%%%%%%%%%%%%%%%%%%%%%%%%%%%%%%%%%%%%%%%%%
%
%%%%%%%%%%%%%%%%%%%%% T A B L E %%%%%%%%%%%%%%%%%%%%%%%%%%%%%%%%%%%%%%%%%%%%%%
\begin{table}[htb!]
\renewcommand{\arraystretch}{1.6}
\BC
\begin{tabular}{|c||c|c|c|c|c|c|c|c||c|}
\hline
Scen. & 
process & $\si_{0,0}[{\rm fb}]$
& $\si_{\rm pol}[{\rm fb}]$ & 
stat.\ prec.$_{0,0}$ & stat.\ prec$_{\rm pol}$ 
\\ \hline\hline
\Sg & $e^+e^- \to \neu{4}\neu{1}$  & 5.2 & 15.0 & $ 4 \% $ &{$ 3 \% $}  
\\ \hline
\Sh & $e^+e^- \to \neu{4}\neu{1}$  & 1.0 & 1.6  & $ 10 \% $ & $ 8\% $  
\\ \hline
\Sg & $e^+e^- \to \neu{4}\neu{2}$  & 1.0 & 2.9 &{$ 10 \% $} &{$ 6 \% $}  
\\ \hline
\Sh & $e^+e^- \to \neu{4}\neu{2}$  & 0.4 & 1.1 &{$ 16 \% $} &{$ 10 \% $}  
\\ \hline
\Sg & $e^+e^- \to \neu{4}\neu{3}$  & 0.5 & 0.4 &{$ 14 \% $} &{$ 16 \% $} 
\\ \hline\hline
\Sg & $e^+e^- \to \sum\neu{4}\neu{j}$  & 6.7 & 18.3 &{$ 4 \% $} &{$ 2 \% $}
\\ \hline
\Sh & $e^+e^- \to \sum\neu{4}\neu{j}$  & 1.4 & 2.7 &{$ 8 \% $} &{$  6\% $}
\\ \hline
\Sg & $e^+e^- \to \sum\neu{3}\neu{j}$  & 39.9 & 71.8 &{$ 2 \% $} &{$ 1 \% $}
\\ \hline
\Sh & $e^+e^- \to \sum\neu{3}\neu{j}$  & 3.2 & 6.1 &{$ 6 \% $} &{$ 4 \% $}
\\ \hline
\Sg & $e^+e^- \to \sum\neu{2}\neu{j}$  & 13.5 & 11.0 &{$ 3 \% $} &{$ 3 \% $}
\\ \hline
\Sh & $e^+e^- \to \sum\neu{2}\neu{j}$  & 46.2 & 132.4 &{$ 1 \% $} &{$ 1 \% $}
\\ \hline
\end{tabular}
\caption{
Neutralino production cross sections at the ILC 1000. 
Here $\si_{0,0}$ denotes the cross section for
unpolarized beams, while $\si_{\rm pol}$ 
denotes that with electron and positron polarization $-80\%$ and
$+60\%$, respectively. 
The sums of $\neu{i}\neu{j}$, for $i=2,3,4$, are performed over $j\le i$.
The two right-most columns show the statistical precision for a
  (hypothetical) branching ratio of $10\%$ assuming an integrated
  luminosity of $1\, \iab$, rounded to $1\%$.
}
\label{tab:ILCXS}
\EC
\renewcommand{\arraystretch}{1.0}
\end{table}
%%%%%%%%%%%%%%%%%%%%%%%%%%%%%%%%%%%%%%%%%%%%%%%%%%%%%%%%%%%%%%%%%%%%%%%%%%%%%%%
%
We have calculated the decay width at tree-level (``tree'') and at the one-loop
level (``full''), including {\em all} one-loop 
contributions as described in \refse{sec:calc}, and in addition the
relative size of this one-loop correction via, 
\begin{align}
\Gtree \equiv \Gtree(\decayNxy)~, \qquad
\Gfull \equiv \Ga^{\rm full}(\decayNxy)~, \qquad
\De\Ga/\Ga \equiv \frac{\Gfull - \Gtree}{\Gtree}~.
\end{align}
In the figures below we show the absolute
value of the various decay widths, $\Ga(\DecayNxy{4})$ on the left and
the relative correction from the full one-loop contributions on the right. 
The total decay width is defined as the sum of all 
kinematically open two-body
decay widths, 
\begin{align}
\Ga_{\rm tot}^{\rm tree} \equiv \sum_{{\rm xy}} \Gtree(\decayNxy)~, \qquad
\Ga_{\rm tot}^{\rm full} \equiv \sum_{{\rm xy}} \Gfull(\decayNxy)~.
\end{align}
The absolute and relative changes of the branching ratios are defined as follows,
\begin{align}
\br^{\rm tree} \equiv \frac{\Ga^{\rm tree}(\decayNxy)}
                     {\Ga_{\rm tot}^{\rm tree}}~, \quad
\br^{\rm full} \equiv \frac{\Ga^{\rm full}(\decayNxy)}
                     {\Ga_{\rm tot}^{\rm full}}~, \quad
\frac{\De\br}{\br} \equiv \frac{\br^{\rm full} - \br^{\rm tree}}{\br^{\rm full}}~. 
\label{brrel}
\end{align}
The last quantity is crucial in order to analyze the impact of the one-loop
corrections on the phenomenology at the LHC and the ILC, 
see below.
Since decays to a light Higgs and the LSP are of particular importance, 
for the decay $\DecayNNh{4}{1}{1}$, we also show, in the lower panels, the branching ratio $\br(\DecayNxy{4})$ 
(left) and the relative size of the one-loop correction (right). The corresponding branching ratios for the other channels can be inferred from these plots.

In order to distinguish the results evaluated in the two schemes we denote those of
scheme~II with a `tilde', i.e.\ $\SgII$ and $\ShII$ for the scenarios $\Sg$ and $\Sh$, respectively.
It should be noted that the tree-level results obtained for the two schemes fully agree, as our two renormalization schemes 
differ only in the treatment of complex parts.
The difference of one-loop results for real parameters is negligible, and is only due to
a different handling of the corrections in the squark sector, which is not highly relevant for the electroweak decays.
Therefore the results for both schemes are only shown on the right panels for the relative corrections.
In \refse{sec:rendiff} we will summarize and discuss the differences between these schemes, highlighting those channels where the deviations are largest.

The numerical results we show in this section are of course dependent on the
choice of the 
MSSM parameters. 
Nevertheless, they give an idea of the relevance of the full one-loop corrections.
Decay widths (and their respective one-loop corrections) that may appear to be 
unobservable due to the small size of their BR, 
could become important if other channels are kinematically forbidden.
Consequently, the one-loop corrections to {\em all} channels are evaluated analytically, 
but in the numerical analysis we only show the channels that are kinematically open in our numerical scenarios,
except for the decays into leptons of the first two families, which are closely related to the decays into third family leptons.

\medskip

% %%%%%%%%%%%%%%%%%%%%%%%%%%%%%%%%%%%%%%%%%%%%%%%%%%%%%%%%%%%%%%%%%%%%%%%%%%%%%
\subsection{Decays into charged Higgs and \boldmath{$W$} bosons}
\label{sec:DecayHpW}
\begin{figure}[t!]
\begin{center}
\begin{tabular}{c}
\includegraphics[width=0.49\textwidth,height=7.5cm]{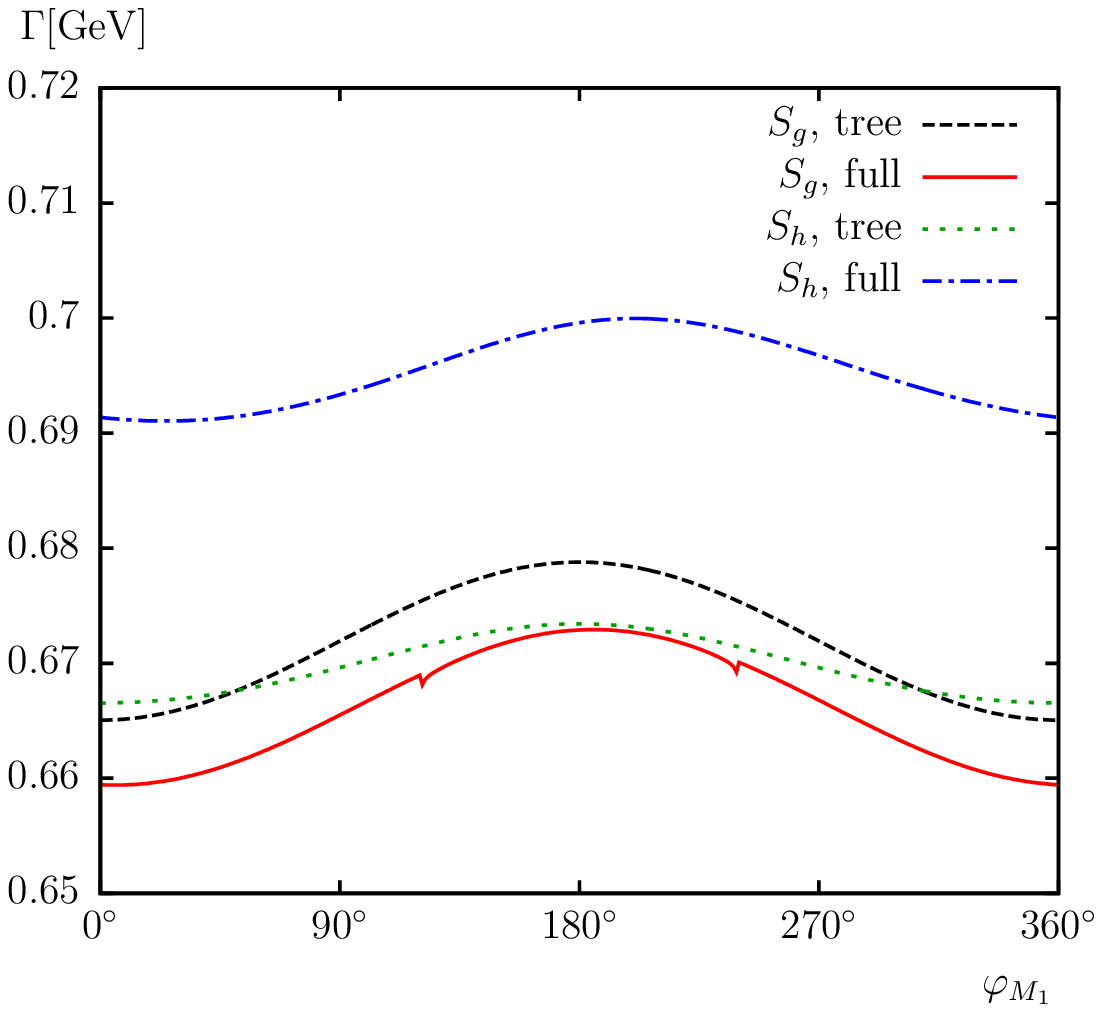}
\hspace{-4mm}
\includegraphics[width=0.49\textwidth,height=7.5cm]{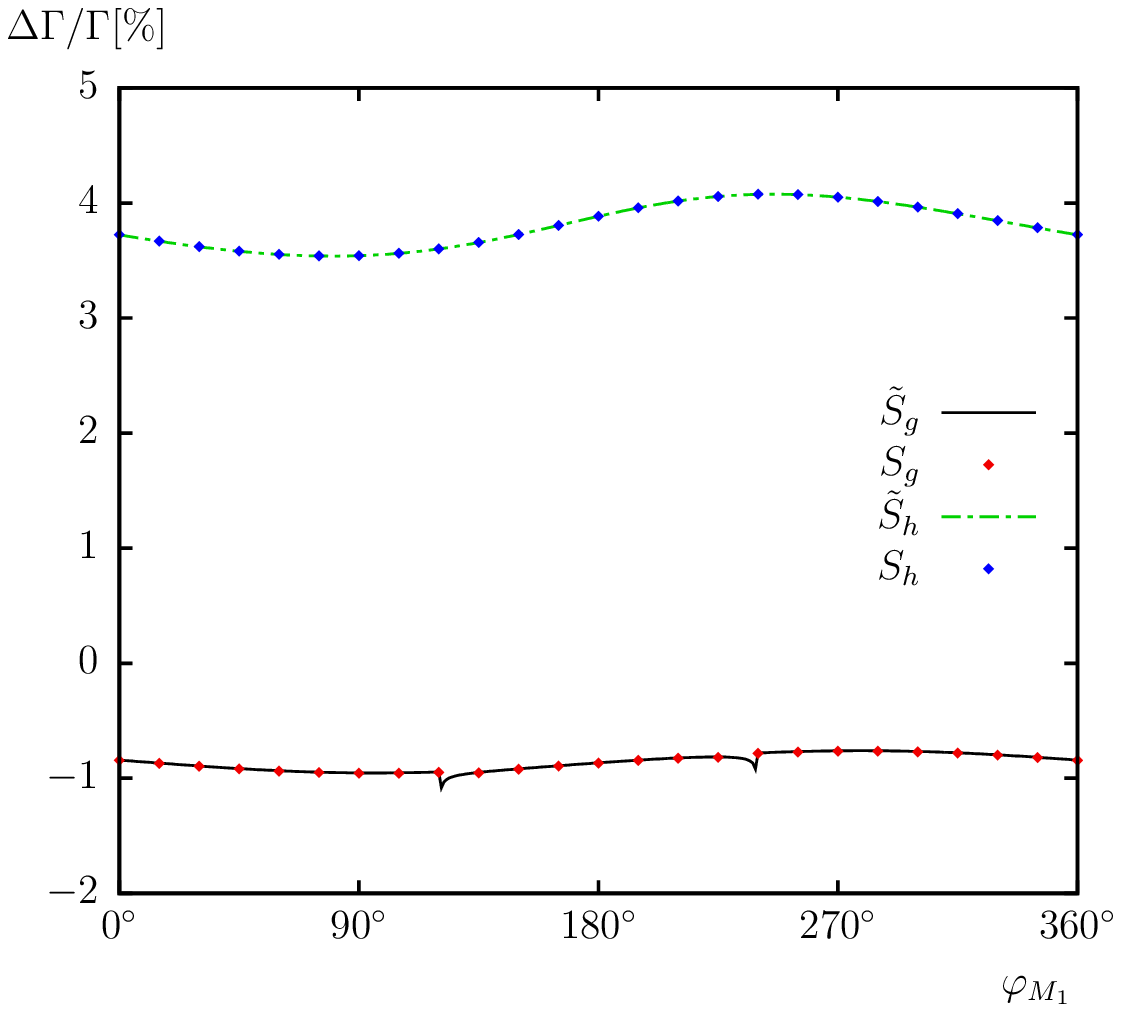} 
\end{tabular}
\caption{
  $\Ga(\DecayNCmH{4}{1})$. 
  Tree-level (``tree'') and full one-loop (``full'') corrected 
  decay widths are shown with the parameters chosen according to 
  \refta{tab:para}, with $\phiMe$ varied.
  The left plot shows the decay width, the right plot shows the relative size of the corrections.
}
\label{fig:PhiM1.neu4cha1hp}
\end{center}
\end{figure}

We start our numerical analysis with the decays 
$\DecayNCmH{4}{1}$, 
presented in \reffi{fig:PhiM1.neu4cha1hp}. 
The partial decay width for the charge conjugated process $\DecayNCpH{4}{1}$ can be obtained
by taking the charge conjugate of all the couplings. 
Since in our analysis  only $\MOne$ is complex
 this is obtained with the transformation 
$\phiMe\to 2\pi - \phiMe$. 
This argument will also be valid for all the decays into $W$-bosons and lepton-slepton pairs described below.
All these decays have also been computed in order to evaluate the total decay width at one-loop level.

This channel yields decay widths of around $0.7\gev$ in both scenarios, corresponding to 
BR's of $\sim 4.5\% $ for $\Sg$ and $\sim 11\% $ for $\Sh$.
The tree-level partial widths are almost equal in both scenarios due to 
the symmetry of the Higgs-gaugino-Higgsino couplings under $\MTwo \leftrightarrow \mu$ exchange, 
as well as to the similar phase space.
The relative corrections, shown in the right plot, are of the order of a few percent.
The  mild dependence on $\phiMe$ for this process is due to our choice of parameters,
in particular the GUT relation on the gaugino parameters $\MOne$ and $\MTwo$,
which leads to a weak dependence of the heavier neutralinos on $\MOne$.
The dips best visible in the right panel 
are due to the $\DecayCmNW{1}{1}$ threshold,
at $\phiMe\simeq 125^\circ$ and $\phiMe\simeq 235^\circ$.
Due to ${\cal CPT}$-invariance the masses are invariant under $\phiMe\to -\phiMe $, 
resulting in mirrored threshold effects. 
These effects are further discussed in \refse{sec:DecayNjHk} for the decays into the second lightest neutralino.

The results for the two schemes are shown for the relative corrections in the right panel.
The agreement between the relative corrections 
is at the level of $10^{-5}$ for both scenarios, 
see \refta{tab:rendiff}, 
and therefore cannot be visibly distinguished here. 
It should be noted that the differences between the schemes are particularly small due to mild dependence of the tree level decay width on $\phiMe$. 
As the schemes are identical in the real case, only decays with a stronger dependence on the phase $\phiMe$ are sensitive to the differences between them.

\begin{figure}[htb!]
\begin{center}
\begin{tabular}{c}
\includegraphics[width=0.49\textwidth,height=7.5cm]{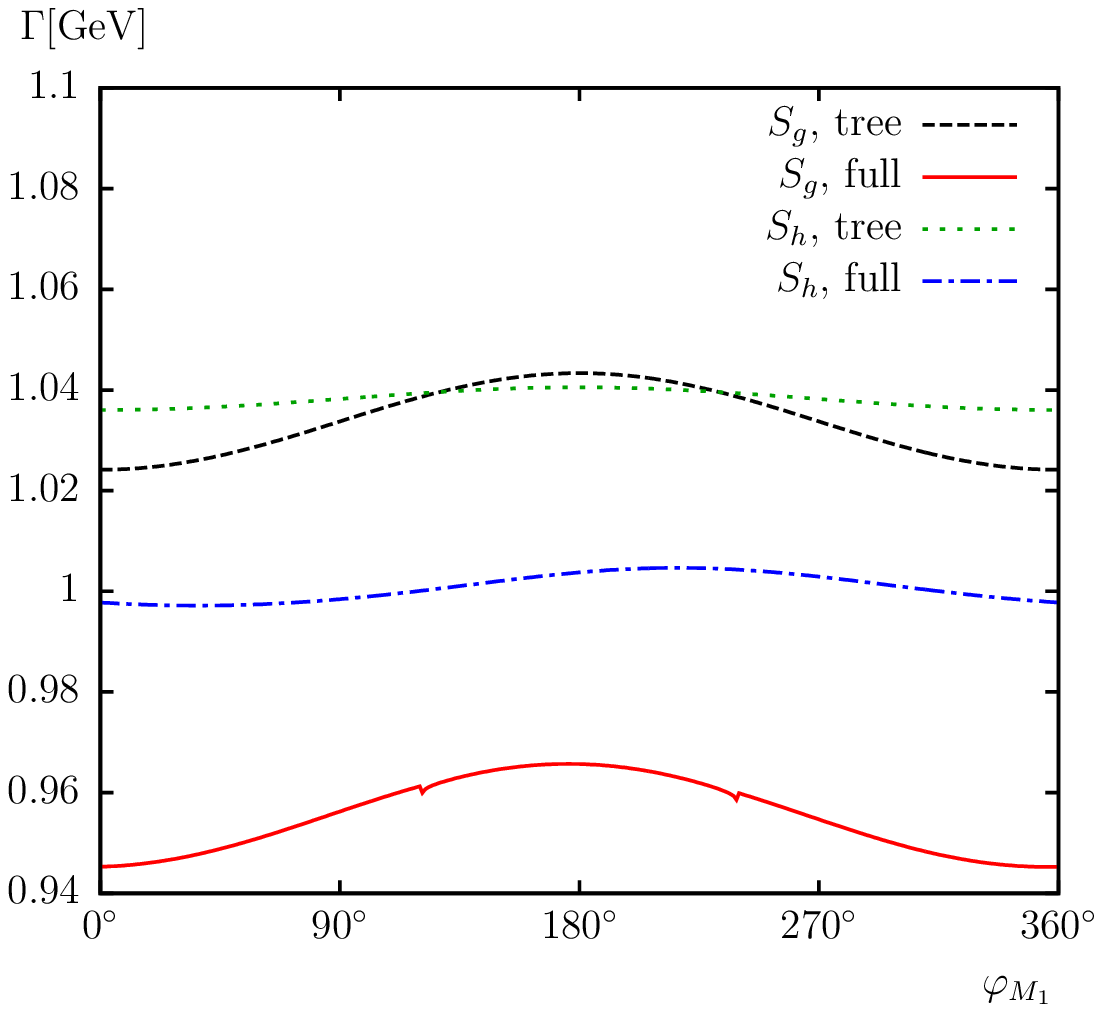}
\hspace{-4mm}
\includegraphics[width=0.49\textwidth,height=7.5cm]{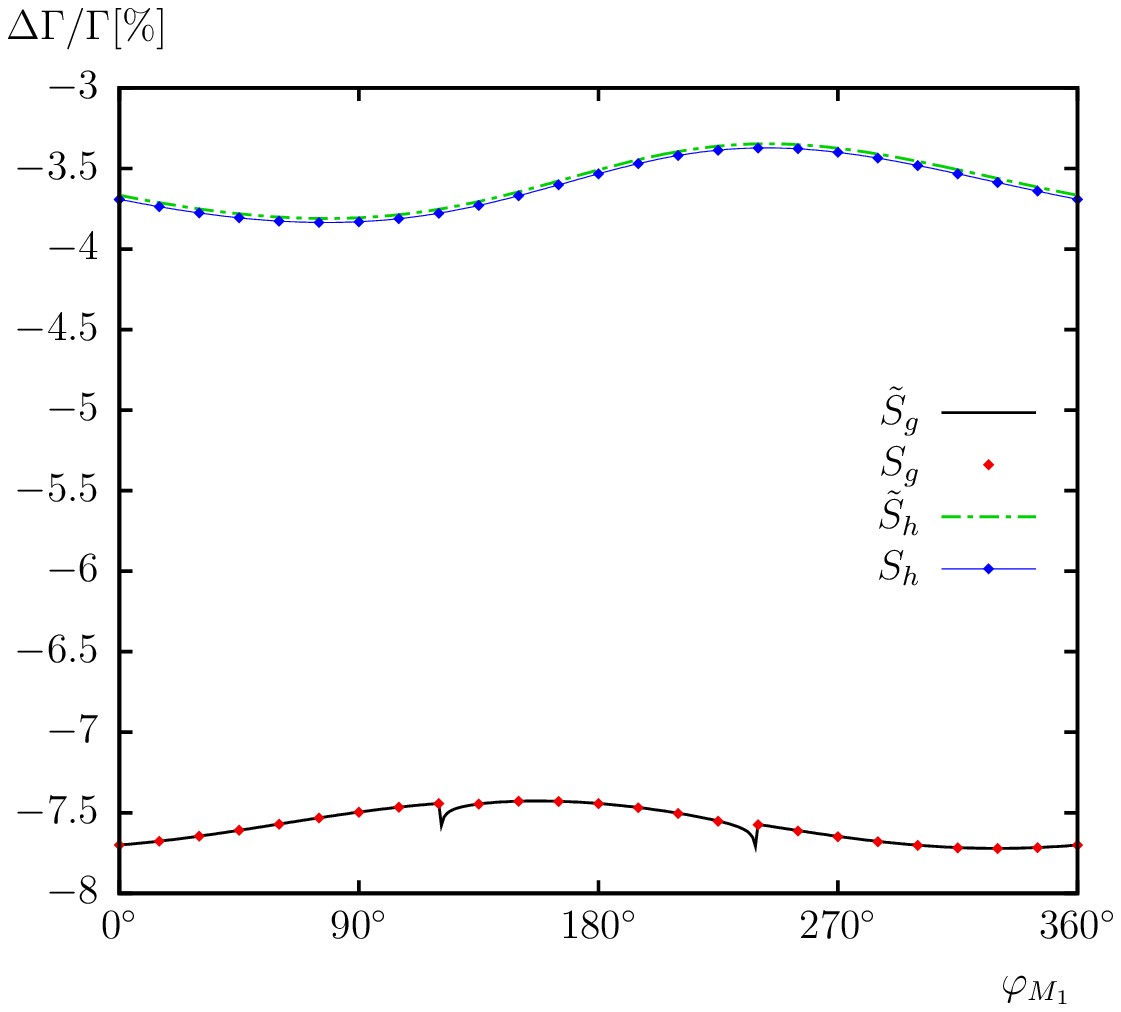} 
\end{tabular}
\caption{
  $\Ga(\DecayNCmW{4}{1})$. 
  Tree-level (``tree'') and full one-loop (``full'') corrected 
  decay widths are shown with the parameters chosen according to 
  \refta{tab:para}, with $\phiMe$ varied.
  The left plot shows the decay width, the right plot shows the relative size of the corrections.
}
\label{fig:PhiM1.neu4cha1w}
\end{center}
\end{figure}

\medskip
Next we analyze the decays $\DecayNCmW{4}{1}$ shown in
\reffis{fig:PhiM1.neu4cha1w}. 
The general
behavior of the decays into $W^+$ is very similar to those into $H^+$
discussed above. This decay yields decay widths around $\sim 1\gev$ in both scenarios, corresponding to
BR's of $\sim 7\% $ and  $\sim 17\% $ for, respectively, $\Sg$ and $\Sh$, 
with a mild dependence on  $\phiMe$.
The one-loop effects are found to be, respectively, of $\sim -8\%$ and $\sim -4\%$
and the same thresholds as in the previous process can be observed as dips in the right panels.
The agreement between the relative corrections is 
at the level of $10^{-5}$ for $\Sg$ and $10^{-4}$ for $\Sh$,
see \refta{tab:rendiff}, and therefore is again too high to be observed.

%%%%%%%%%%%%%%%%%%%%%%%%% F I G U R E %%%%%%%%%%%%%%%%%%%%%%%%%%%%%%%%%%%%%%%%%
%%%%%%%%%%%%%%%%%%%%%%%%% F I G U R E %%%%%%%%%%%%%%%%%%%%%%%%%%%%%%%%%%%%%%%%%
%
%%%%%%%%%%%%%%%%%%%%%%%%% F I G U R E %%%%%%%%%%%%%%%%%%%%%%%%%%%%%%%%%%%%%%%%%
%%%%%%%%%%%%%%%%%%%%%%%%% F I G U R E %%%%%%%%%%%%%%%%%%%%%%%%%%%%%%%%%%%%%%%%%

% %%%%%%%%%%%%%%%%%%%%%%%%%%%%%%%%%%%%%%%%%%%%%%%%%%%%%%%%%%%%%%%%%%%%%%%%%%%%%
\subsection{Decays into neutral Higgs bosons}
\label{sec:DecayNjHk}
Now we turn to the decays involving neutral Higgs bosons. 
The channels  $\DecayNNh{4}{j}{k}$ 
($j=1,2,3;\ k = 1,2,3$) can serve as sources for Higgs
production from SUSY cascades at the LHC, and are therefore of
particular interest. 

\begin{figure}[t!]
\begin{center}
\begin{tabular}{c}
\includegraphics[width=0.49\textwidth,height=7.5cm]{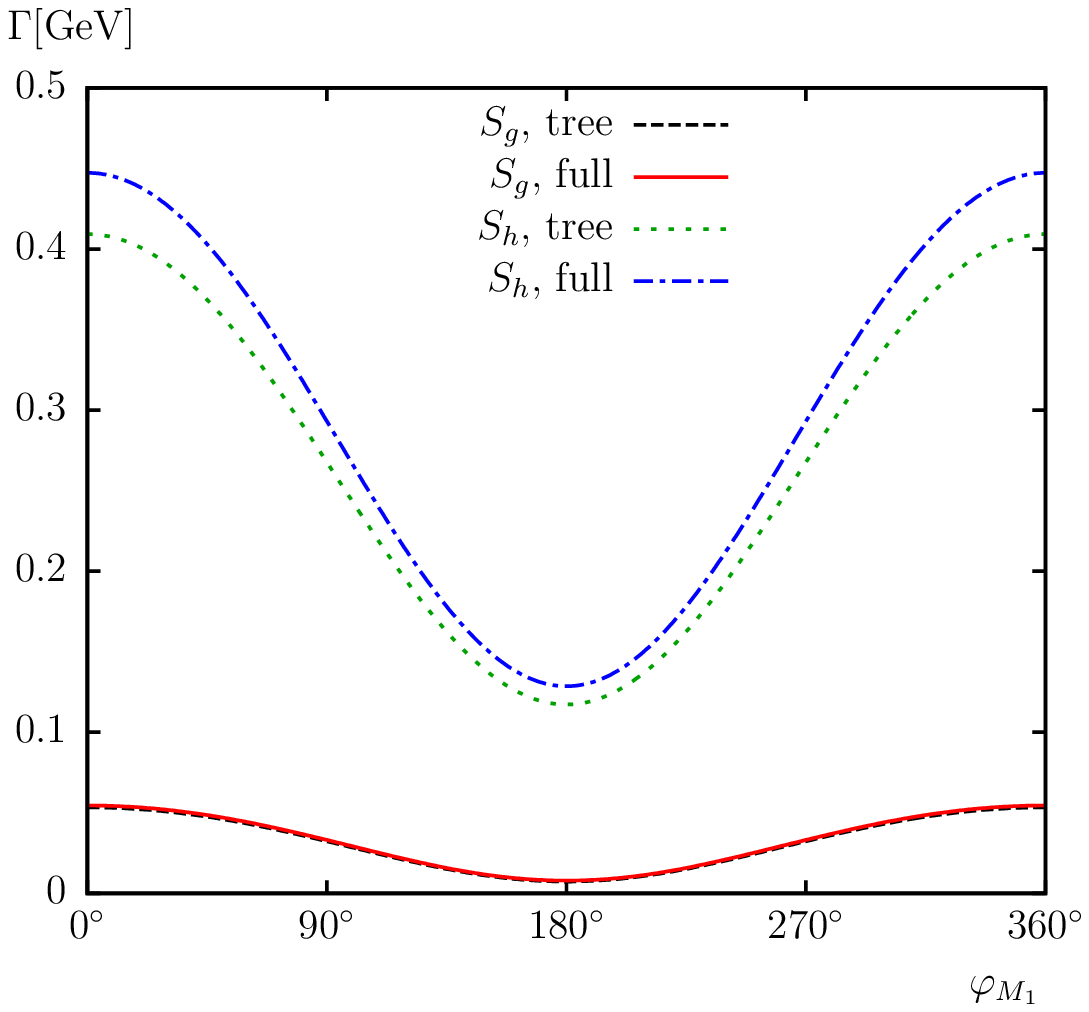}
\hspace{-4mm}
\includegraphics[width=0.49\textwidth,height=7.5cm]{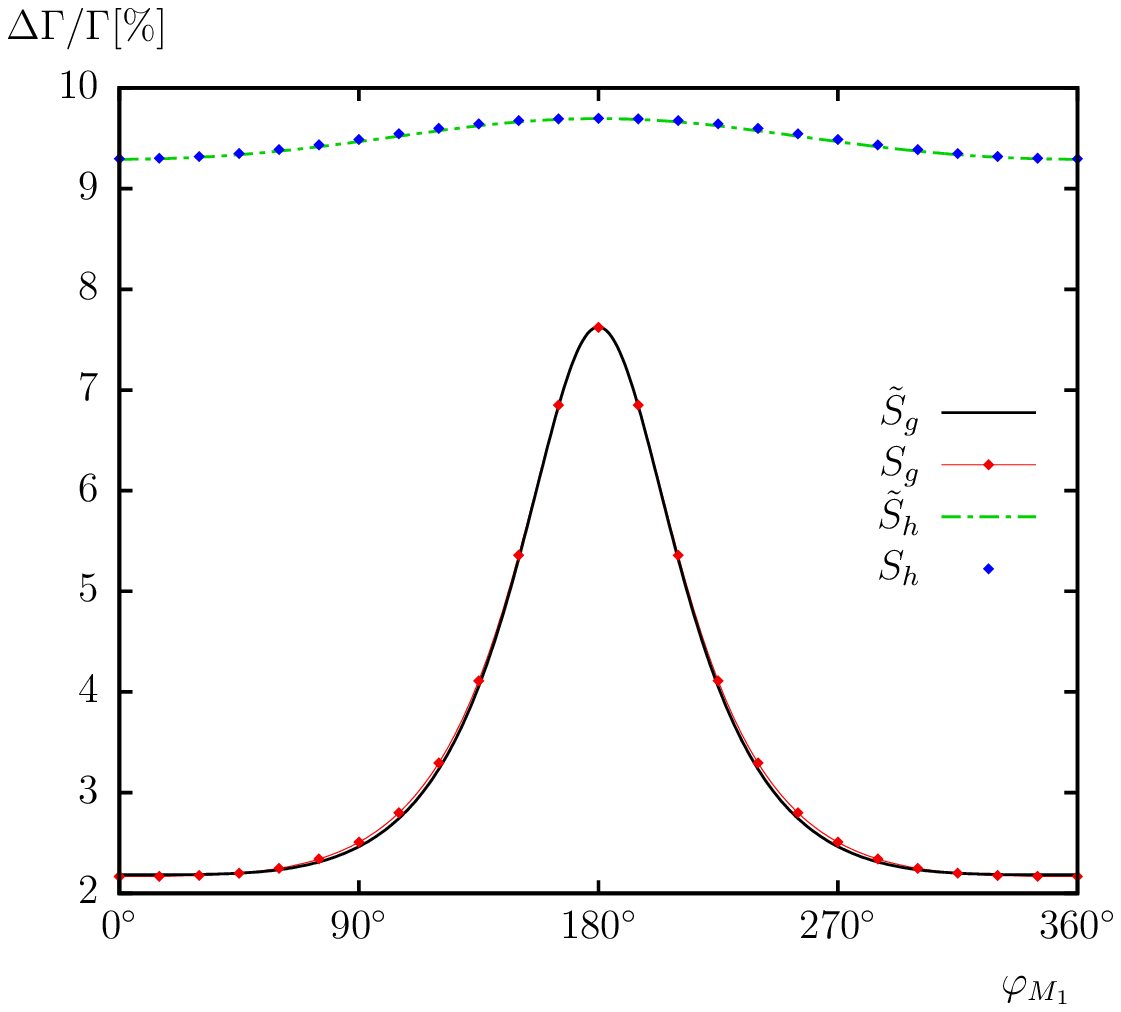} 
\\[1.5em]
\includegraphics[width=0.49\textwidth,height=7.5cm]{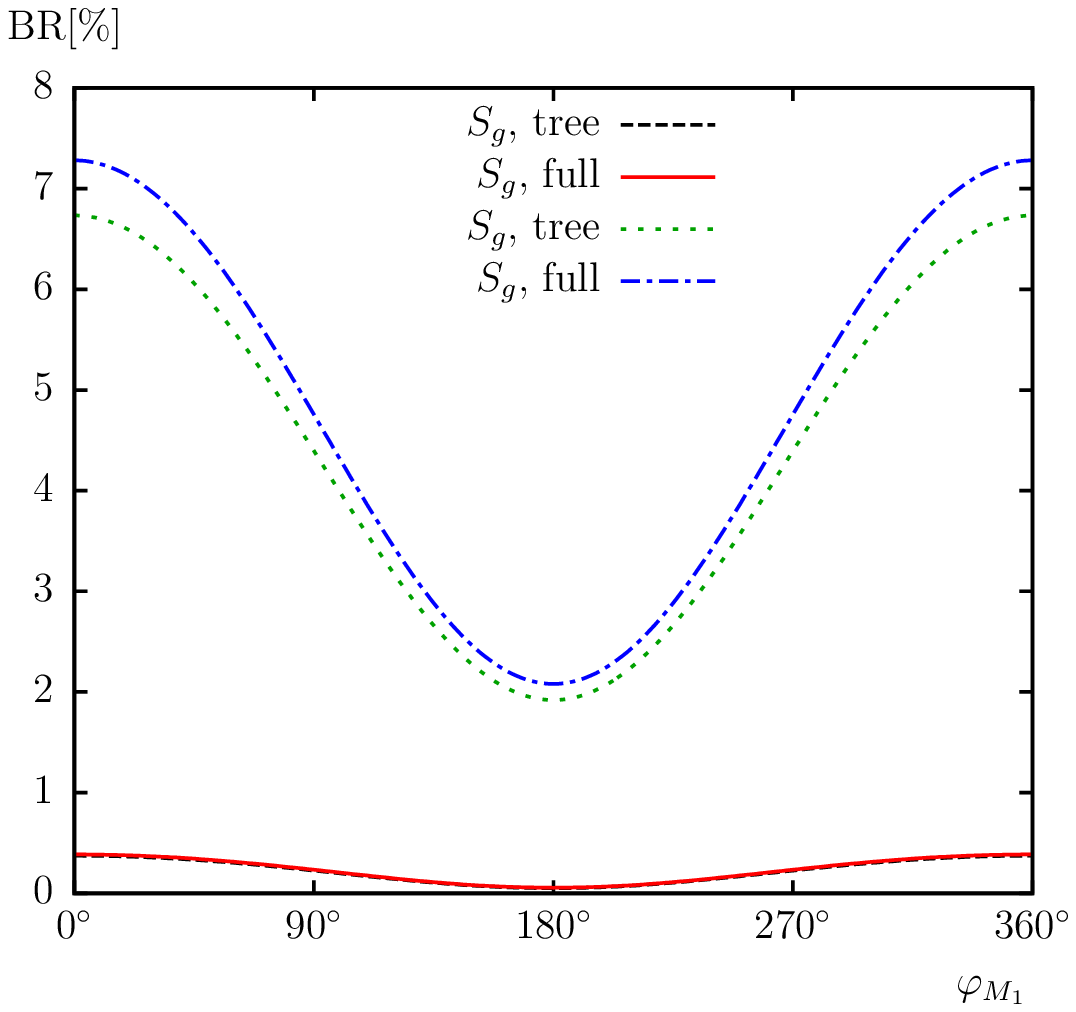}
\hspace{-4mm}
\includegraphics[width=0.49\textwidth,height=7.5cm]{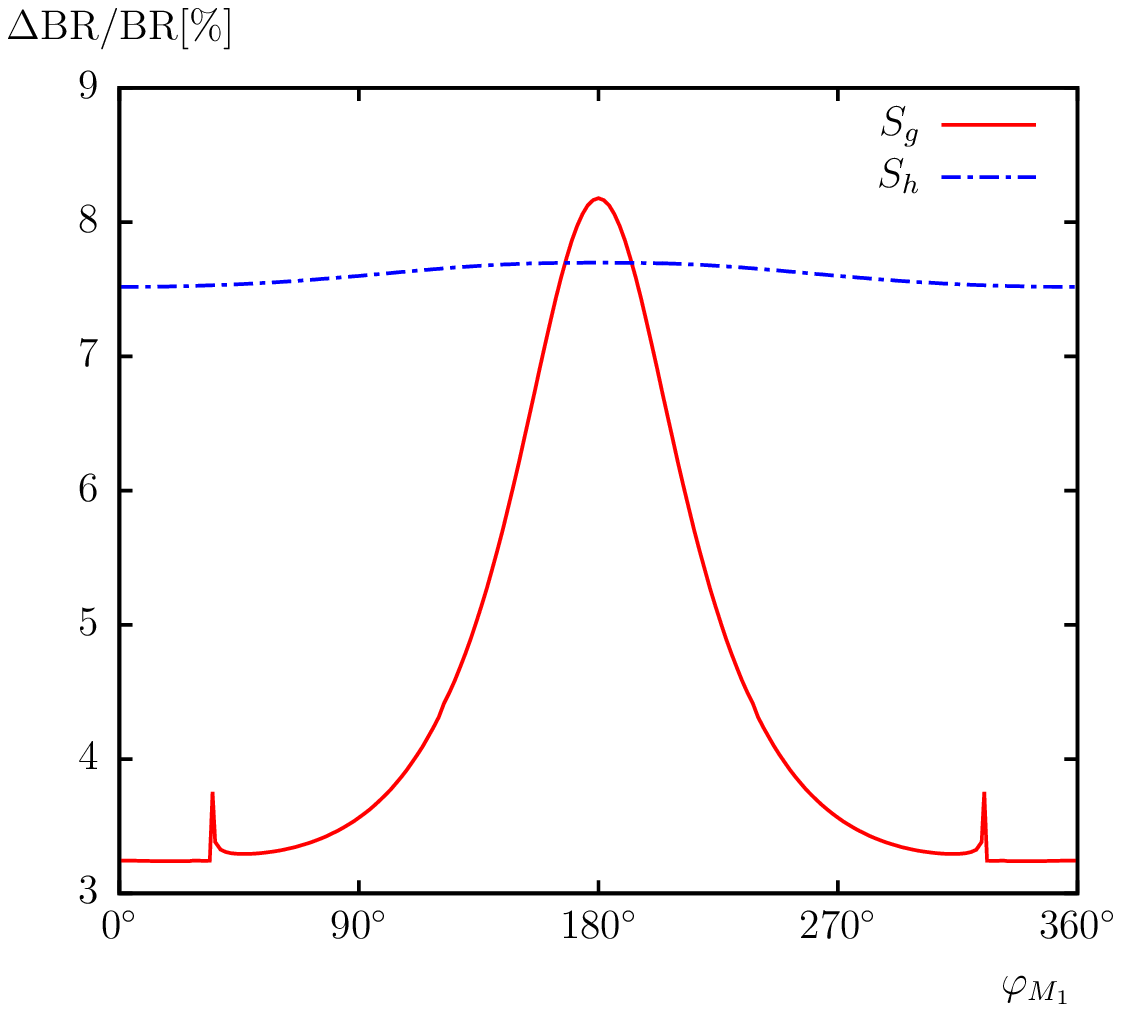}
\end{tabular}
\caption{
  $\Ga(\DecayNNh{4}{1}{1})$. 
  Tree-level (``tree'') and full one-loop (``full'') corrected 
  decay widths are shown with the parameters chosen according to
  \refta{tab:para}, with $\phiMe$ varied.
  The upper left plot shows the decay width, the upper right plot shows the relative size of the corrections.
  The lower left plot shows the BR, the lower right plot shows the relative size of the BR.
}
\label{fig:PhiM1.neu4neu1h1}
\end{center}
\end{figure}

%%% 411
The decay $\DecayNNh{4}{1}{1}$ is shown in \reffi{fig:PhiM1.neu4neu1h1}. 
%%%%%%%%%%%%%%%%%%%%%%%%% F I G U R E %%%%%%%%%%%%%%%%%%%%%%%%%%%%%%%%%%%%%%%%%
%%%%%%%%%%%%%%%%%%%%%%%%% F I G U R E %%%%%%%%%%%%%%%%%%%%%%%%%%%%%%%%%%%%%%%%%
Contrary to what we observed for the decays into charginos, the two scenarios result in 
very different decay widths and both show a strong dependence on $\phiMe$, with partial widths
varying between $ 0.45\gev $ and $ 0.13\gev $ for $\Sh$ and between $0.06 \gev $ and $0.01 \gev$ for $\Sg$.
The strong dependence on the phase $\phiMe$ is a consequence of the change in the relative $\cp$-phase of $\neu{4}$ and $\neu{1}$, 
while the corresponding $\cp$-parity of the light Higgs, which is $\cp$-even here, is not strongly affected. 
For $\phiMe=0$, $\neu{4}$ and $\neu{1}$ have the same relative $\cp$-parity, while for $\phiMe=\pi$ they have the opposite one. 
Therefore, at  $\phi_{\MOne}=\pi$ 
  the decay is p-wave suppressed while at 
   $\phi_{\MOne}=0$ 
   the s-wave mode is allowed. 
In $\Sh$ the partial decay widths are only partially suppressed, due to the relatively large phase space.
   In $\Sg$ the  suppression is stronger for the tree-level amplitude, 
leading to larger relative corrections.
The relative corrections, 
shown in the upper right panel, are $\sim 10\%$ for $\Sh$~and are between $2\% $ and $8\%$ for $\Sg$.
For {$\Sg$}~we observe a small difference between the two schemes 
at $\phiMe\approx \pm 90^\circ$ , of the order of $0.1\%$ 
(see also \refta{tab:rendiff}). 
It should be noted that here, as opposed to the case of the previous decays, 
the fact that the tree-level decay width depends strongly on $\phiMe$ leads to a noticeable difference between the two schemes.
This difference has been highlighted in \reffi{fig:PhiM1.rendiff}.
In the lower left panel we show the branching ratios and in the lower right panel its relative corrections.
Since the difference between the schemes is here very small we only show these results in scheme~I.
The peaks at $\phiMe=35^\circ$ and $325^\circ$ are due to the threshold for the decay $\DecayNNh{2}{1}{1}$, 
which leads to a singularity which affects the total width
(see the discussion below on these threshold effects).
It should be noted that the decay widths and the corresponding branching ratios, as well as their relative corrections, are roughly proportional here because the total width of $\neu{4}$,
shown in \reffi{fig:phiM1.neu4tot} below, is almost independent of $\phiMe$ in both scenarios,
see the discussion in \refse{sec:rendiff}.

\begin{figure}[t!]
\begin{center}
\begin{tabular}{c}
\includegraphics[width=0.49\textwidth,height=7.5cm]{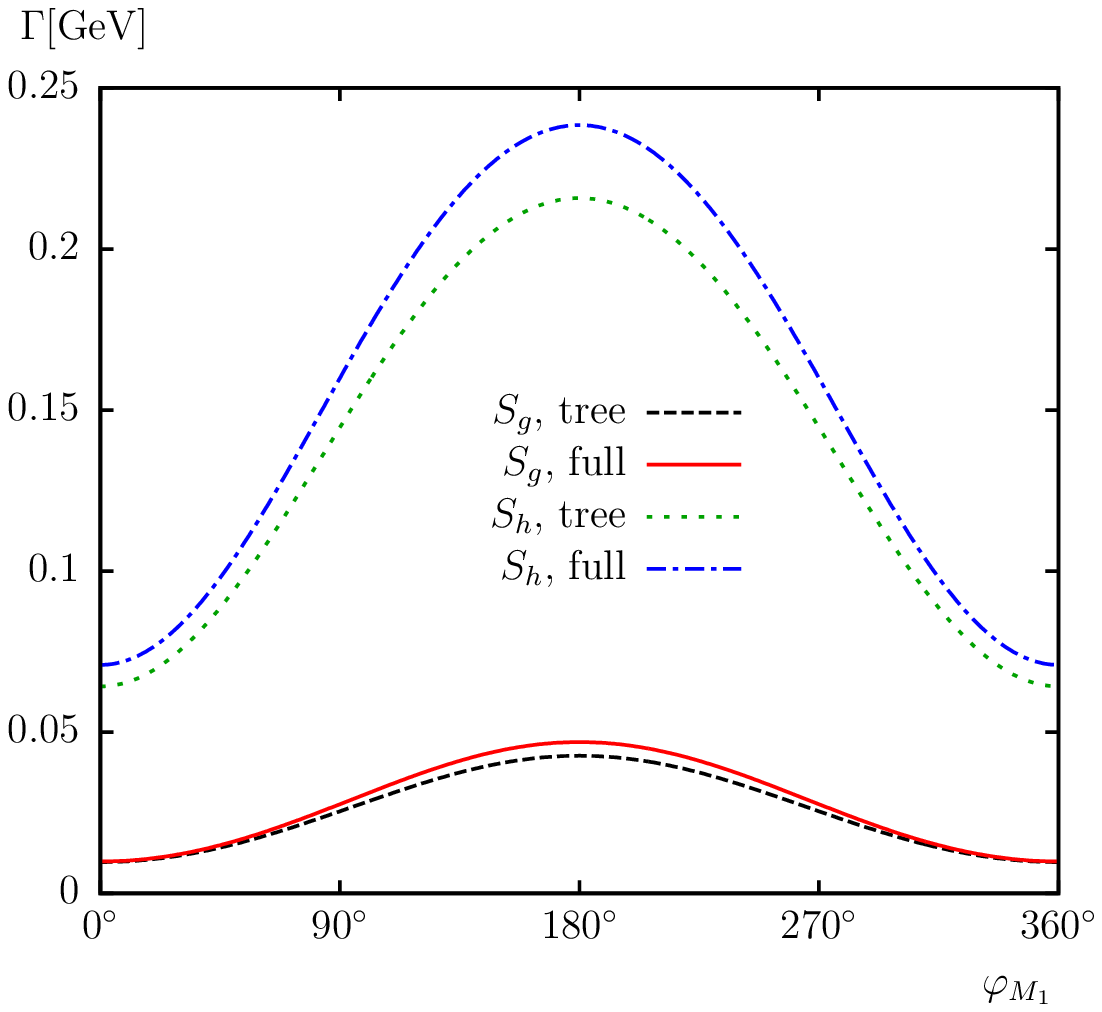}
\hspace{-4mm}
\includegraphics[width=0.49\textwidth,height=7.5cm]{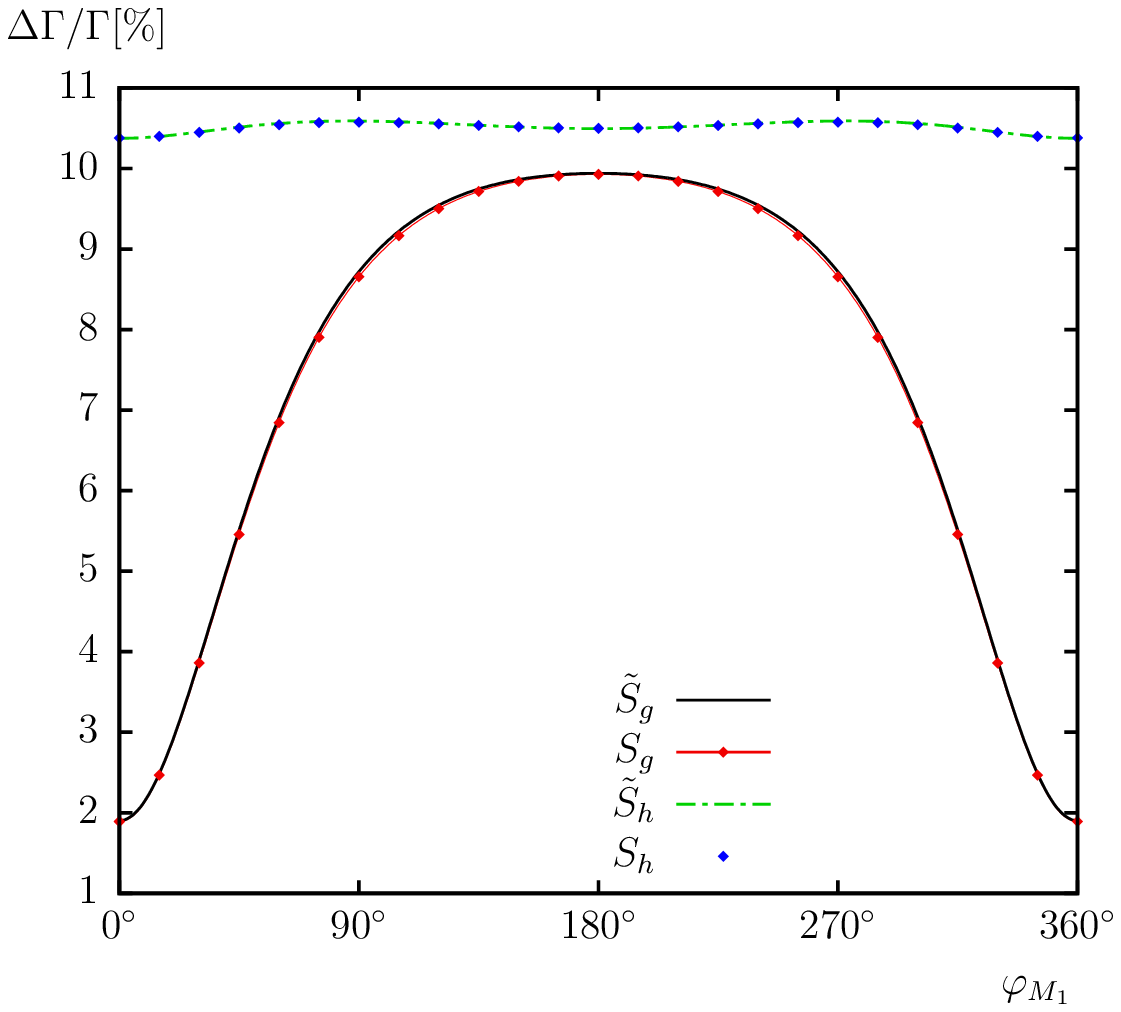} 
\end{tabular}
\caption{
  $\Ga(\DecayNNh{4}{1}{2})$. 
  Tree-level (``tree'') and full one-loop (``full'') corrected 
  decay widths are shown with the parameters chosen according to 
  \refta{tab:para}, with $\phiMe$ varied.
  The  left plot shows the decay width, the right plot shows the relative size of the corrections.
}
\label{fig:PhiM1.neu4neu1h2}
\end{center}
\end{figure}

%%% 412
The results for the decay $\DecayNNh{4}{1}{2}$ are shown in \reffi{fig:PhiM1.neu4neu1h2}.
Since in both scenarios $h_2$ tends to the $\cp$-odd Higgs boson for real couplings,
the dependence on $\phiMe$ is opposite to that for the decay $\DecayNNh{4}{1}{1}$ discussed above,
with a p-wave suppression at $\phiMe=0$ and with the s-wave mode being allowed at $\phi_{\MOne}=\pi$. 
The decay widths are a factor two smaller than those for the decay into the lightest Higgs for $\Sh$, and of the same order for $\Sg$.
The relative corrections are also similar, of order $\sim 10\% $ in $\Sh$ and between $2\% $ and $10\% $ for $\Sg$.
We can also observe a small difference at large $\phiMe$ between the two schemes in $\Sg$, of the order of $0.05\% $, 
(see \refta{tab:rendiff}).

%%%%%%%%%%%%%%%%%%%%%%%%% F I G U R E %%%%%%%%%%%%%%%%%%%%%%%%%%%%%%%%%%%%%%%%%
%%%%%%%%%%%%%%%%%%%%%%%%% F I G U R E %%%%%%%%%%%%%%%%%%%%%%%%%%%%%%%%%%%%%%%%%

\begin{figure}[t!]
\begin{center}
\begin{tabular}{c}
\includegraphics[width=0.49\textwidth,height=7.5cm]{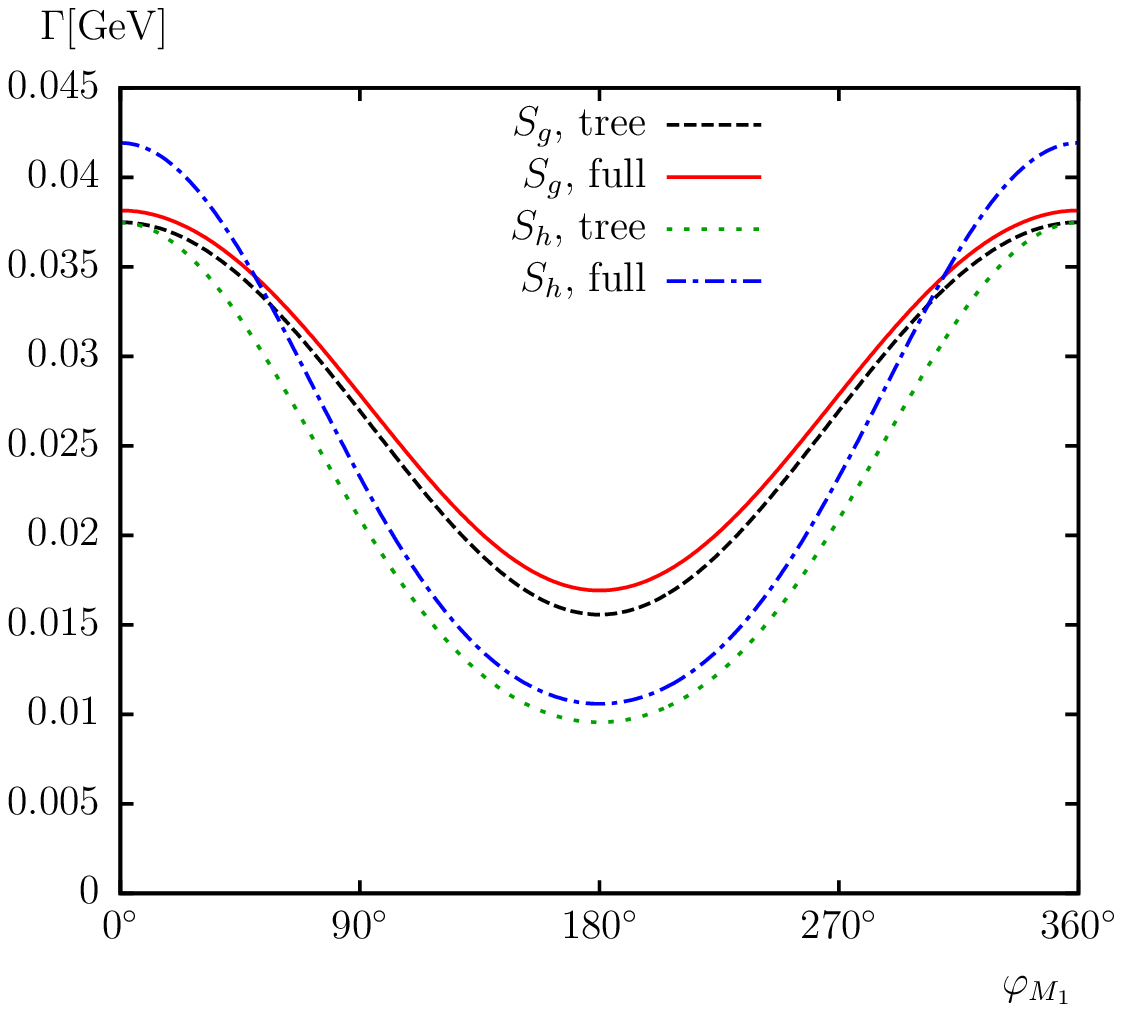}
\hspace{-4mm}
\includegraphics[width=0.49\textwidth,height=7.5cm]{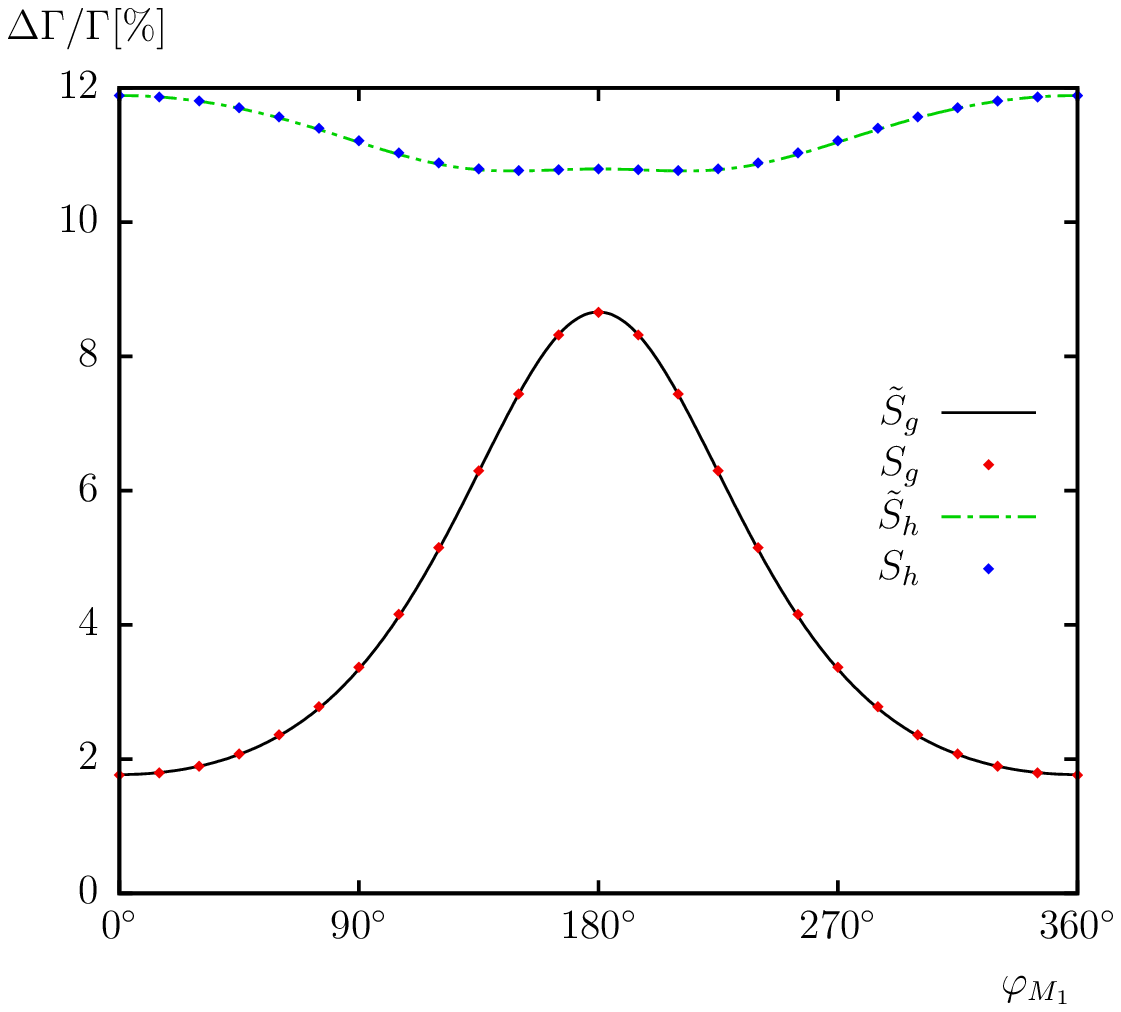} 
\end{tabular}
\caption{
  $\Ga(\DecayNNh{4}{1}{3})$. 
  Tree-level (``tree'') and full one-loop (``full'') corrected 
  decay widths are shown with the parameters chosen according to
  \refta{tab:para}, with $\phiMe$ varied.
  The  left plot shows the decay width, the right plot shows the relative size of the corrections.
}
\label{fig:PhiM1.neu4neu1h3}
\end{center}
\end{figure}

%%% 413
The decay $\DecayNNh{4}{1}{3}$ is shown in \reffi{fig:PhiM1.neu4neu1h3}. 
The $\phiMe$ dependence of the partial width and its relative correction for this process 
is qualitatively similar to that of $\DecayNNh{4}{1}{1}$.
However, for $\Sh$ the partial width is much smaller, between $0.01 \gev $ and $0.04 \gev $.
As the difference between the two schemes is of the order of $0.01\% $, 
(see \refta{tab:rendiff}), 
they can not be visibly distinguished here. 

%%%%%%%%%%%%%%%%%%%%%%%%% F I G U R E %%%%%%%%%%%%%%%%%%%%%%%%%%%%%%%%%%%%%%%%%
%%%%%%%%%%%%%%%%%%%%%%%%% F I G U R E %%%%%%%%%%%%%%%%%%%%%%%%%%%%%%%%%%%%%%%%%

\begin{figure}[t!]
\begin{center}
\begin{tabular}{c}
\includegraphics[width=0.49\textwidth,height=7.5cm]{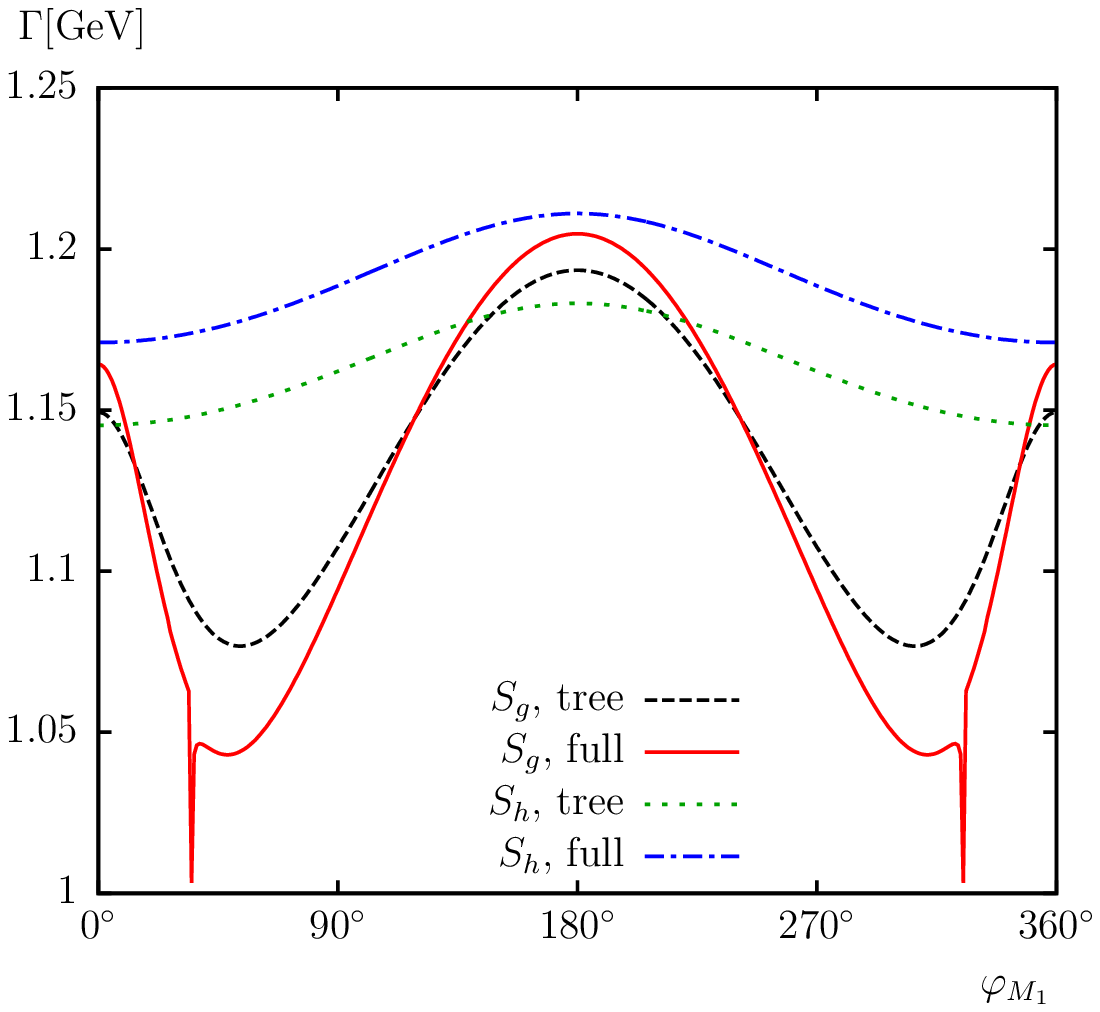}
\hspace{-4mm}
\includegraphics[width=0.49\textwidth,height=7.5cm]{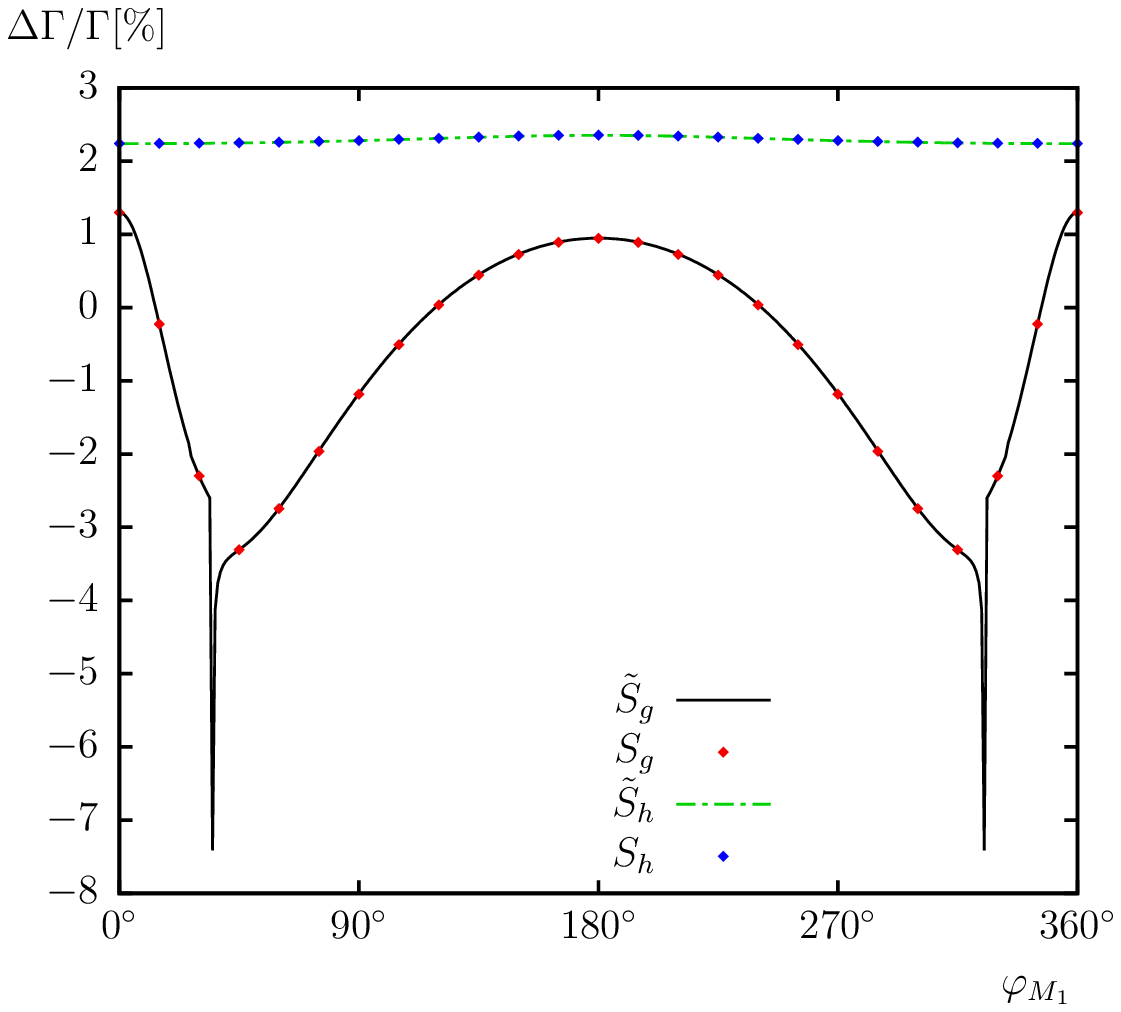} 
\end{tabular}
\caption{
  $\Ga(\DecayNNh{4}{2}{1})$. 
  Tree-level (``tree'') and full one-loop (``full'') corrected 
  decay widths are shown with the parameters chosen according to 
  \refta{tab:para}, with $\phiMe$ varied.
  The left plot shows the decay width, the right plot shows the relative size of the corrections.
}
\label{fig:PhiM1.neu4neu2h1}
\end{center}
\end{figure}

\begin{figure}[t!]
\begin{center}
\begin{tabular}{c}
\includegraphics[width=0.49\textwidth,height=7.5cm]{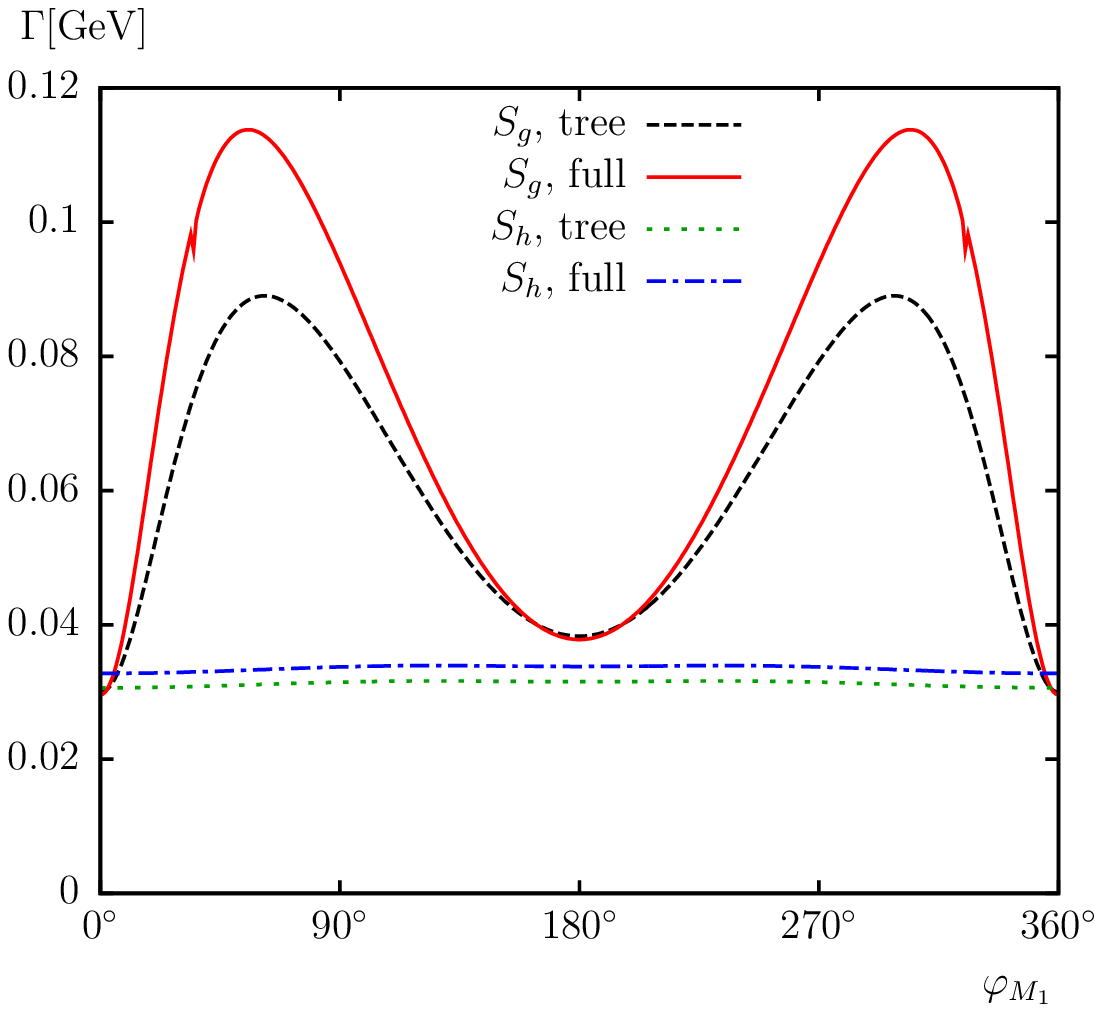}
\hspace{-4mm}
\includegraphics[width=0.49\textwidth,height=7.5cm]{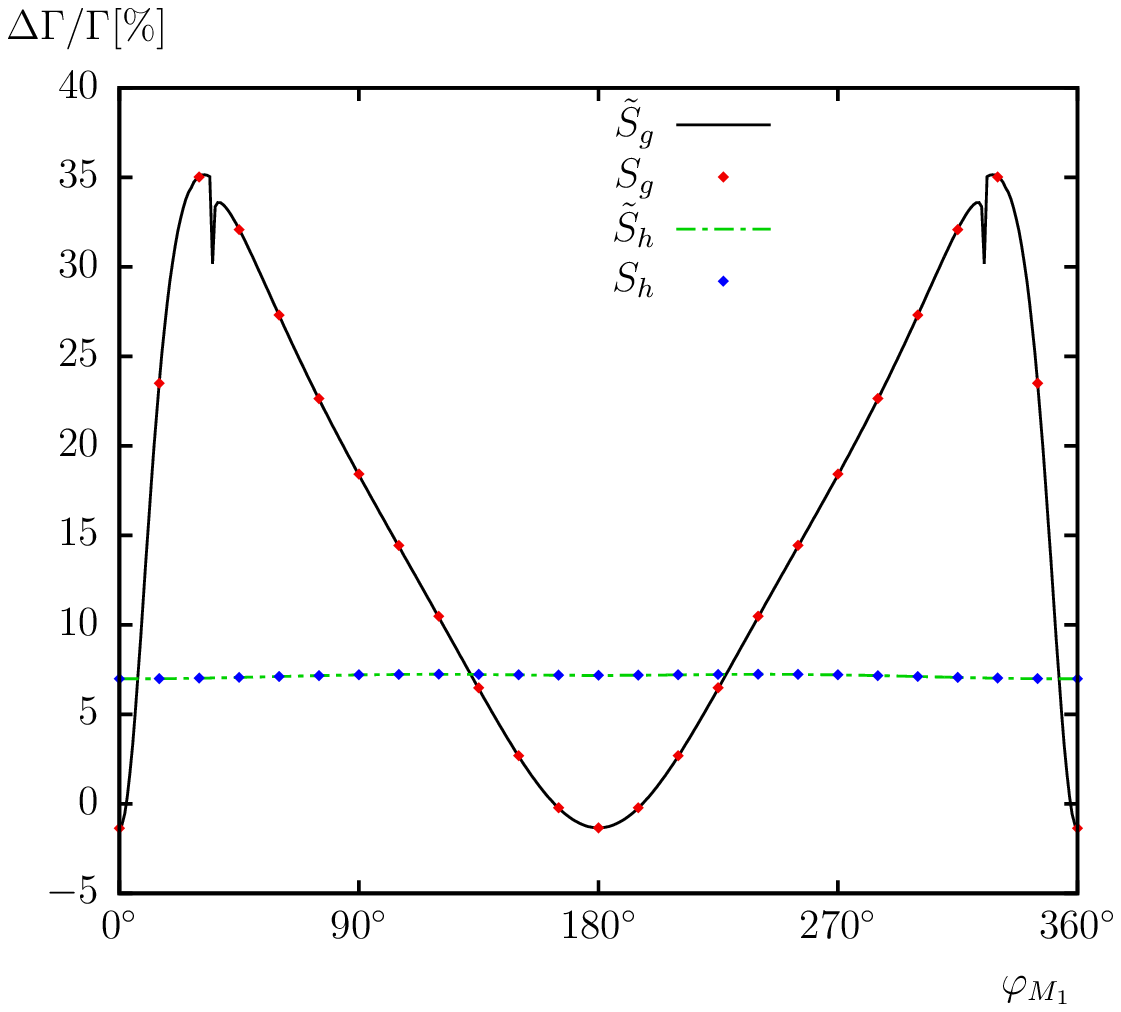} 
\end{tabular}
\caption{
  $\Ga(\DecayNNh{4}{2}{2})$. 
  Tree-level (``tree'') and full one-loop (``full'') corrected 
  decay widths are shown with the parameters chosen according to 
  \refta{tab:para}, with $\phiMe$ varied.
  The left plot shows the decay width, the right plot shows the relative size of the corrections.
}
\label{fig:PhiM1.neu4neu2h2}
\end{center}
\end{figure}

\begin{figure}[t!]
\begin{center}
\begin{tabular}{c}
\includegraphics[width=0.49\textwidth,height=7.5cm]{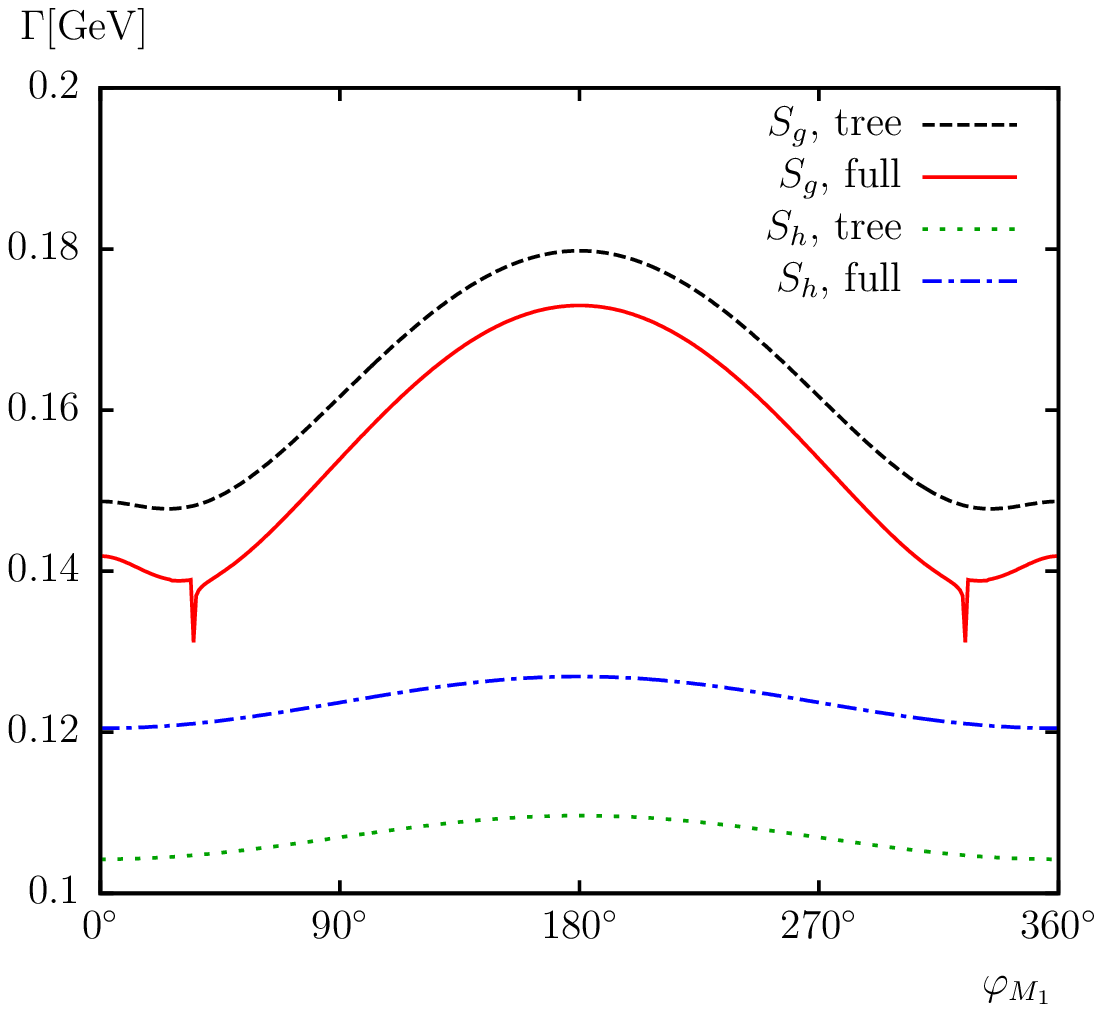}
\hspace{-4mm}
\includegraphics[width=0.49\textwidth,height=7.5cm]{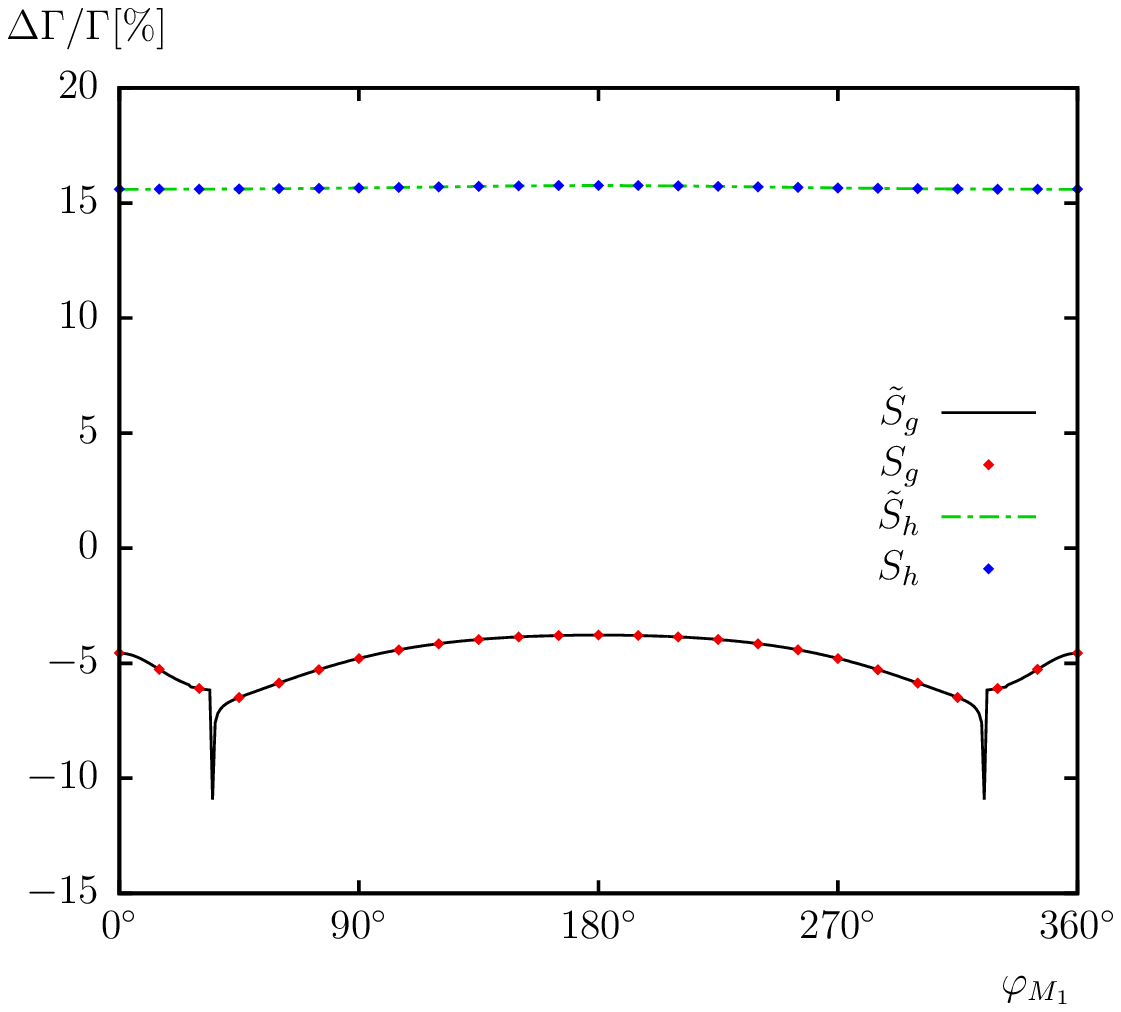} 
\end{tabular}
\caption{
  $\Ga(\DecayNNh{4}{2}{3})$. 
  Tree-level (``tree'') and full one-loop (``full'') corrected 
  decay widths are shown with the parameters chosen according to 
  \refta{tab:para}, with $\phiMe$ varied.
  The left plot shows the decay width, the right plot shows the relative size of the corrections.
}
\label{fig:PhiM1.neu4neu2h3}
\end{center}
\end{figure}

\medskip
The decays $\DecayNNh{4}{2}{1}$, $\DecayNNh{4}{2}{2}$,  $\DecayNNh{4}{2}{3}$ are shown in \reffis{fig:PhiM1.neu4neu2h1}, \ref{fig:PhiM1.neu4neu2h2}  and \ref{fig:PhiM1.neu4neu2h3}, respectively. 
In scenario $\Sh$ the second lightest neutralino is mainly wino-like, with a small mixing with the bino component, leading to a  weak dependence on $\phiMe$. 
For $\Sg$~the dependence on $\phiMe$ is much larger since the second lightest neutralino has both large Higgsino and bino components.

%%%%%%%%%%%%%%%%%%%%%%%%% F I G U R E %%%%%%%%%%%%%%%%%%%%%%%%%%%%%%%%%%%%%%%%%
%%%%%%%%%%%%%%%%%%%%%%%%% F I G U R E %%%%%%%%%%%%%%%%%%%%%%%%%%%%%%%%%%%%%%%%%
%%% 421
For $\DecayNNh{4}{2}{1}$, in $\Sg$,  as it is shown in \reffi{fig:PhiM1.neu4neu2h1}, 
the decay width oscillates from $\sim1.05$ to $1.2\gev$, 
and for $\Sh$ the decay width is $\sim 1.2\gev $. 
It should be noted that this decay is s-wave mode allowed, and therefore 
the decay widths are $2.5$ to $9$ times larger than the decay into the lightest Higgs and neutralino. 
This turns out to be the dominating process, 
with branching ratios of up to $\sim 8\%$ and $\sim 19\%$ for, respectively, $\Sg$ and $\Sh$.
The larger $\phiMe$-dependence in  $\Sg$  is due to the strong bino-Higgsino mixing of $\neu{2}$ in this scenario. 
This feature will be equally relevant for the remaining decays into either $\neu{2}$ or $\neu{3}$ discussed below.
For $\Sg$, we observe %, 
the effect of the threshold for $\DecayNNh{2}{1}{1}$ at $\phiMe=35^\circ$ and $325^\circ$. 
The dips are due to the resulting singular behavior of the derivatives of the self energies entering the field renormalization constants. 
This effect will also be observed in the other decays to a $\neu{2}$, 
see Figs.~\ref{fig:PhiM1.neu4neu2h2}, \ref{fig:PhiM1.neu4neu2h3}, and \ref{fig:PhiM1.neu4neu2z},
in the total width of $\neu{4}$, Fig.~\ref{fig:phiM1.neu4tot}, and in the branching ratios, see e.g.\ \reffi{fig:PhiM1.neu4neu1h1}.
It should be noted that both schemes have the same dips at $\phiMe=35^\circ$ and $325^\circ$.
This will be true as well for the other decays to $\neu{2}$ described below.
The corrections are relatively small, $\sim 1\%$ to $-3\%$ for $\Sh$~and $\sim 2\%$ for $\Sh$. 
In \reffi{fig:PhiM1.neu4neu2h1}, 
the renormalization schemes cannot be visibly distinguished from each other. 
The difference 
is below $0.01\% $ for $\Sg$, and below $10^{-5} $ for $\Sh$, 
see \refta{tab:rendiff} for details.

%%% 422
For $\DecayNNh{4}{2}{2}$, in $\Sg$, as shown in \reffi{fig:PhiM1.neu4neu2h2}, 
the decay width oscillates from $\sim0.03$ to $0.11\gev$, 
while for  $\Sh$  it is $\sim 0.03\gev $.
The corresponding branching ratios in $\Sg$ and $\Sh$ are, respectively,   $\sim0.2-0.8\%$ and $\sim 0.5\%$.
It should be noted that this decay is {s-wave} mode suppressed, 
and the decay width is therefore smaller than that for the decays into $h_1$ and $h_3$. 
Again for $\Sg$ the effect of the threshold for $\DecayNNh{2}{1}{1}$ at $\phiMe=35^\circ$ and $325^\circ$ is visible. 
The corrections are comparatively large for $\Sg$, $\sim 0$ to $35\%$, and for $\Sh$~are $\sim 6\%$. 
Despite a relatively significant difference between the schemes of $0.05\%$ for $\Sg$, in \reffi{fig:PhiM1.neu4neu2h2} 
this remains invisible.
%(see \refta{tab:rendiff})
The large relative corrections in $\Sg$ are a consequence of both the suppressed tree-level result, 
as well as the strong effect of the corrections on the mixing of the second lightest neutralino.

%%%%%%%%%%%%%%%%%%%%%%%%% F I G U R E %%%%%%%%%%%%%%%%%%%%%%%%%%%%%%%%%%%%%%%%%
%%%%%%%%%%%%%%%%%%%%%%%%% F I G U R E %%%%%%%%%%%%%%%%%%%%%%%%%%%%%%%%%%%%%%%%%

%%% 423
The decay $\DecayNNh{4}{2}{3}$, shown in \reffi{fig:PhiM1.neu4neu2h3}, 
 is s-wave mode allowed, and qualitatively similar to $\DecayNNh{4}{2}{1}$. 
However, its decay width is smaller due to phase space suppression, 
oscillating from $0.14$ to $0.18\gev$ for $\Sg$ and $0.12\gev$ for $\Sh$, 
and corrections are sizeable, $\sim-5\%$ and $15\%$ for $\Sg$~and $\Sh$,
 respectively. 
The two schemes cannot be visibly distinguished here.
The difference between the schemes reaches $0.01\%$ for $\Sg$, and again 
this remains invisible (see \refta{tab:rendiff}).

%%%%%%%%%%%%%%%%%%%%%%%%% F I G U R E %%%%%%%%%%%%%%%%%%%%%%%%%%%%%%%%%%%%%%%%%
%%%%%%%%%%%%%%%%%%%%%%%%% F I G U R E %%%%%%%%%%%%%%%%%%%%%%%%%%%%%%%%%%%%%%%%%

\begin{figure}[t!]
\begin{center}
\begin{tabular}{c}
\includegraphics[width=0.49\textwidth,height=7.5cm]{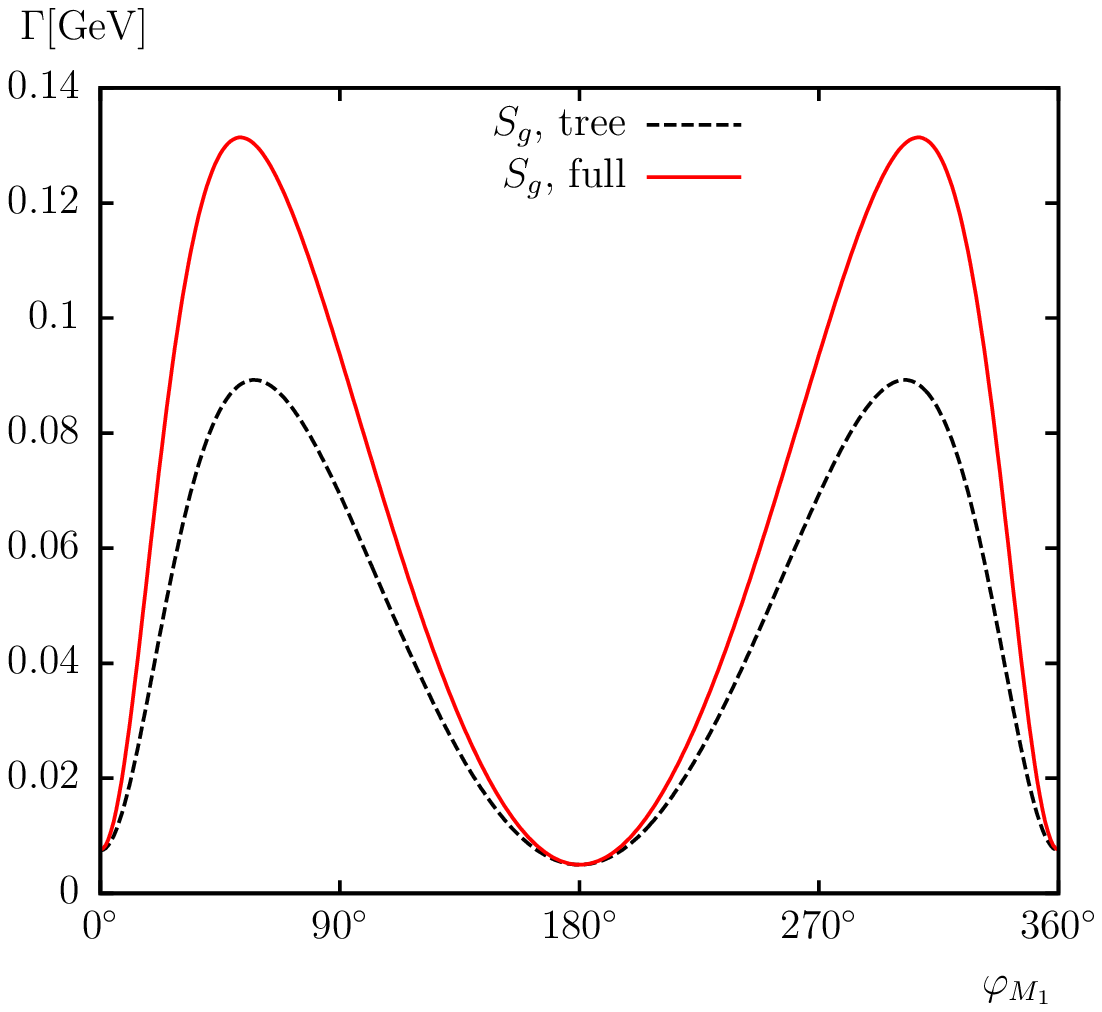}
\hspace{-4mm}
\includegraphics[width=0.49\textwidth,height=7.5cm]{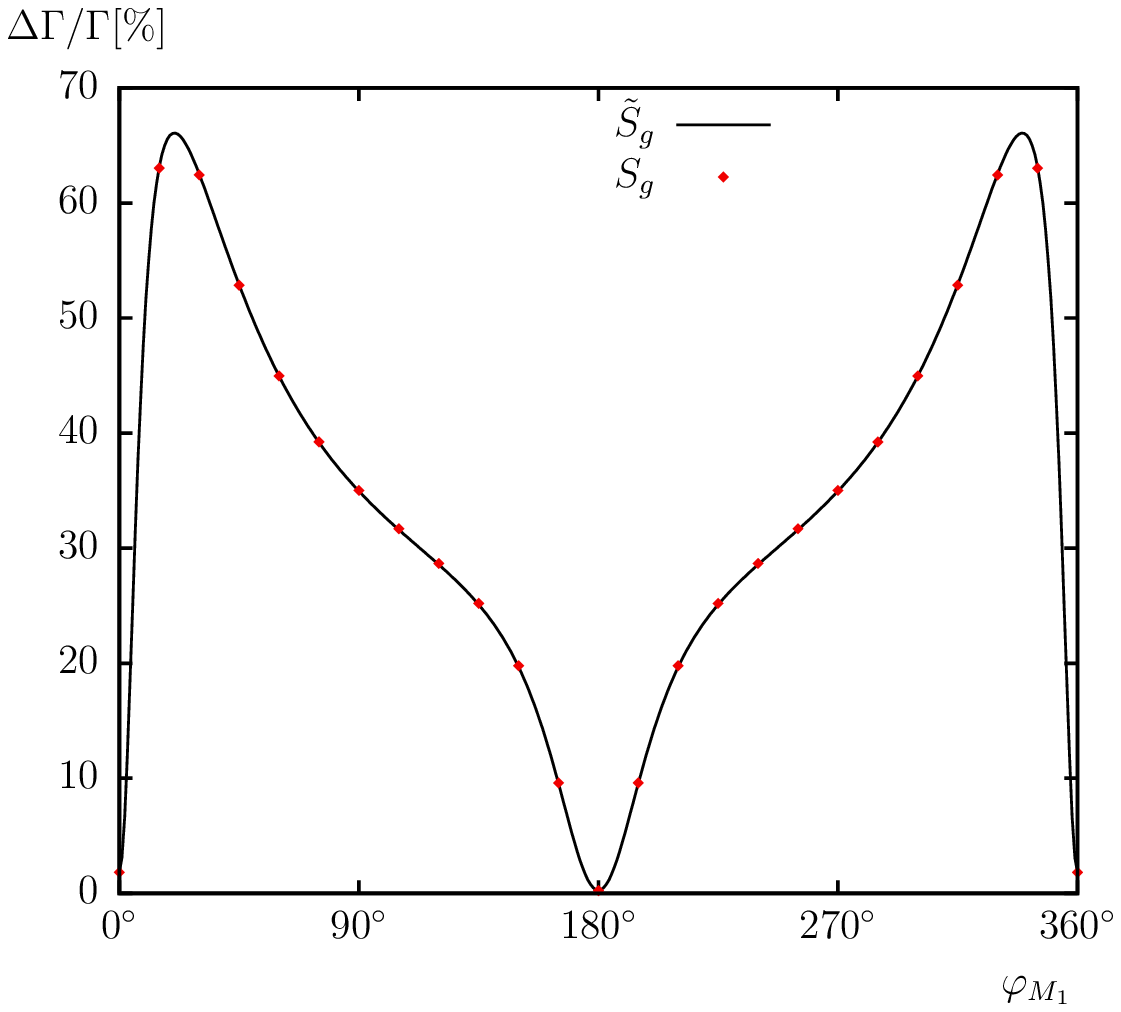} 
\end{tabular}
\caption{
  $\Ga(\DecayNNh{4}{3}{1})$. 
  Tree-level (``tree'') and full one-loop (``full'') corrected 
  decay widths are shown with the parameters chosen according to 
  \refta{tab:para}, with $\phiMe$ varied.
  The left plot shows the decay width, the right plot shows the relative size of the corrections.
}
\label{fig:PhiM1.neu4neu3h1}
\end{center}
\end{figure}

\begin{figure}[t!]
\begin{center}
\begin{tabular}{c}
\includegraphics[width=0.49\textwidth,height=7.5cm]{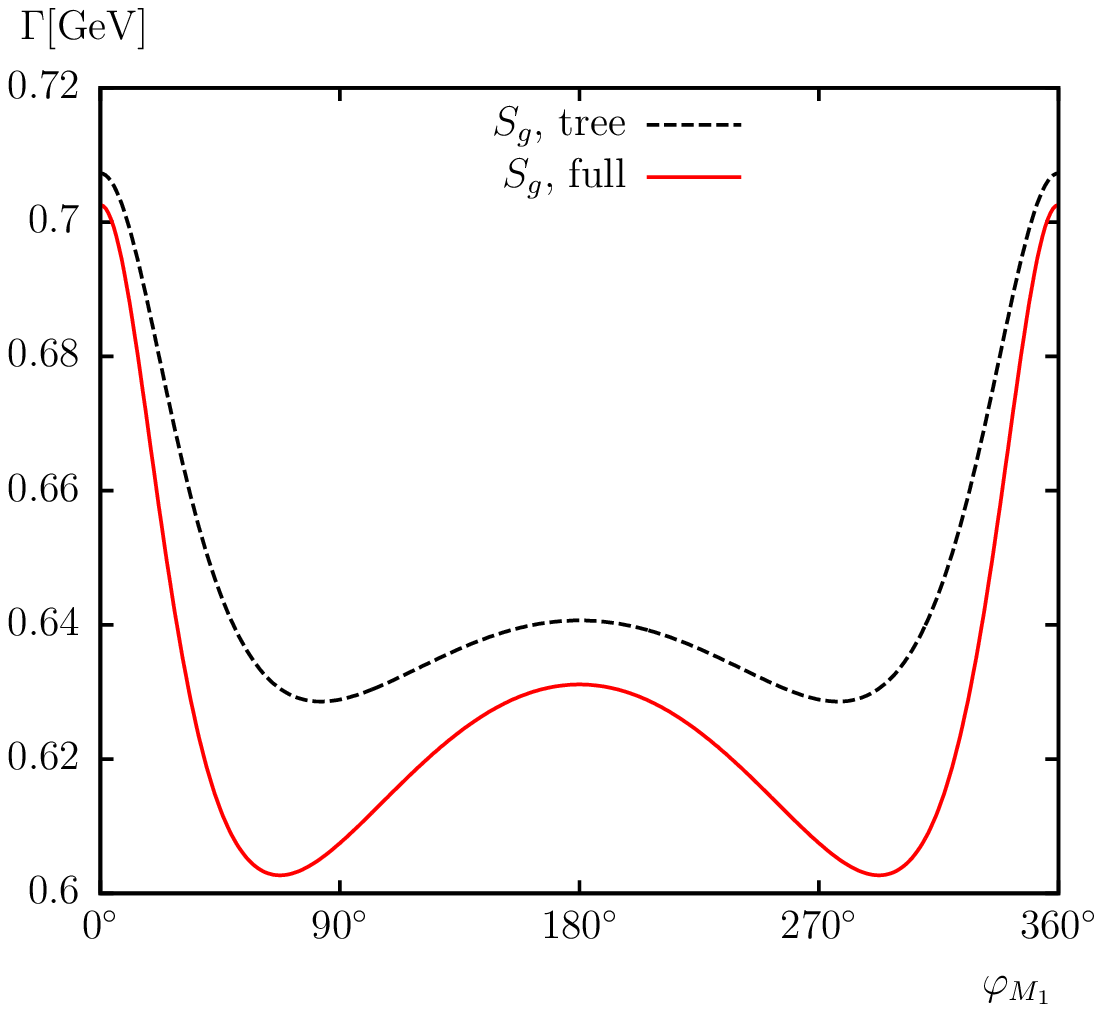}
\hspace{-4mm}
\includegraphics[width=0.49\textwidth,height=7.5cm]{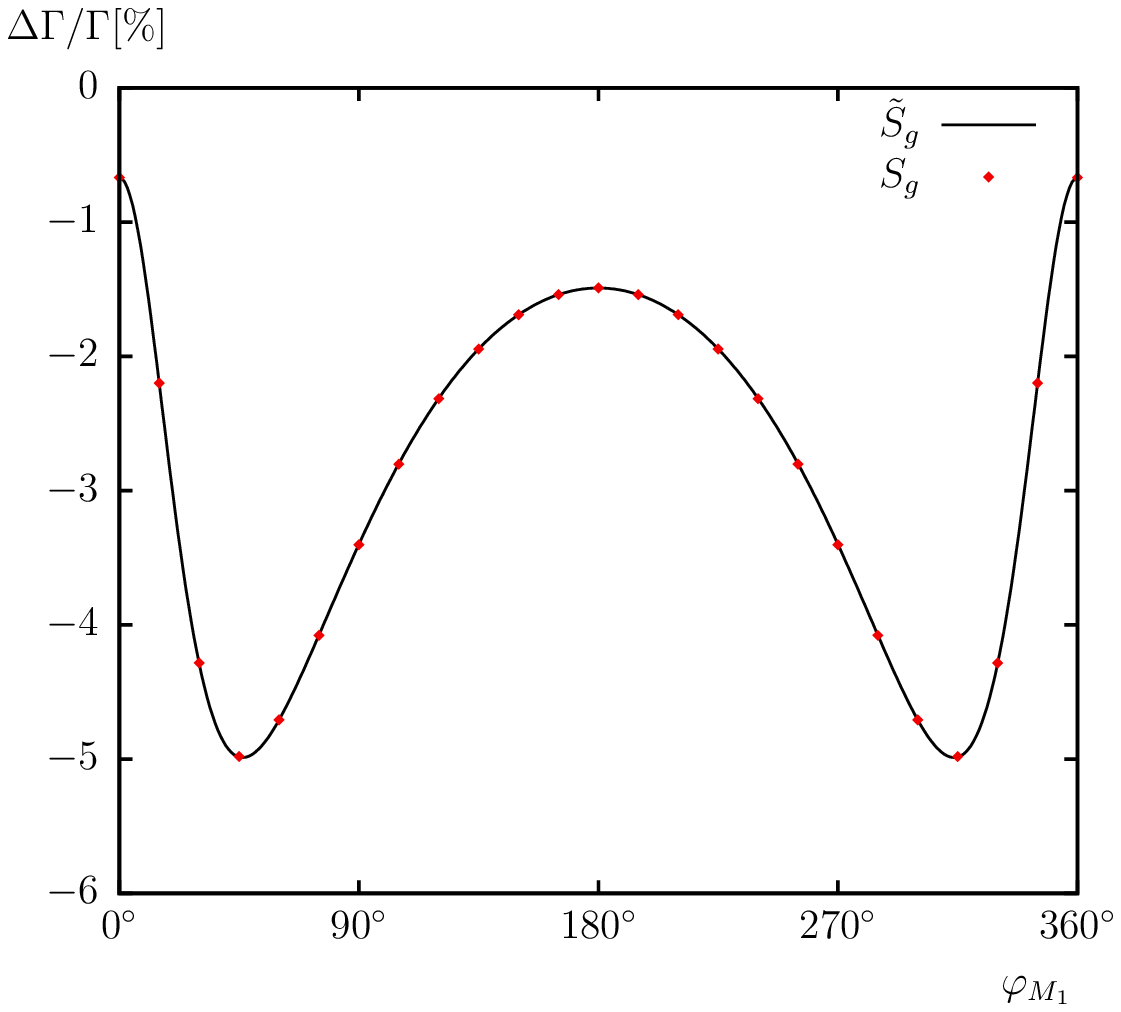} 
\end{tabular}
\caption{
  $\Ga(\DecayNNh{4}{3}{2})$. 
  Tree-level (``tree'') and full one-loop (``full'') corrected 
  decay widths are shown with the parameters chosen according to 
  \refta{tab:para}, with $\phiMe$ varied.
  The left plot shows the decay width, the right plot shows the relative size of the corrections.
}
\label{fig:PhiM1.neu4neu3h2}
\end{center}
\end{figure}

\begin{figure}[t!]
\begin{center}
\begin{tabular}{c}
\includegraphics[width=0.49\textwidth,height=7.5cm]{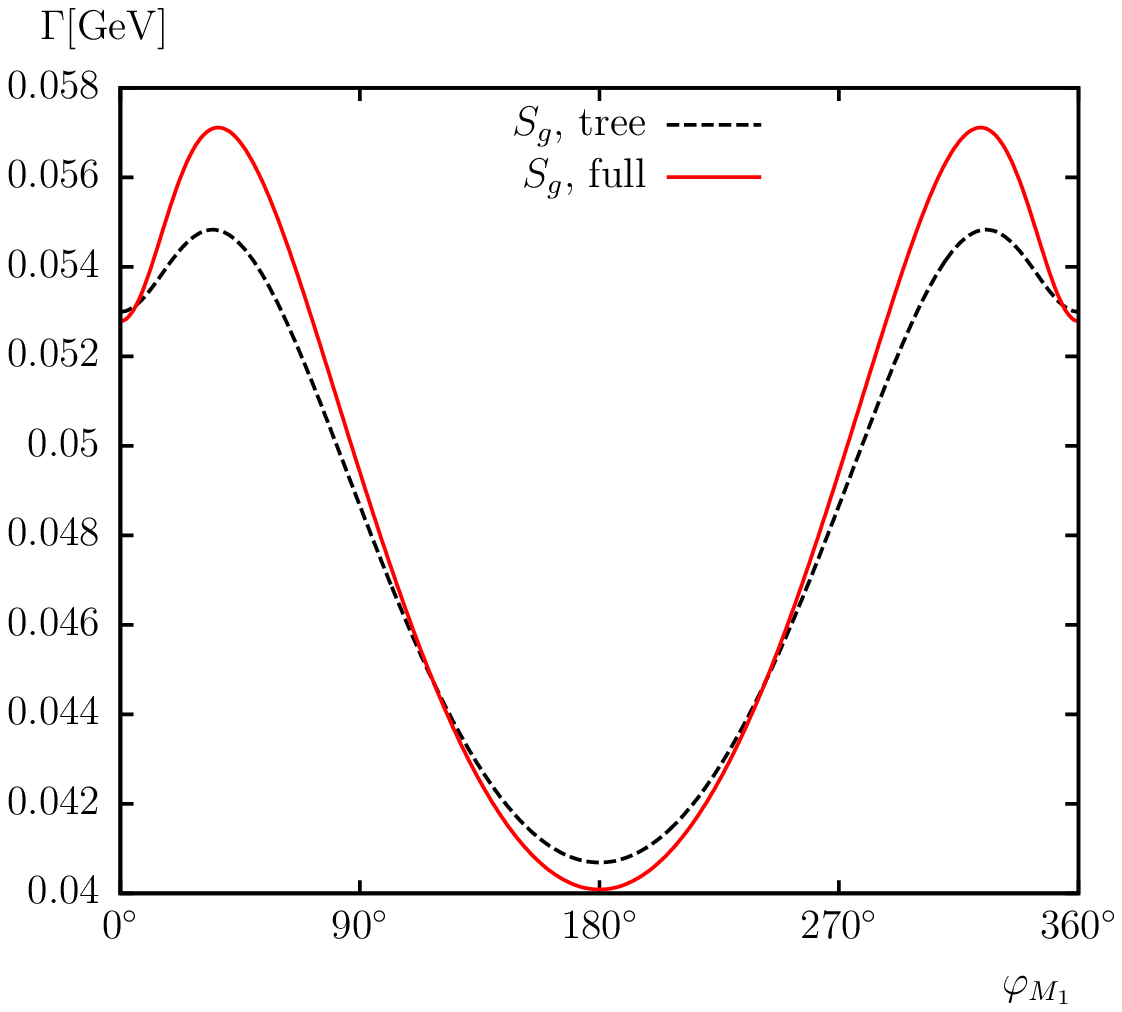}
\hspace{-4mm}
\includegraphics[width=0.49\textwidth,height=7.5cm]{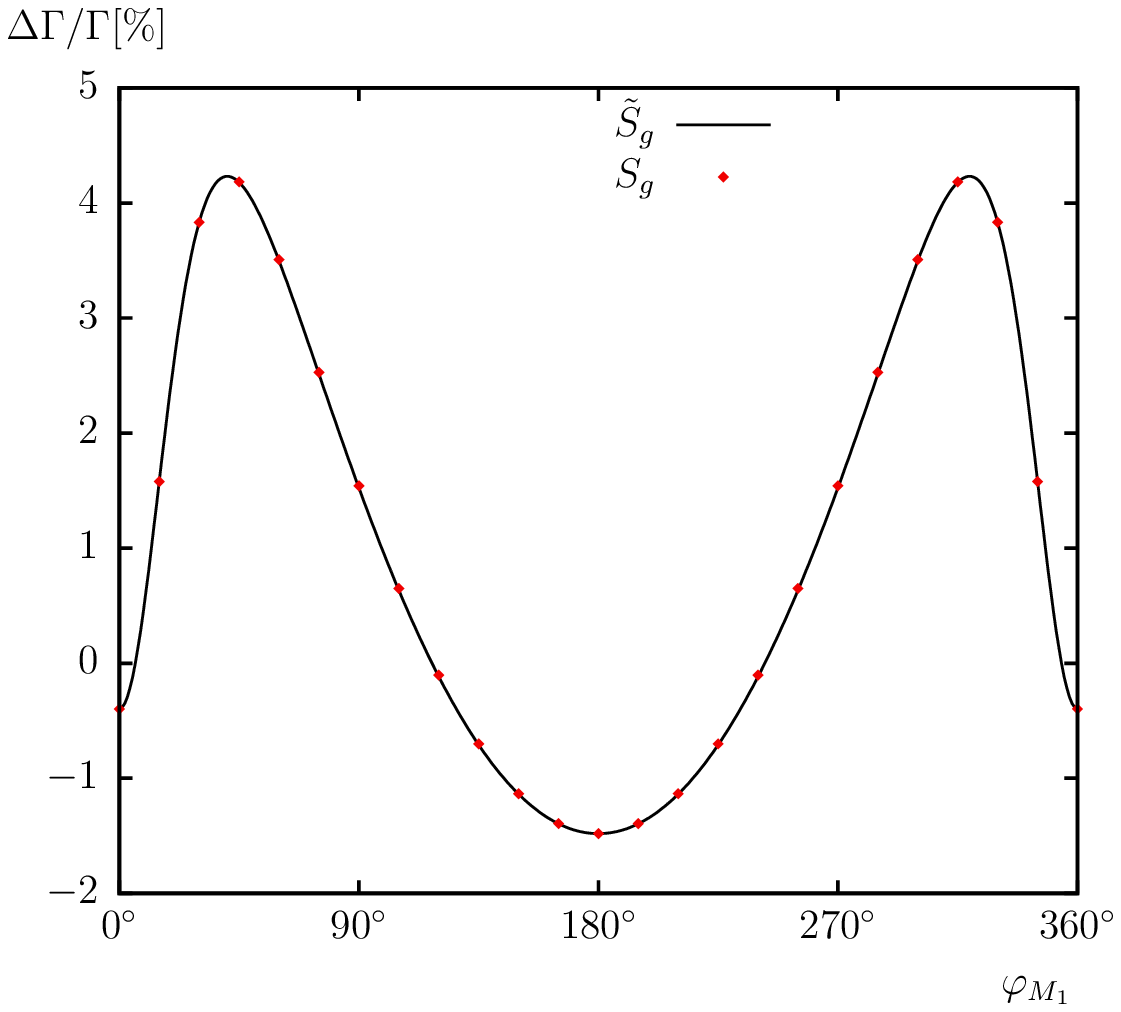} 
\end{tabular}
\caption{
  $\Ga(\DecayNNh{4}{3}{3})$. 
  Tree-level (``tree'') and full one-loop (``full'') corrected 
  decay widths are shown with the parameters chosen according to 
  \refta{tab:para}, with $\phiMe$ varied.
  The left plot shows the decay width, the right plot shows the relative size of the corrections.
}
\label{fig:PhiM1.neu4neu3h3}
\end{center}
\end{figure}

The decays $\DecayNNh{4}{3}{1}$, $\DecayNNh{4}{3}{2}$,  $\DecayNNh{4}{3}{3}$, 
shown in \reffis{fig:PhiM1.neu4neu3h1}, \ref{fig:PhiM1.neu4neu3h2}  and \ref{fig:PhiM1.neu4neu3h3}, respectively, 
are kinematically closed
in scenario $\Sh$. 
For $\Sg$, there is a strong dependence on $\phiMe$ 
since $\neu{3}$ has both large Higgsino and bino components.
However,  contrary to what we observed for the decays to $\neu{2}$ in $\Sg$, the decay to $h_2$ is s-wave mode allowed for $\phiMe=0$, while the other two decays are suppressed. 
This is due to the opposite relative $\cp$-parity of the $\neu{3}-\neu{4}$ pair relative to the $\neu{2}-\neu{4}$ pair.

%%% 431
For $\DecayNNh{4}{3}{1}$, as shown in \reffi{fig:PhiM1.neu4neu3h1}, 
an oscillating behaviour similar to that for $\DecayNNh{4}{2}{2}$ but even more enhanced  
(going from $\sim0.01$ to $0.13\gev$) 
 results  in even larger relative corrections. 
In fact the suppression of the tree level is now larger, due to the smaller phase space for this decay,
and mixing with the unsuppressed states of the neutralinos has a dramatic effect, with corrections approaching $70\%$.
Notice, however, that the second and third lightest neutralinos are roughly degenerate in $\Sg$, see \refta{tab:para}. 
This leads naturally to a large mixing character for these two mass eigenstates,
supported by the fact that the sum of the decay widths is much less sensitive to $\phiMe$.
Therefore, the large corrections should not be regarded as a breakdown of the renormalization procedure 
but rather as an indication that one should consider all the neutralino states simultaneously.
For this particular decay the branching ratio does not reach $1\%$.
In \reffi{fig:PhiM1.neu4neu3h1}, 
 the difference between the renormalization schemes reaches  $\sim 0.02\%$ and 
they cannot be visibly distinguished from each other, 
see \refta{tab:rendiff} for details.

%%%%%%%%%%%%%%%%%%%%%%%%% F I G U R E %%%%%%%%%%%%%%%%%%%%%%%%%%%%%%%%%%%%%%%%%
%%%%%%%%%%%%%%%%%%%%%%%%% F I G U R E %%%%%%%%%%%%%%%%%%%%%%%%%%%%%%%%%%%%%%%%%

%%% 432
For the decays into the heavier Higgs bosons the corrections are mild,  at the level of a few percent.
In $\DecayNNh{4}{3}{2}$, as shown in \reffi{fig:PhiM1.neu4neu3h2},  
the unsuppressed decay, with widths between $\sim0.6$ and $0.7\gev$, 
receives small corrections via mixing with the p-wave suppressed states.

%%%%%%%%%%%%%%%%%%%%%%%%% F I G U R E %%%%%%%%%%%%%%%%%%%%%%%%%%%%%%%%%%%%%%%%%
%%%%%%%%%%%%%%%%%%%%%%%%% F I G U R E %%%%%%%%%%%%%%%%%%%%%%%%%%%%%%%%%%%%%%%%%

%%% 433
In $\DecayNNh{4}{3}{3}$, as shown in \reffi{fig:PhiM1.neu4neu3h3}, 
 the $\phiMe$ dependence is small due to the combination of one
 p-wave suppressed amplitude with an s-wave allowed one in which the couplings are small,
resulting in corrections of a few percent.
For these decays the small difference between the two schemes cannot be observed in the figures.

%%%%%%%%%%%%%%%%%%%%%%%%% F I G U R E %%%%%%%%%%%%%%%%%%%%%%%%%%%%%%%%%%%%%%%%%
%%%%%%%%%%%%%%%%%%%%%%%%% F I G U R E %%%%%%%%%%%%%%%%%%%%%%%%%%%%%%%%%%%%%%%%%

% %%%%%%%%%%%%%%%%%%%%%%%%%%%%%%%%%%%%%%%%%%%%%%%%%%%%%%%%%%%%%%%%%%%%%%%%%%%%%
\subsection{Decays into \boldmath{$Z$} bosons}
\label{sec:DecayNjZ}
\begin{figure}[htb!]
\begin{center}
\begin{tabular}{c}
\includegraphics[width=0.49\textwidth,height=7.5cm]{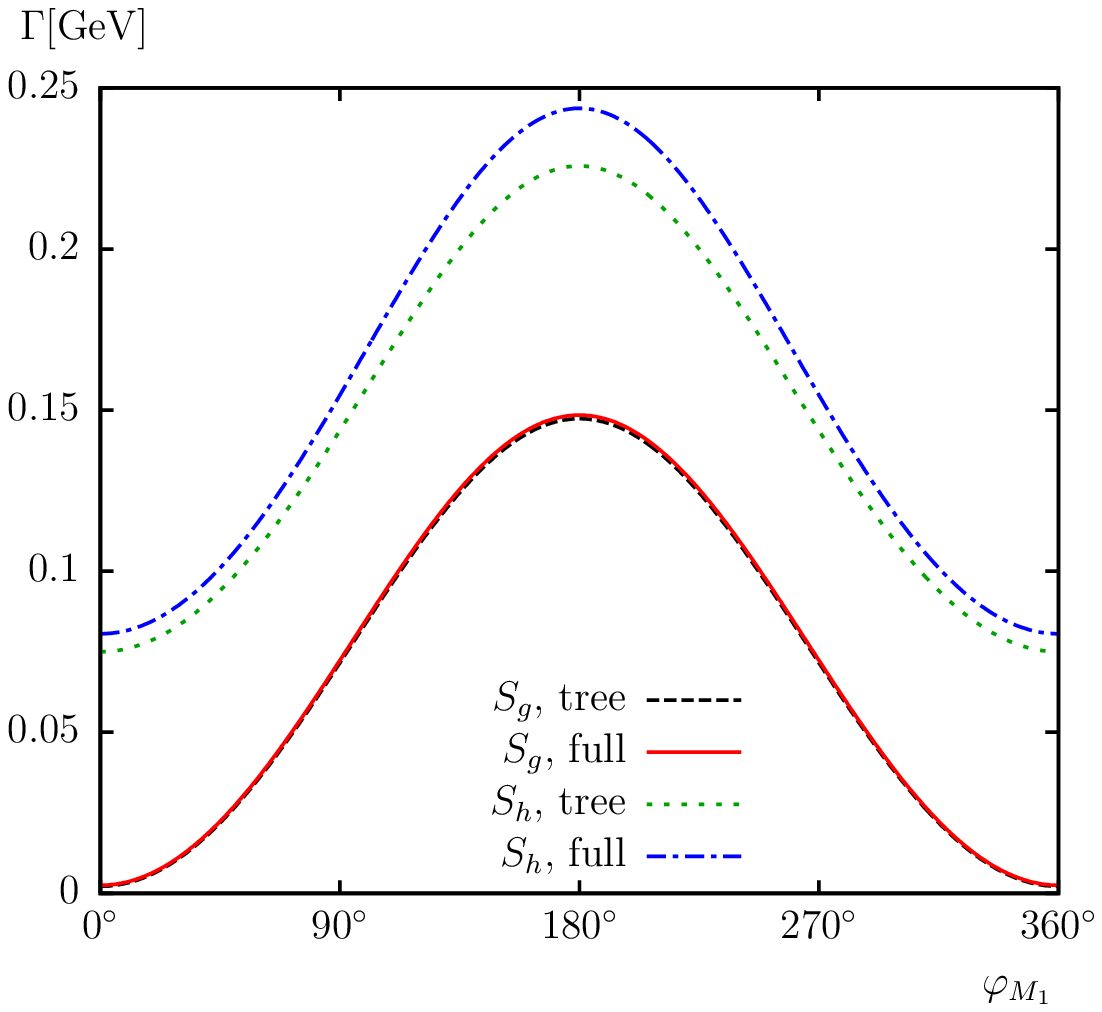}
\hspace{-4mm}
\includegraphics[width=0.49\textwidth,height=7.5cm]{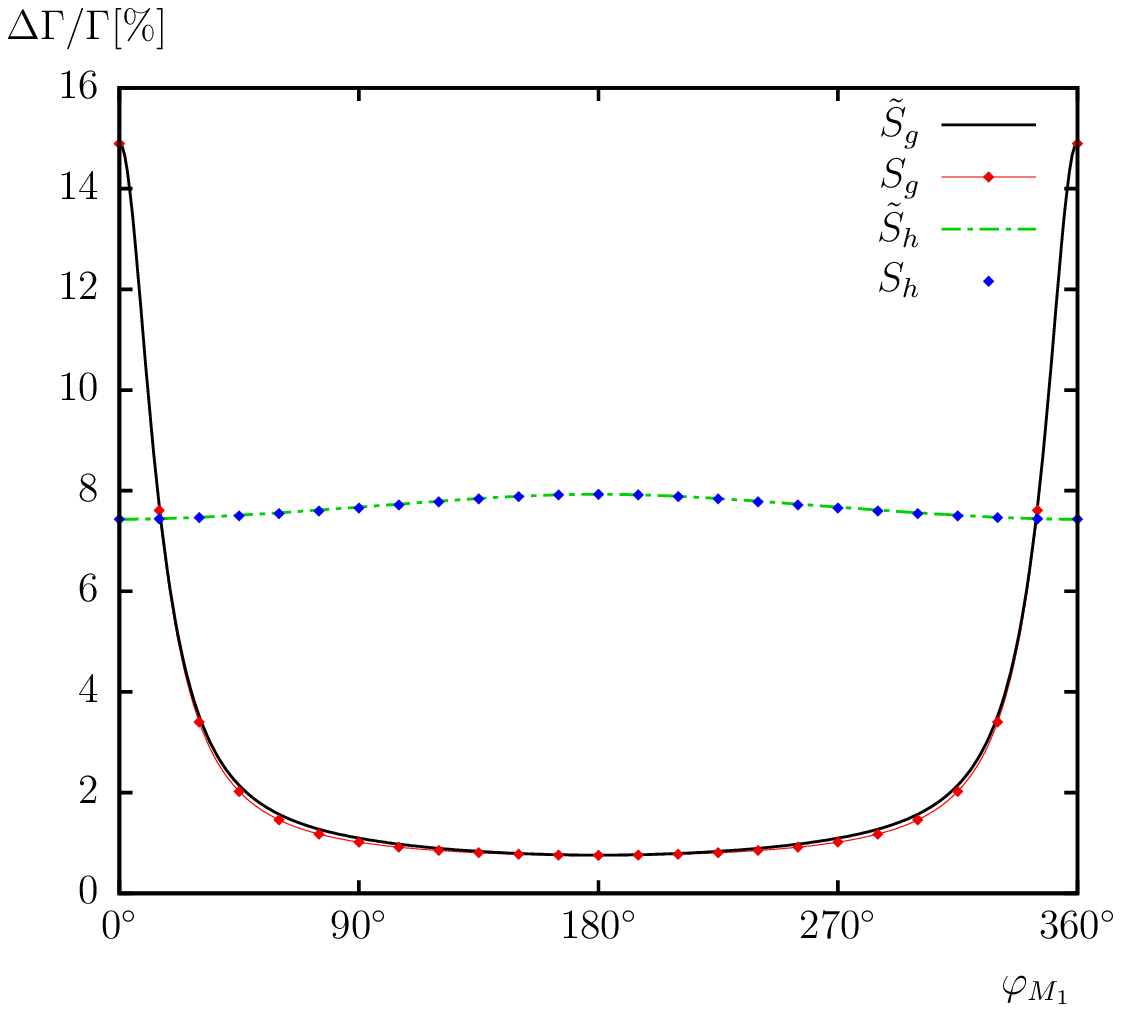} 
\end{tabular}
\caption{
  $\Ga(\DecayNNZ{4}{1})$. 
  Tree-level (``tree'') and full one-loop (``full'') corrected 
  decay widths are shown with the parameters chosen according to 
  \refta{tab:para}, with $\phiMe$ varied.
  The left plot shows the decay width, the right plot shows the relative size of the corrections.
}
\label{fig:PhiM1.neu4neu1z}
\end{center}
\end{figure}

\begin{figure}[t!]
\begin{center}
\begin{tabular}{c}
\includegraphics[width=0.49\textwidth,height=7.5cm]{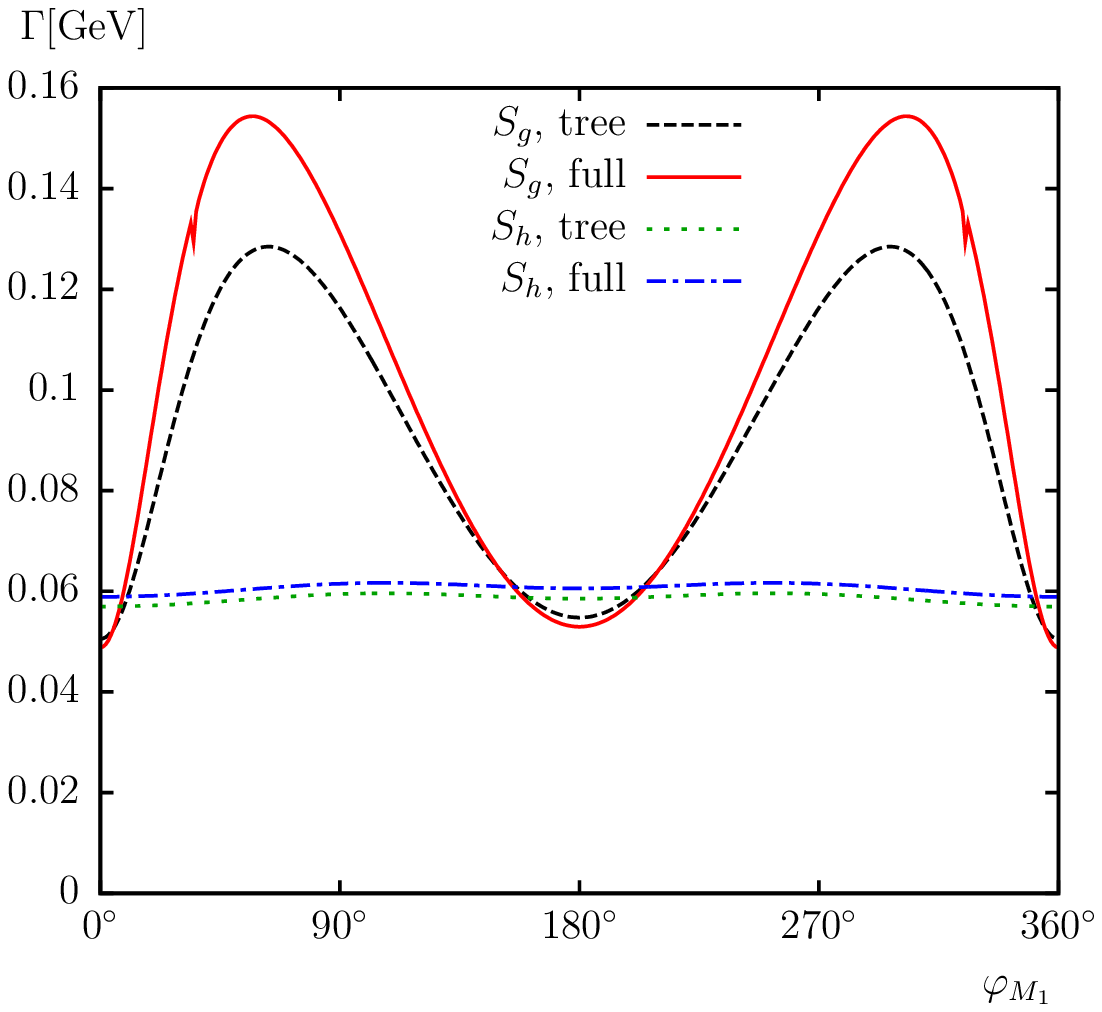}
\hspace{-4mm}
\includegraphics[width=0.49\textwidth,height=7.5cm]{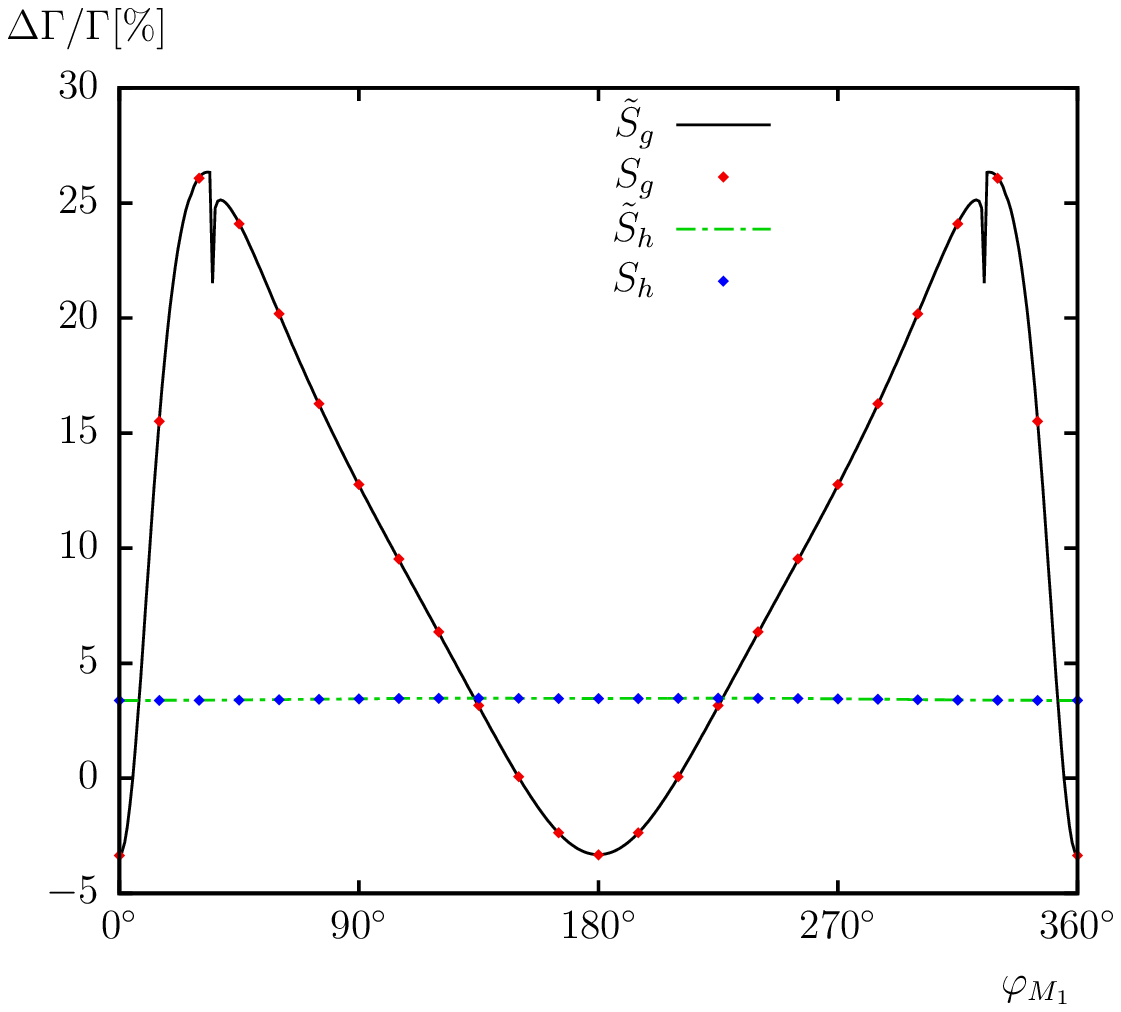} 
\end{tabular}
\caption{
  $\Ga(\DecayNNZ{4}{2})$. 
  Tree-level (``tree'') and full one-loop (``full'') corrected 
  decay widths are shown with the parameters chosen according to 
  \refta{tab:para}, with $\phiMe$ varied.
  The left plot shows the decay width, the right plot shows the relative size of the corrections.
}
\label{fig:PhiM1.neu4neu2z}
\end{center}
\end{figure}

\begin{figure}[t!]
\begin{center}
\begin{tabular}{c}
\includegraphics[width=0.49\textwidth,height=7.5cm]{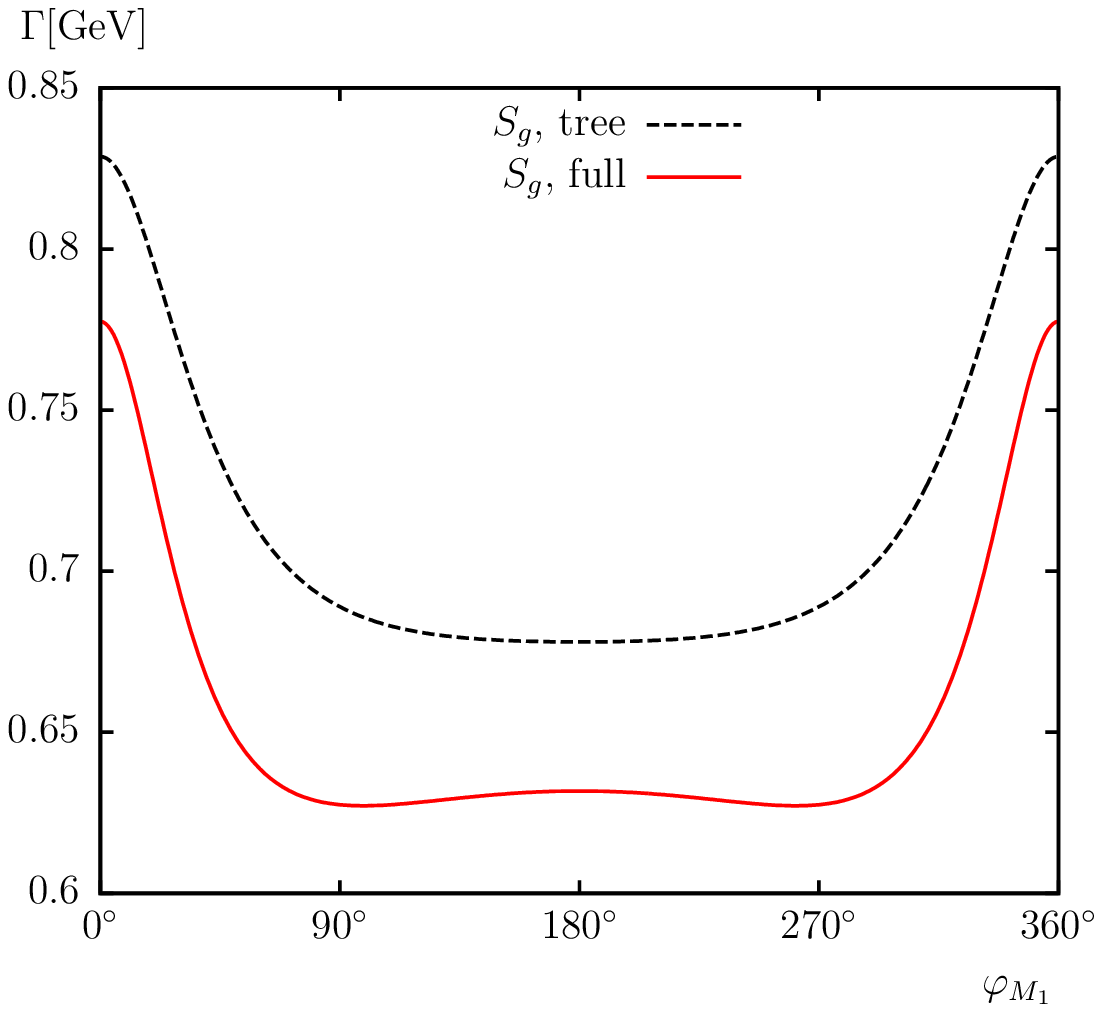}
\hspace{-4mm}
\includegraphics[width=0.49\textwidth,height=7.5cm]{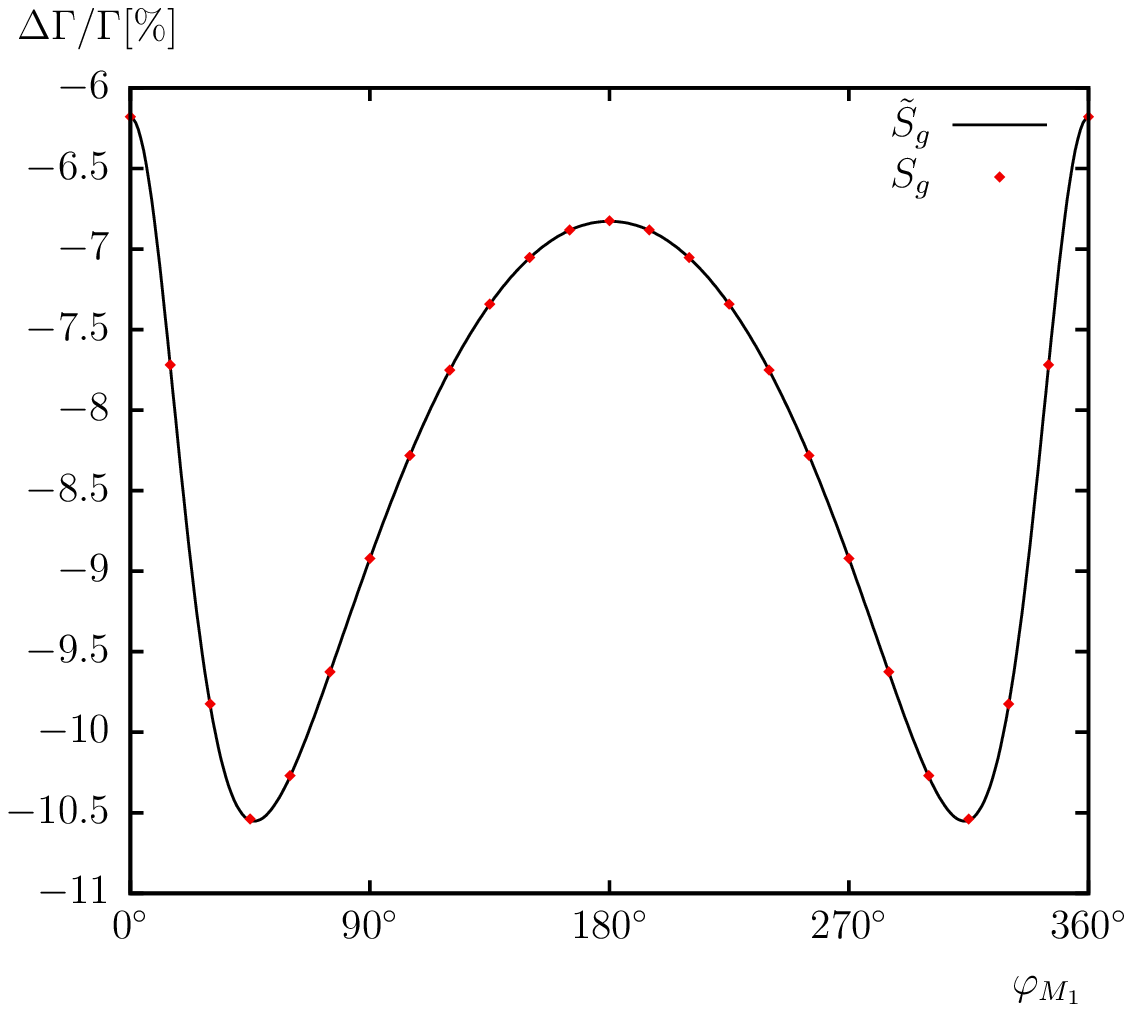} 
\end{tabular}
\caption{
  $\Ga(\DecayNNZ{4}{3})$. 
  Tree-level (``tree'') and full one-loop (``full'') corrected 
  decay widths are shown with the parameters chosen according to 
  \refta{tab:para}, with $\phiMe$ varied.
  The left plot shows the decay width, the right plot shows the relative size of the corrections.
}
\label{fig:PhiM1.neu4neu3z}
\end{center}
\end{figure}

The channels involving the $Z$ boson,
$\DecayNNZ{4}{j}$, are presented
in \reffis{fig:PhiM1.neu4neu1z}-\ref{fig:PhiM1.neu4neu3z}. 
The strong resemblance in the $\phiMe$-dependence between these plots and those for the decay into $h_2$ 
is due to fact that gauge bosons are $\cp$-odd, 
while in our scenarios the second Higgs boson 
has a very small $\cp$-even component and tends to the $\cp$-odd state
 for $\phiMe=0,\pi $.
There is a strong dependence on the relative $\cp$-phase of the neutralinos 
in the initial and final states, 
which lead to visible differences between the renormalization schemes, 
given explicitly in \refta{tab:rendiff}.

%%%%%%%%%%%%%%%%%%%%%%%%% F I G U R E %%%%%%%%%%%%%%%%%%%%%%%%%%%%%%%%%%%%%%%%%
%%%%%%%%%%%%%%%%%%%%%%%%% F I G U R E %%%%%%%%%%%%%%%%%%%%%%%%%%%%%%%%%%%%%%%%%

%%% 41
In both scenarios the lightest neutralino is mainly bino-like, therefore $\DecayNNZ{4}{1}$ depends strongly on $\phiMe$,
as it is shown in \reffi{fig:PhiM1.neu4neu1z}.
The decay is both qualitatively and quantitatively similar to the case of $\DecayNNh{4}{1}{2}$: 
due to the relative $\cp$-phase of the neutralinos, and the fact that the $Z$ is $\cp$-odd, 
the decay is suppressed at $\phiMe=0$, and maximal at $\phiMe=180^\circ$, 
and the decay width ranging from 0 to $0.14\gev$ and 0.08 to $0.25\gev$ for $\Sg$~and $\Sh$~respectively. 
For $\Sg$~the dependence of the relative size of the corrections on $\phiMe$ is much larger, 
from $1\%$ to $15\%$, with visible differences between the schemes at large $\phiMe$
of up to $0.2\%$, 
while the loop corrections for $\Sh$ are almost independent of $\phiMe$, of $\sim 8\%$. 
The difference between the schemes in $\Sg$ has been highlighted in \reffi{fig:PhiM1.rendiff}.
%%%%%%%%%%%%%%%%%%%%%%%%% F I G U R E %%%%%%%%%%%%%%%%%%%%%%%%%%%%%%%%%%%%%%%%%
%%%%%%%%%%%%%%%%%%%%%%%%% F I G U R E %%%%%%%%%%%%%%%%%%%%%%%%%%%%%%%%%%%%%%%%%

%%% 42
For $\DecayNNZ{4}{2}$, in $\Sg$ the decay width oscillates from $\sim0.05$ to $0.15\gev$ and for $\Sh$ the decay width is $\sim 0.06\gev$, 
 as it is shown in \reffi{fig:PhiM1.neu4neu2z}. 
For $\Sg$ the effect of the threshold for $\DecayNNh{2}{1}{1}$ at $\phiMe=35^\circ$ and $325^\circ$
is visible as a small dip in the decay width and a marked dip in the relative corrections. 
The corrections are comparatively large for $\Sg$, $\sim0$ to $25\%$, and for $\Sh$~are $\sim3\%$. 
The two schemes are not visibly distinguished from each other, with the largest differences in $\Sg$ of $\sim 4\times 10^{-4}$.
%%% 43
Although in $\Sh$ it is below threshold, in $\Sg$ the 
decay $\DecayNNZ{4}{3}$, shown in  \reffi{fig:PhiM1.neu4neu3z},
 is s-wave mode allowed in $\Sg$, 
resulting in the largest branching ratio into $Z$, 
while it is  below threshold in $\Sh$.
The decay width goes from  $0.78\gev$ at $\phiMe\approx 0$
to $0.62\gev$, with 
 the corrections between  $\sim 6.2$ and $10.5\%$.
The two schemes differ by up to 
$\sim 5\times 10^{-5}$ and cannot be visibly distinguished.

%%%%%%%%%%%%%%%%%%%%%%%%% F I G U R E %%%%%%%%%%%%%%%%%%%%%%%%%%%%%%%%%%%%%%%%%
%%%%%%%%%%%%%%%%%%%%%%%%% F I G U R E %%%%%%%%%%%%%%%%%%%%%%%%%%%%%%%%%%%%%%%%%

% %%%%%%%%%%%%%%%%%%%%%%%%%%%%%%%%%%%%%%%%%%%%%%%%%%%%%%%%%%%%%%%%%%%%%%%%%%%%%
\subsection{Decays into (s)leptons}
\label{sec:Decaylepton}
Now we turn to the decays involving (scalar) leptons. 
The expressions for all these decay widths follow the same pattern, 
see the expressions for the tree level widths in Appendix~\ref{sec:treeresults}. 
The dependence on $\phiMe$ is small, 
although the results for $\Sh$ do show some dependence due to the small bino-like component of the decaying neutralino.
We have chosen $\MslL < \MslR$, leading to lighter left-handed and heavier right-handed sleptons, 
and significant mixing in the scalar tau sector. 

\begin{figure}[t!]
\begin{center}
\begin{tabular}{c}
\includegraphics[width=0.49\textwidth,height=7.5cm]{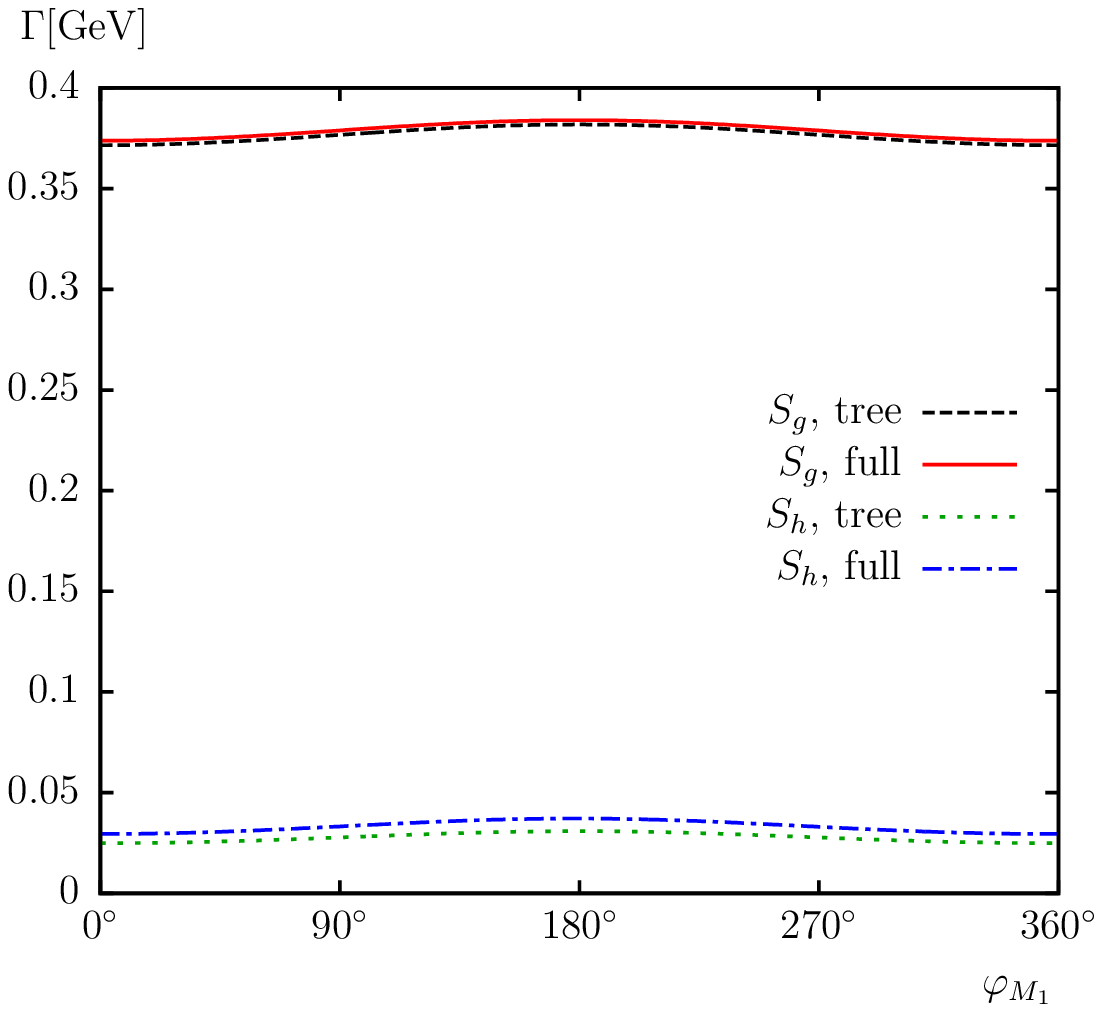}
\hspace{-4mm}
\includegraphics[width=0.49\textwidth,height=7.5cm]{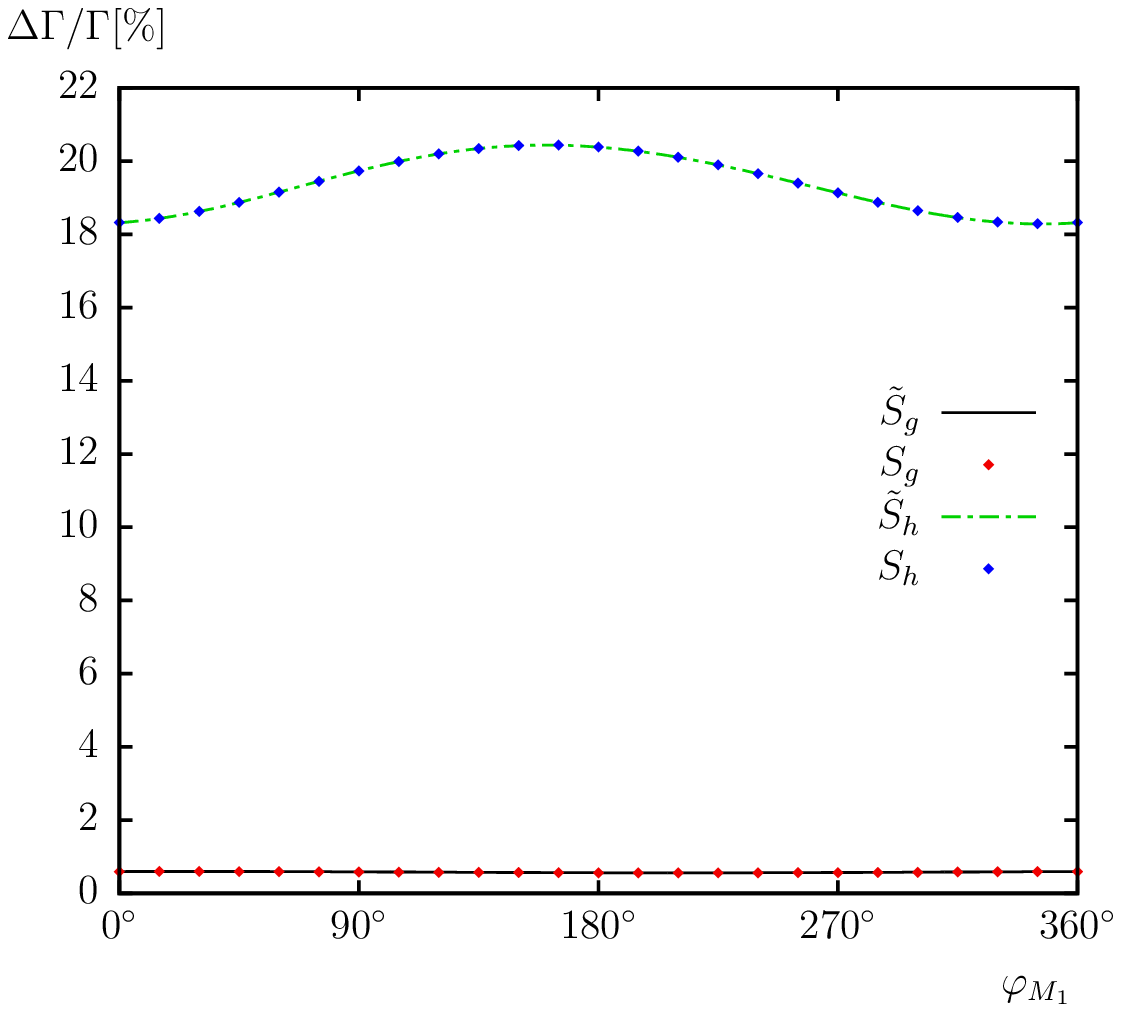} 
\end{tabular}
\caption{
  $\Ga(\DecayNlpSl{4}{\tau}{1})$. 
  Tree-level (``tree'') and full one-loop (``full'') corrected 
  decay widths are shown with the parameters chosen according to 
  \refta{tab:para}, with $\phiMe$ varied.
  The left plot shows the decay width, the right plot shows the relative size of the corrections.
}
\label{fig:PhiM1.neu4stau1taum}
\end{center}
\end{figure}

%%% 41
In \reffi{fig:PhiM1.neu4stau1taum}
we show the results for the decay $\DecayNlSl{4}{\tau}{1}$. 
The decay widths are found to be an order of magnitude larger in  $\Sg$ ($\sim 0.37\gev$) than in $\Sh$ ($\sim 0.03\gev$), 
since the gaugino-like neutralino has an unsuppressed coupling to the large left component of the stau,
while the Higgsino-like neutralino couples to the suppressed Yukawa coupling.
This pattern is even more significant for the decays into the lower generation sleptons not shown here.
In the right panel we observe that the one-loop corrections are very small in $\Sg$, while they are around $\sim 20\%$ in $\Sh$, due to the larger dependence on the stau mixing 
\medskip

%%%%%%%%%%%%%%%%%%%%%%%%% F I G U R E %%%%%%%%%%%%%%%%%%%%%%%%%%%%%%%%%%%%%%%%%
%%%%%%%%%%%%%%%%%%%%%%%%% F I G U R E %%%%%%%%%%%%%%%%%%%%%%%%%%%%%%%%%%%%%%%%%

\begin{figure}[t!]
\begin{center}
\begin{tabular}{c}
\includegraphics[width=0.49\textwidth,height=7.5cm]{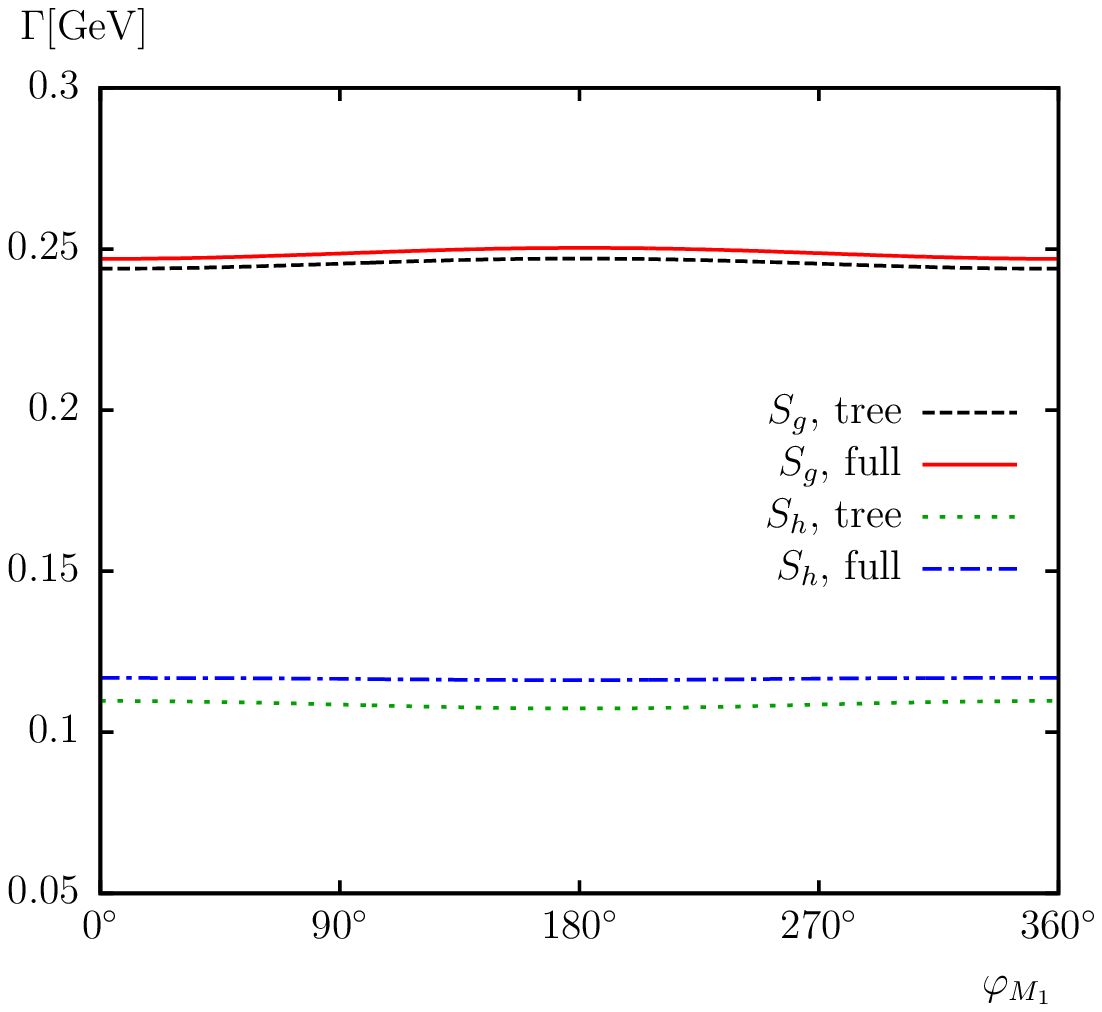}
\hspace{-4mm}
\includegraphics[width=0.49\textwidth,height=7.5cm]{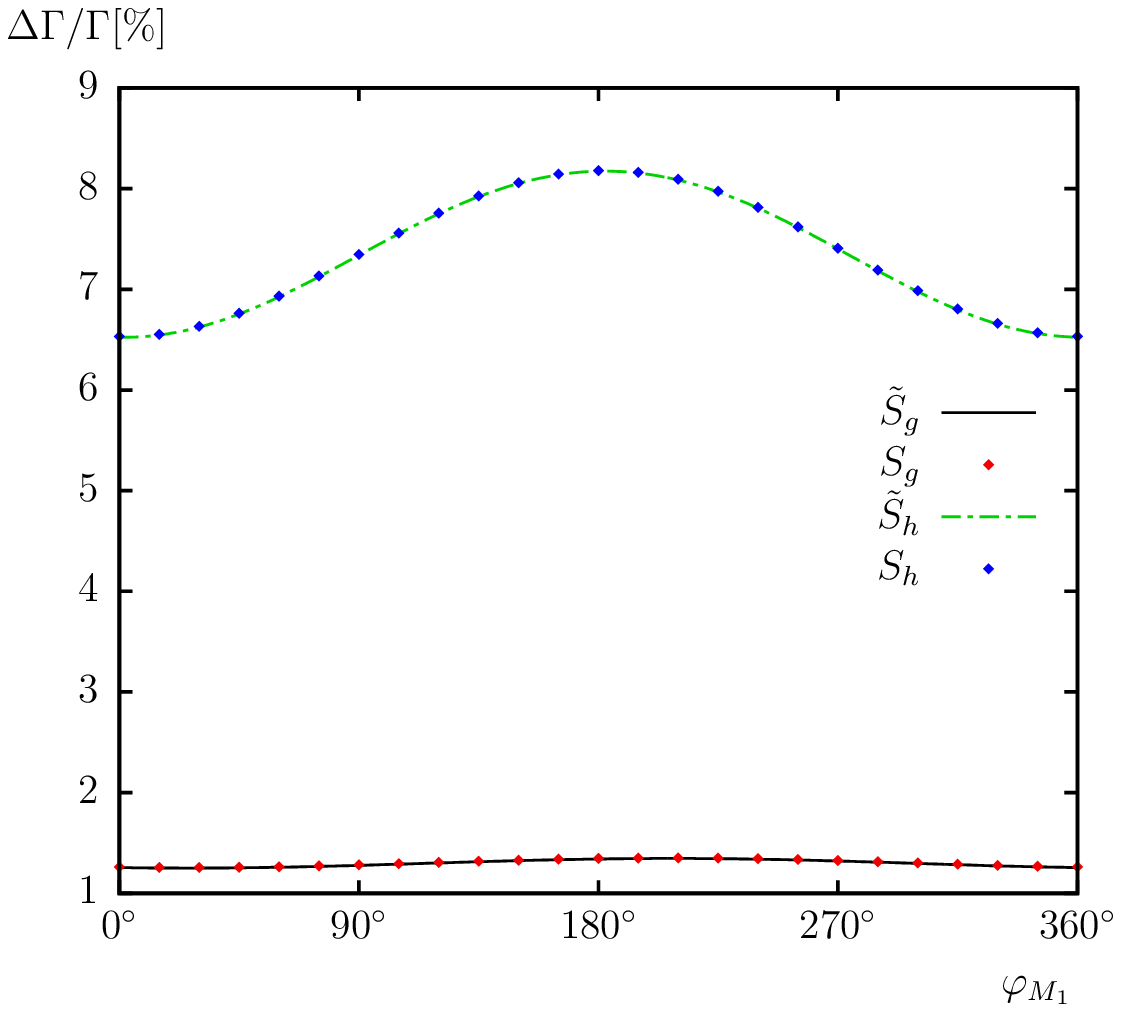} 
\end{tabular}
\caption{
  $\Ga(\DecayNlpSl{4}{\tau}{2})$. 
  Tree-level (``tree'') and full one-loop (``full'') corrected 
  decay widths are shown with the parameters chosen according to 
  \refta{tab:para}, with $\phiMe$ varied.
  The left plot shows the decay width, the right plot shows the relative size of the corrections.
}
\label{fig:PhiM1.neu4stau2taum}
\end{center}
\end{figure}

%%% 42
The results for the decay into the heavier
scalar tau 
are shown 
in \reffi{fig:PhiM1.neu4stau2taum}. 
The pattern is similar to the preceding decay, 
the difference being that here the right-handed component of the heavier stau is larger, resulting in a decay width which is four times larger for $\Sh$, and $30\%$ smaller for $\Sg$. 
The corrections in $\Sg$ remain very small while those in $\Sh$ are now $\sim 7\%$. 
For the decay into the first two generations, where the slepton mixing is usually negligible, 
both scenarios have very small partial widths.

%%%%%%%%%%%%%%%%%%%%%%%%% F I G U R E %%%%%%%%%%%%%%%%%%%%%%%%%%%%%%%%%%%%%%%%%
%%%%%%%%%%%%%%%%%%%%%%%%% F I G U R E %%%%%%%%%%%%%%%%%%%%%%%%%%%%%%%%%%%%%%%%%

\begin{figure}[t!]
\begin{center}
\begin{tabular}{c}
\includegraphics[width=0.49\textwidth,height=7.5cm]{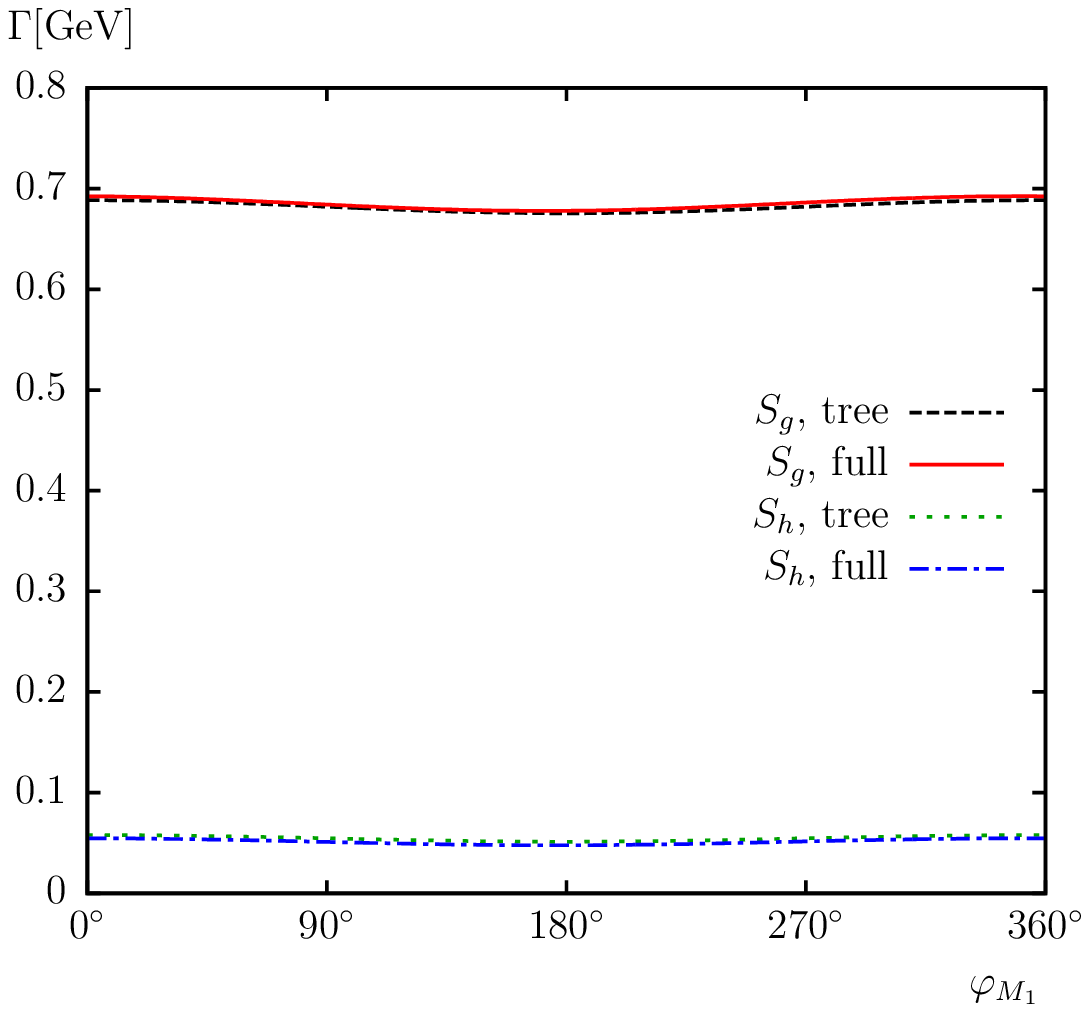}
\hspace{-4mm}
\includegraphics[width=0.49\textwidth,height=7.5cm]{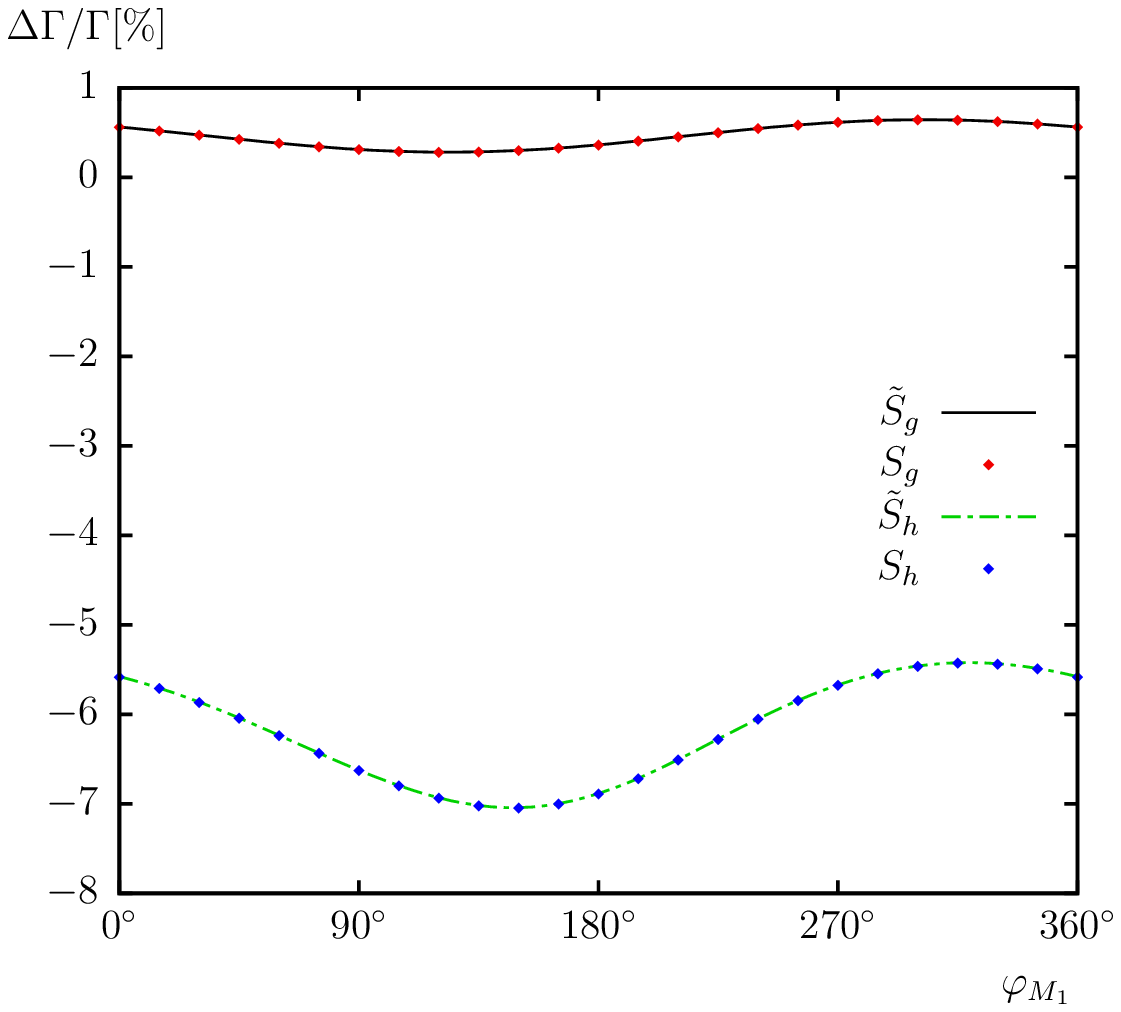} 
\end{tabular}
\caption{
  $\Ga(\DecayNnSn{4}{\tau})$. 
  Tree-level (``tree'') and full one-loop (``full'') corrected 
  decay widths are shown with the parameters chosen according to 
  \refta{tab:para}, with $\phiMe$ varied.
  The left plot shows the decay width, the right plot shows the relative size of the corrections.
}
\label{fig:PhiM1.neu4snutaunu}
\end{center}
\end{figure}

\medskip
%%% 40
 The results for the decay $\DecayNnSn{4}{\tau}$, are shown 
in \reffi{fig:PhiM1.neu4snutaunu}. 
Here, the gaugino-like neutralino has a decay width which is roughly the sum of decay widths for $\DecayNlSl{4}{\tau}{1}$ and $\DecayNlSl{4}{\tau}{2}$, 
i.e. $\sim 0.7\gev$ for $\Sg$ and $\sim 0.05\gev$ for $\Sh$. 
The radiative corrections are $\sim0.5\%$ in $\Sg$ and $\sim -6\%$ in $\Sh$. 

In $\Sg$, where the neutralino is gaugino-like, the branching ratios for these tree processes are, respectively $\sim 2.5\% $, $\sim 2\% $, and $\sim 5\%$. 
Taking into account the charged conjugated processes this results in a branching ratio of almost $20\%$ for the third lepton family.
For the first two generations the branching ratios into every left-handed slepton or sneutrino is $\sim 4.5$ and $5\%$, respectively, 
while the decays into the right-handed sleptons are negligible.
Therefore, in this class of scenarios the leptonic decays could be the dominant ones.
In $\Sh$, on the other hand, the leptonic decays are subdominant, especially for the first to generations, 
and mainly due to the small gaugino component of the decaying neutralino.

In all the decays to leptons, the difference between the renormalization
schemes is negligible, as 
shown in \reffis{fig:PhiM1.neu4stau1taum}, \ref{fig:PhiM1.neu4stau2taum} and \ref{fig:PhiM1.neu4snutaunu}, 
and summarized in \refta{tab:rendiff}.

%%%%%%%%%%%%%%%%%%%%%%%%% F I G U R E %%%%%%%%%%%%%%%%%%%%%%%%%%%%%%%%%%%%%%%%%
%%%%%%%%%%%%%%%%%%%%%%%%% F I G U R E %%%%%%%%%%%%%%%%%%%%%%%%%%%%%%%%%%%%%%%%%

% %%%%%%%%%%%%%%%%%%%%%%%%%%%%%%%%%%%%%%%%%%%%%%%%%%%%%%%%%%%%%%%%%%%%%%%%%%%%%
\subsection{Full one-loop results: total decay widths}
\label{sec:gatot}

\begin{figure}[t!]
\begin{center}
\begin{tabular}{c}
\includegraphics[width=0.49\textwidth,height=7.5cm]{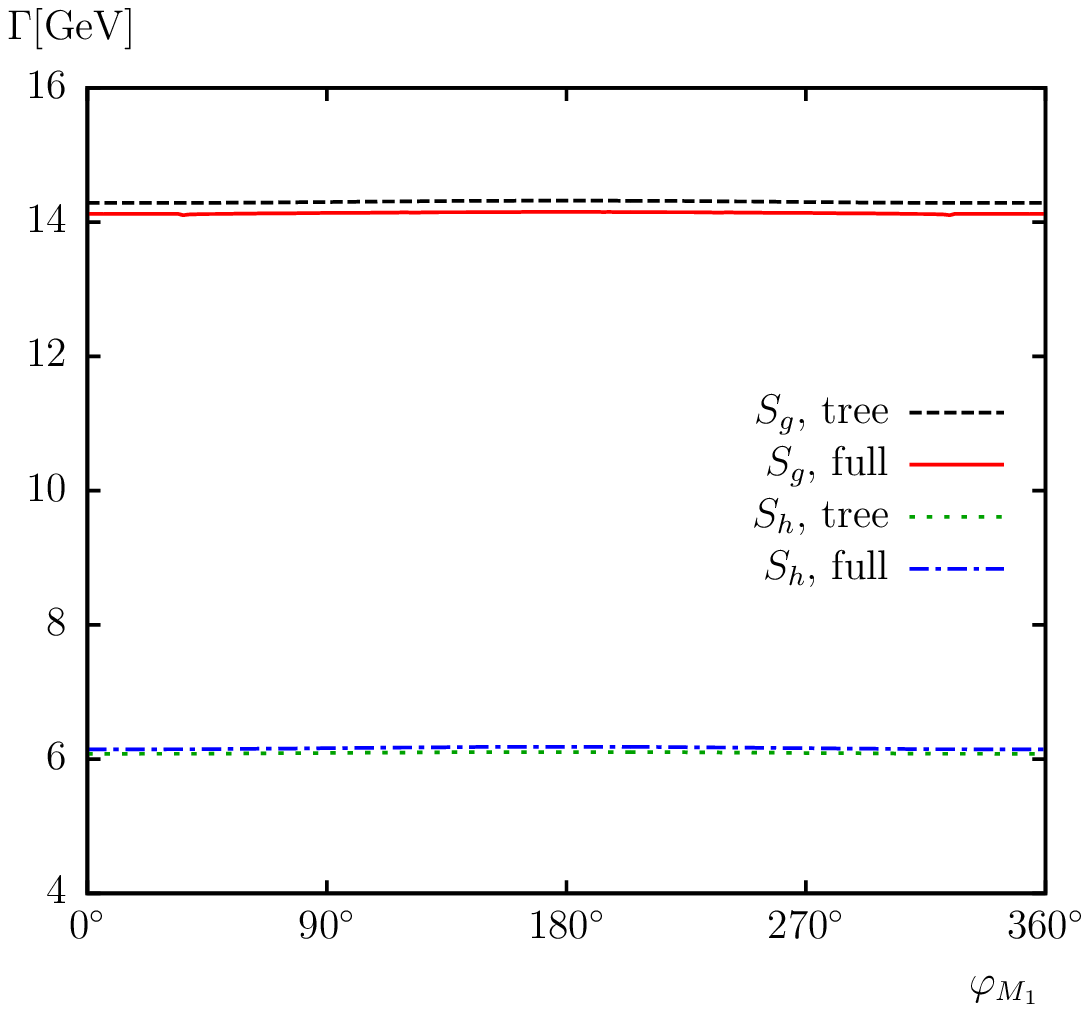}
\hspace{-4mm}
\includegraphics[width=0.49\textwidth,height=7.5cm]{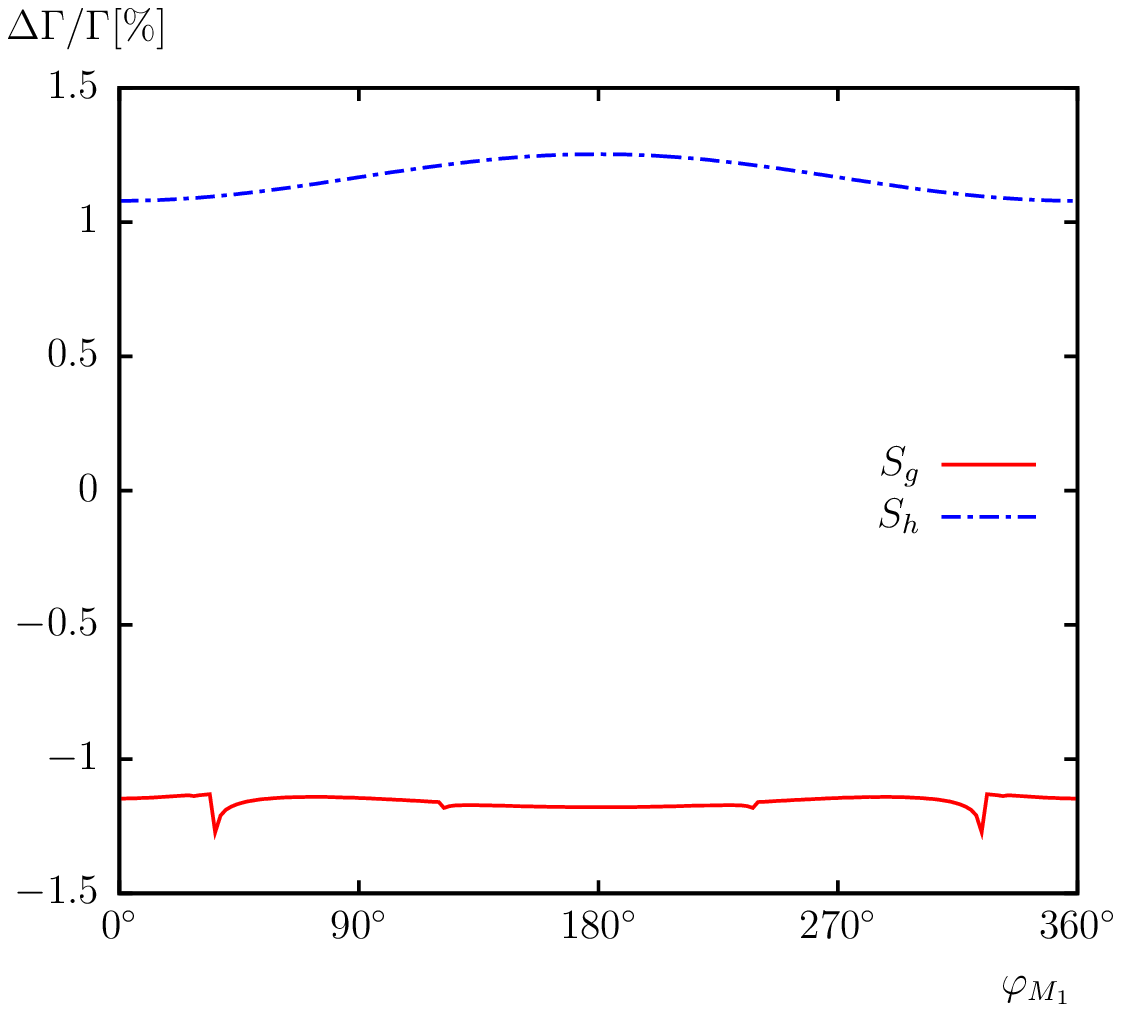}
\end{tabular}
\caption{  $\Ga_{\rm tot}(\neu{4})$.
  Tree-level (``tree'') and full one-loop (``full'') corrected total 
  decay widths are shown with the parameters chosen according to 
  \refta{tab:para}, with $\phiMe$ varied.
  The left panel shows the decay width, 
  the right panel shows the corresponding relative size of the 
  corrections.
}
%\vspace{1em}
\label{fig:phiM1.neu4tot}
\end{center}
\end{figure}

In this subsection we briefly show the results for the total decay
widths and its relative correction of the three heaviest neutralinos in $\Sg$ and $\Sh$.
%%% 4
The decay width of $\neu{4}$, shown in the l.h.s.\ of \reffi{fig:phiM1.neu4tot}, is almost independent of $\phiMe$,
as is expected from the heavier neutralino in a GUT-re\-la\-ted scenario.
Therefore, the branching ratios can be easily obtained 
from the partial widths.
The large $\phiMe$ dependence of the single channels is due to the strong effect on the mixing of the different neutralinos, 
as already argued in the preceding subsections.
The corrections are also very small, of $\sim \pm 1\%$, 
with the expected dips due to the thresholds for $\DecayNNh{2}{1}{1}$ at
$\phiMe= 35^\circ$ and $325^\circ$
and $\DecayCmNW{1}{1}$ at  $\phiMe= 125^\circ$ and $235^\circ$.
%%%%%%%%%%%%%%%%%%%%%%%%% F I G U R E %%%%%%%%%%%%%%%%%%%%%%%%%%%%%%%%%%%%%%%%%
%%%%%%%%%%%%%%%%%%%%%%%%% F I G U R E %%%%%%%%%%%%%%%%%%%%%%%%%%%%%%%%%%%%%%%%%

\begin{figure}[t!]
\begin{center}
\begin{tabular}{c}
\includegraphics[width=0.49\textwidth,height=7.5cm]{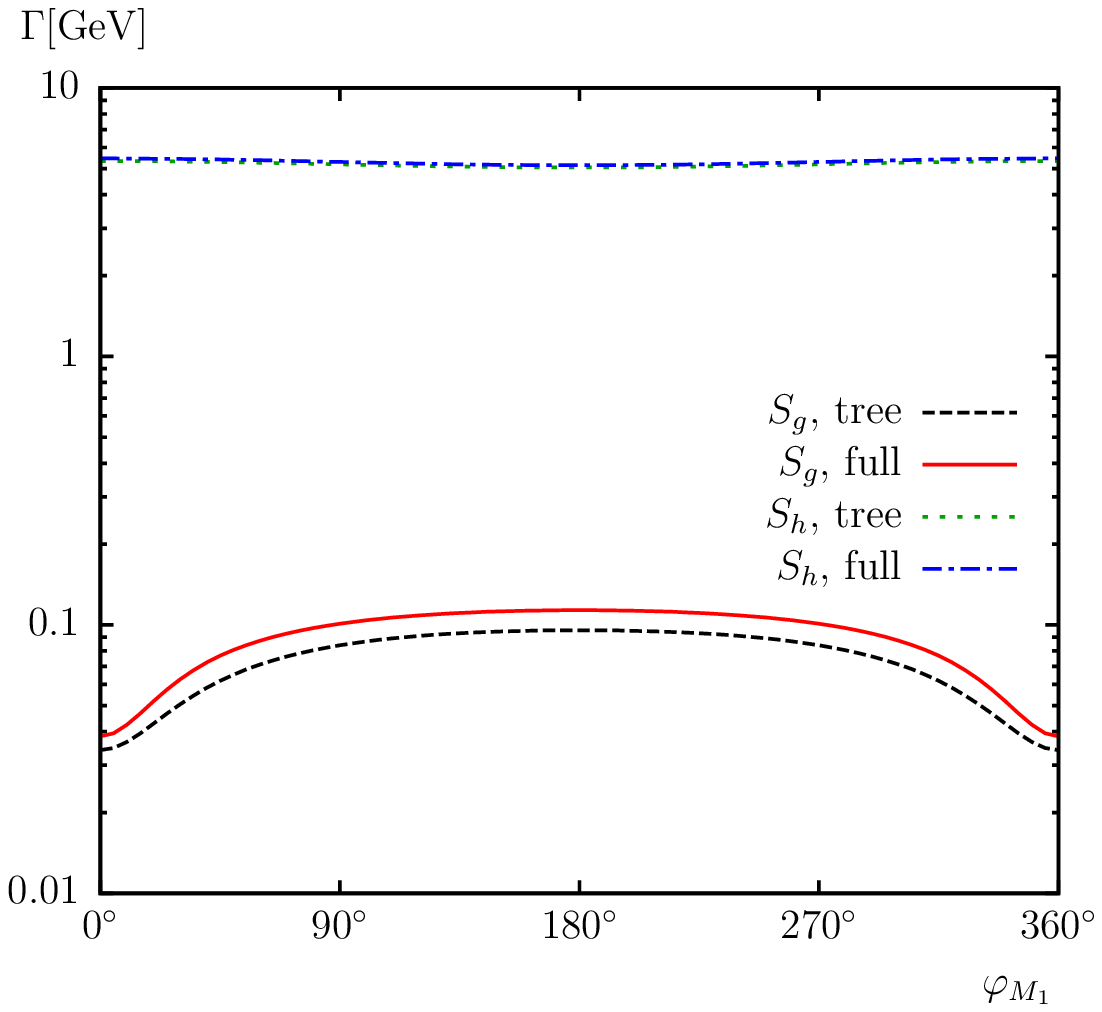}
\hspace{-4mm}
\includegraphics[width=0.49\textwidth,height=7.5cm]{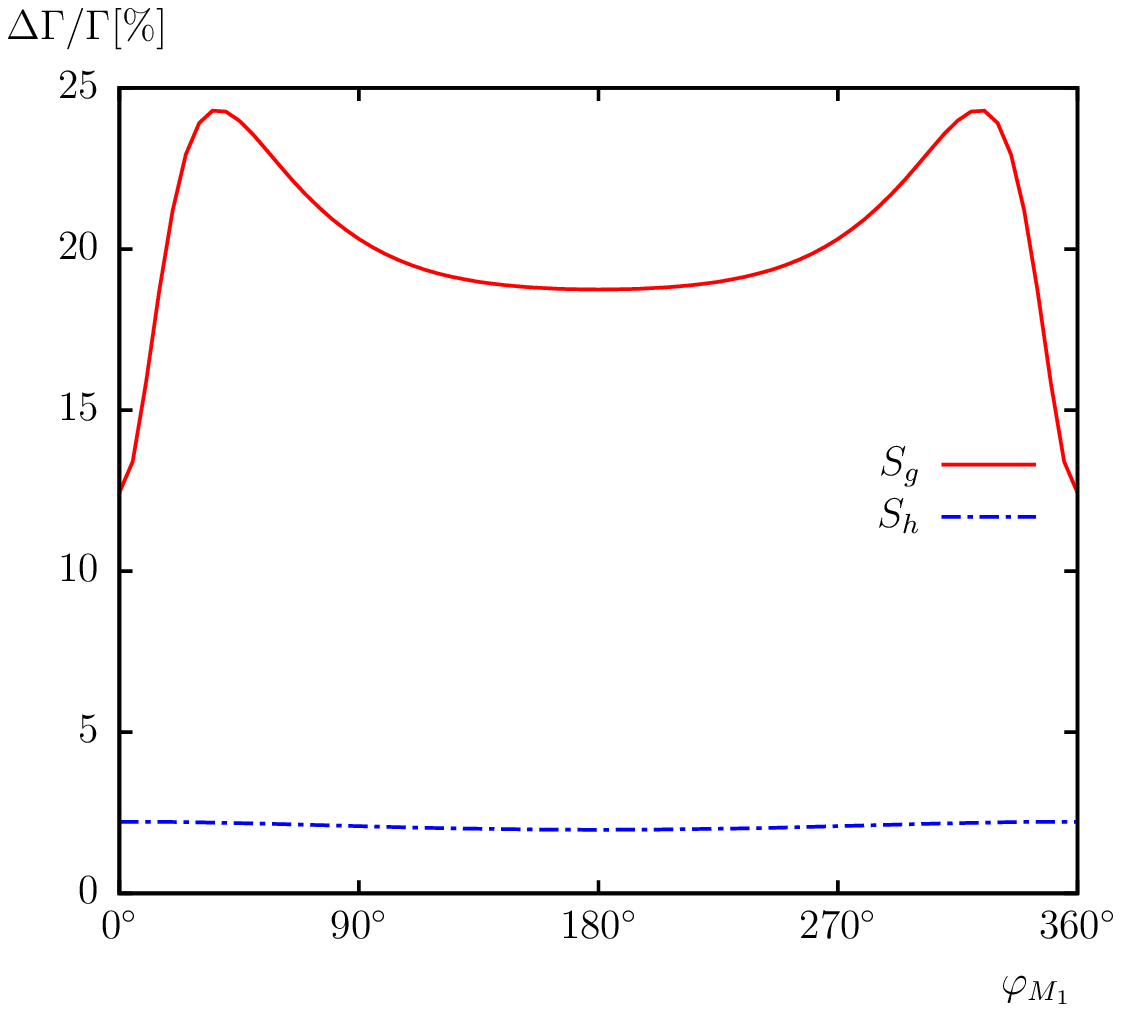}
\end{tabular}
\caption{  $\Ga_{\rm tot}(\neu{3})$.
  Tree-level (``tree'') and full one-loop (``full'') corrected total 
  decay widths are shown with the parameters chosen according to 
  \refta{tab:para}, with $\phiMe$ varied.
  The left panel shows the decay width, 
  the right panel shows the corresponding relative size of the 
  corrections.
}
%\vspace{1em}
\label{fig:phiM1.neu3tot}
\end{center}
\end{figure}

%%% 3
The total decay width of $\neu{3}$ is shown in the l.h.s.\ of \reffi{fig:phiM1.neu3tot}.
As already discussed, in $\Sg$ the second and third lightest neutralinos have similar masses and 
 have a Higgsino-bino mixing character which  strongly depends on $\phiMe$.
This is  both true at the tree-level, where the total width goes from $\sim 0.04\gev$ for $\phiMe=0 $ to almost three times as much for $\phiMe=\pi$,
as well as for the loop corrections, which  are of  $\order{20\%}$.
The total width is also significantly smaller than that of $\neu{4}$, largely
due to the reduced phase space, see \refta{tab:chaneu}.
Here only the leptonic decays  and those to $Z$  and the lightest neutralino are open.
On the contrary, in $\Sh$ both $\neu{3}$ and $\neu{4}$ are Higgsino-dominated and nearly degenerate.
Consequently the same decay channels are open 
and their widths are similar, both at tree-level and at one-loop.

%%%%%%%%%%%%%%%%%%%%%%%%% F I G U R E %%%%%%%%%%%%%%%%%%%%%%%%%%%%%%%%%%%%%%%%%
%\begin{figure}[htb!]
%%%%%%%%%%%%%%%%%%%%%%%%% F I G U R E %%%%%%%%%%%%%%%%%%%%%%%%%%%%%%%%%%%%%%%%%

%%%%%%%%%%%%%%%%%%%%%%%%% F I G U R E %%%%%%%%%%%%%%%%%%%%%%%%%%%%%%%%%%%%%%%%%
\begin{figure}[t!]
\begin{center}
\begin{tabular}{c}
\includegraphics[width=0.49\textwidth,height=7.5cm]{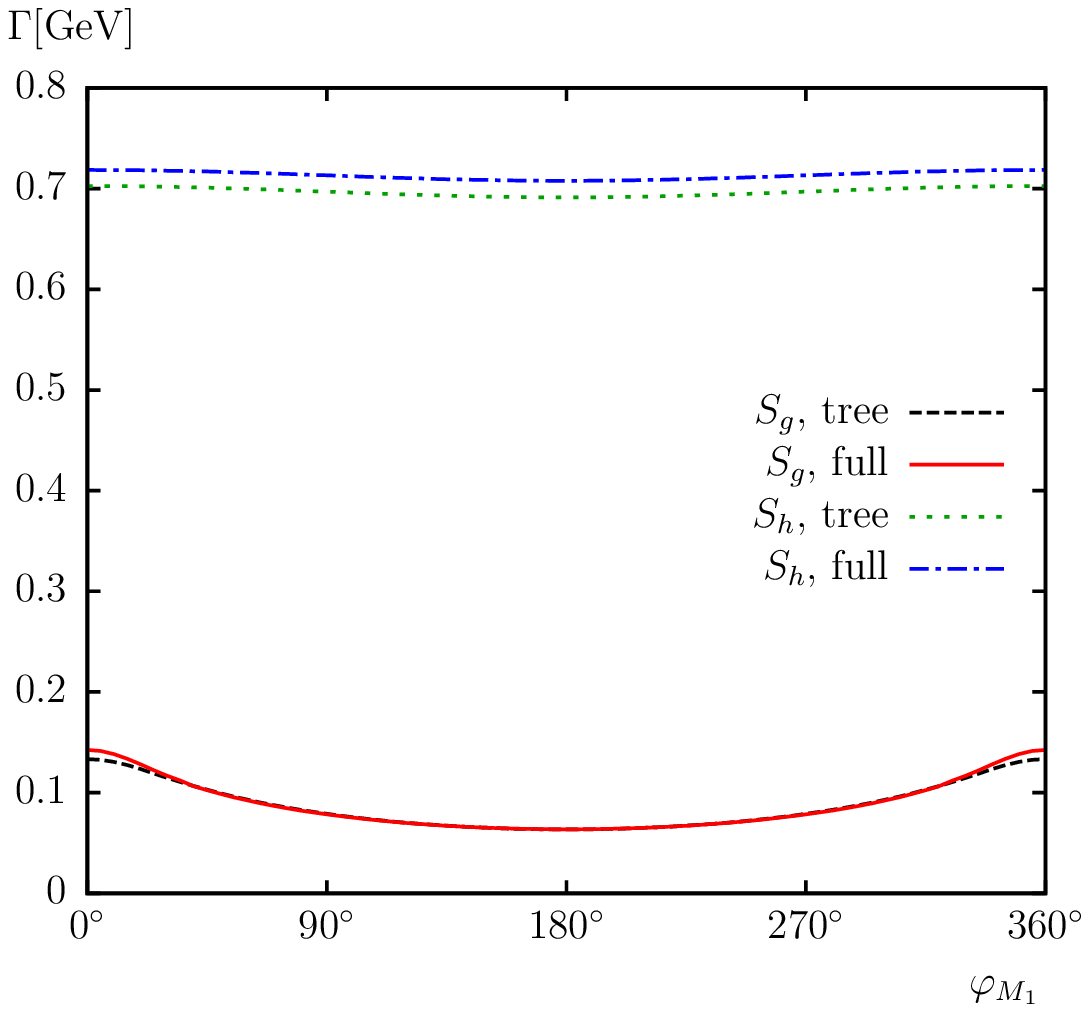}
\hspace{-4mm}
\includegraphics[width=0.49\textwidth,height=7.5cm]{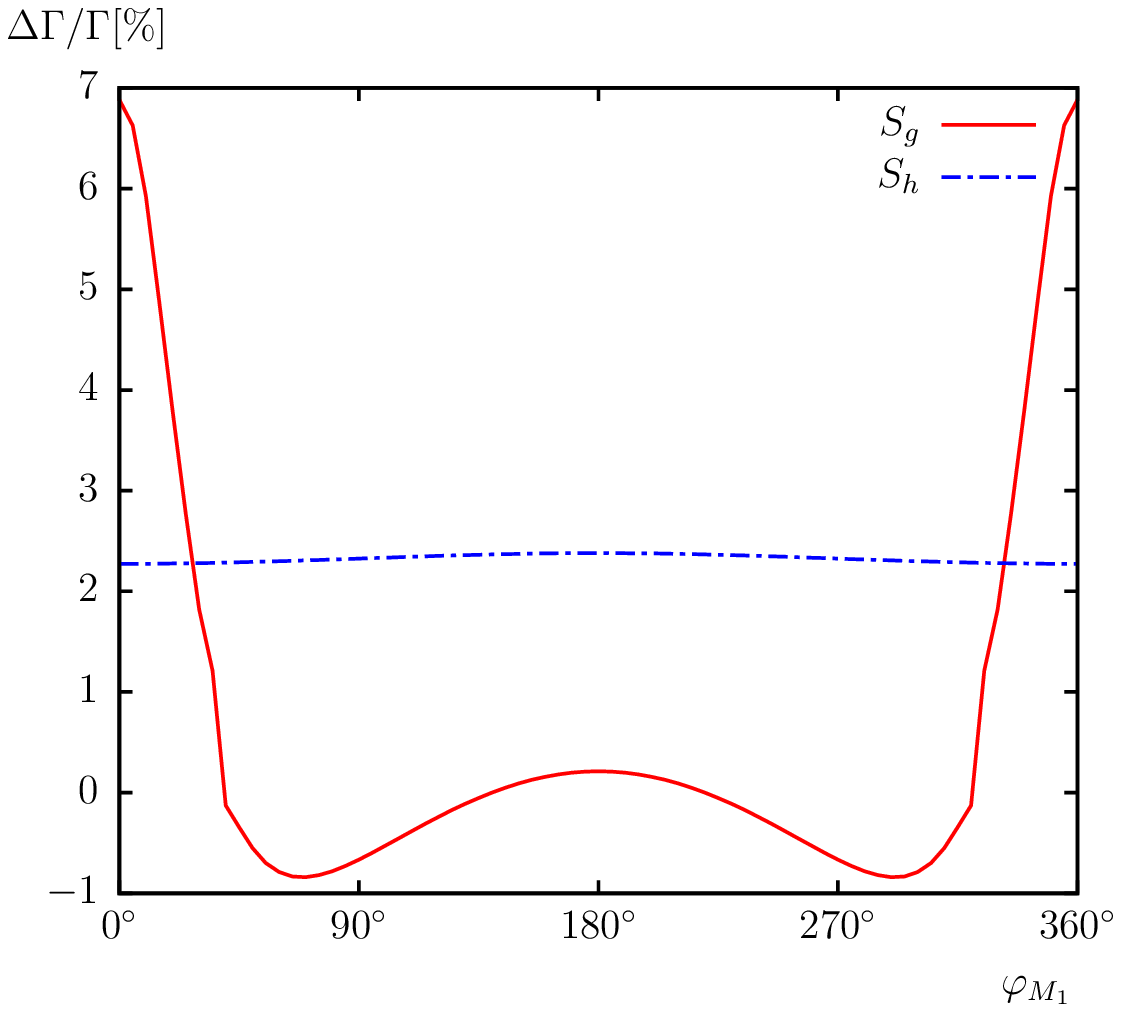}
\end{tabular}
\caption{  $\Ga_{\rm tot}(\neu{2})$.
  Tree-level (``tree'') and full one-loop (``full'') corrected total 
  decay widths are shown with the parameters chosen according to 
  \refta{tab:para}, with $\phiMe$ varied.
  The left panel shows the decay width, 
  the right panel shows the corresponding relative size of the 
  corrections.
}
\label{fig:phiM1.neu2tot}
\end{center}
\end{figure}
%%%%%%%%%%%%%%%%%%%%%%%%% F I G U R E %%%%%%%%%%%%%%%%%%%%%%%%%%%%%%%%%%%%%%%%%

%%% 2
The total decay width of $\neu{2}$ is shown in the l.h.s.\ of \reffi{fig:phiM1.neu2tot}.
The strong mixing of the second and third lightest neutralinos in $\Sg$ 
has already been discussed. 
The same decay channels as for $\neu{3}$
are open, except for that 
 to $Z$  and the lightest neutralino which is  only open for $|\phiMe|< 32^\circ$.
However, the decay to $Z$ is subdominant with a BR smaller than $1\%$. 
The threshold effect we observe in the right panel for the relative corrections 
is due to the decay into the lightest Higgs boson, as already discussed in this section for those decays with final $\neu{2}$.
\footnote{It should be noted that this effect, due to the singularity of the 
wave function renormalization,
is characteristic of on-shell renormalization schemes, which are less precise
when thresholds of external particles are open.
The masses entering these thresholds are the tree-level ones, which in the case of the lightest Higgs boson is close to that of the $Z$ boson.
The renormalized lightest Higgs boson, on the other hand, has a mass of $\sim 126\gev$ and the decay is closed.
}
The decay width and its relative correction show a 
 complementary behavior to the corresponding ones for the third lightest neutralino,
i.e., the dependence on $\phiMe$ of the sum of both widths 
is much weaker, with
corrections of $\sim 10\%$.

In $\Sh$, again, the neutralinos do not strongly mix and $\neu{2}$ is wino-dominated.
Here the  decays into left-handed sleptons or sneutrinos dominate due to the strong wino coupling.
However, the subdominant decays to the lightest neutralino and $Z$ or Higgs bosons
become the only open channels if the sleptons are chosen heavier.
These decays show a strong (and complementary) dependence on $\phiMe$
due to the change in the  relative $\cp$-parity of the two lightest neutralinos.

% %%%%%%%%%%%%%%%%%%%%%%%%%%%%%%%%%%%%%%%%%%%%%%%%%%%%%%%%%%%%%%%%%%%%%%%%%%%%%
\subsection{Differences between the renormalization schemes}
\label{sec:rendiff}

%%%%%%%%%%%%%%%%%%%%%%%%% T A B L E %%%%%%%%%%%%%%%%%%%%%%%%%%%%%%%%%%%%%%%%%%%
\begin{table}[t!]
\renewcommand{\arraystretch}{1.5}
\BC
\begin{tabular}{|l|r|r|r|r|}
\hline
Channel & \multicolumn{2}{c|}{$\Sg$}&  \multicolumn{2}{c|}{$\Sh$} \\
\cline{2-5}
&  $\qquad 45^\circ\qquad $& $\qquad 90^\circ \qquad$&  $45^\circ\qquad$& $90^\circ\qquad$\\
\hline\hline
$\DecayNCmH{4}{1}$       & $ 1.5\times 10^{-5}$ & $ 2.0\times 10^{-5}$ & $-6.9\times 10^{-6}$ & $-4.8\times 10^{-6}$ \\
$\DecayNCmW{4}{1}$       & $ 3.4\times 10^{-6}$ & $ 6.1\times 10^{-6}$ & $ 9.9\times 10^{-5}$ & $ 9.7\times 10^{-5}$ \\
$\DecayNNh{4}{1}{1}$     & $-1.9\times 10^{-4}$ & $-6.1\times 10^{-4}$ & $-6.3\times 10^{-5}$ & $-1.8\times 10^{-4}$ \\
$\DecayNNh{4}{1}{2}$     & $ 4.5\times 10^{-4}$ & $ 5.2\times 10^{-4}$ & $ 1.5\times 10^{-4}$ & $ 1.7\times 10^{-4}$ \\
$\DecayNNh{4}{1}{3}$     & $-1.4\times 10^{-4}$ & $-3.6\times 10^{-4}$ & $-9.1\times 10^{-5}$ & $-2.2\times 10^{-4}$ \\
$\DecayNNh{4}{2}{1}$     & $-1.2\times 10^{-5}$ & $ 5.4\times 10^{-5}$ & $ 2.7\times 10^{-6}$ & $ 6.1\times 10^{-6}$ \\
$\DecayNNh{4}{2}{2}$     & $ 1.3\times 10^{-4}$ & $-4.9\times 10^{-4}$ & $ 6.0\times 10^{-6}$ & $ 7.1\times 10^{-6}$ \\
$\DecayNNh{4}{2}{3}$     & $ 5.5\times 10^{-5}$ & $ 1.0\times 10^{-4}$ & $ 4.5\times 10^{-6}$ & $ 8.6\times 10^{-6}$ \\
$\DecayNNh{4}{3}{1}$     & $ 2.4\times 10^{-4}$ & $-7.2\times 10^{-4}$ & $\qquad -- \qquad$  & $\qquad -- \qquad$ \\
$\DecayNNh{4}{3}{2}$     & $-2.8\times 10^{-5}$ & $ 3.2\times 10^{-5}$ & $\qquad -- \qquad$  & $\qquad -- \qquad$ \\
$\DecayNNh{4}{3}{3}$     & $-5.1\times 10^{-5}$ & $-1.6\times 10^{-4}$ & $\qquad -- \qquad$  & $\qquad -- \qquad$ \\
$\DecayNNZ{4}{1} $       & $ 1.2\times 10^{-3}$ & $ 7.5\times 10^{-4}$ & $ 1.2\times 10^{-4}$ & $ 1.7\times 10^{-4}$ \\
$\DecayNNZ{4}{2} $       & $ 2.2\times 10^{-4}$ & $-3.4\times 10^{-4}$ & $ 9.5\times 10^{-6}$ & $ 7.3\times 10^{-6}$ \\
$\DecayNNZ{4}{3} $       & $-4.5\times 10^{-5}$ & $ 1.2\times 10^{-5}$ & $\qquad -- \qquad$  & $\qquad -- \qquad$ \\
$\DecayNlSl{4}{\tau}{1}$ & $-1.3\times 10^{-6}$ & $ 3.7\times 10^{-6}$ & $-3.5\times 10^{-5}$ & $-1.1\times 10^{-5}$ \\
$\DecayNlSl{4}{\tau}{2}$ & $-6.5\times 10^{-5}$ & $-6.2\times 10^{-5}$ & $-1.1\times 10^{-4}$ & $-1.1\times 10^{-4}$ \\
$\DecayNnSn{4}{\tau}$    & $-3.5\times 10^{-6}$ & $-7.0\times 10^{-6}$ & $-8.9\times 10^{-6}$ & $-1.9\times 10^{-5}$ \\
\hline
\end{tabular}
\caption{Differences between the relative corrections to the decay width for scheme I and II, 
shown in both scenarios $\Sg$ and $\Sh$, i.e.~$\De\Ga/\Ga(\Sg/\Sh)-\De\Ga/\Ga(\SgII/\ShII)$, 
at the specified values of $\phiMe$. 
The missing results correspond to those channels for which the decays are below threshold $\Sh$.}
\label{tab:rendiff}
\EC
\renewcommand{\arraystretch}{1.0}
\vspace{-1em}
\end{table}
%%%%%%%%%%%%%%%%%%%%%%%%%%%%%%%%%%%%%%%%%%%%%%%%%%%%%%%%%%%%%%%%%%%%%%%%%%%%%%%

%%%%%%%%%%%%%%%%%%%%%%%%% F I G U R E %%%%%%%%%%%%%%%%%%%%%%%%%%%%%%%%%%%%%%%%%
\begin{figure}[htb!]
\begin{center}
\begin{tabular}{c}
\includegraphics[width=0.49\textwidth,height=7.5cm]{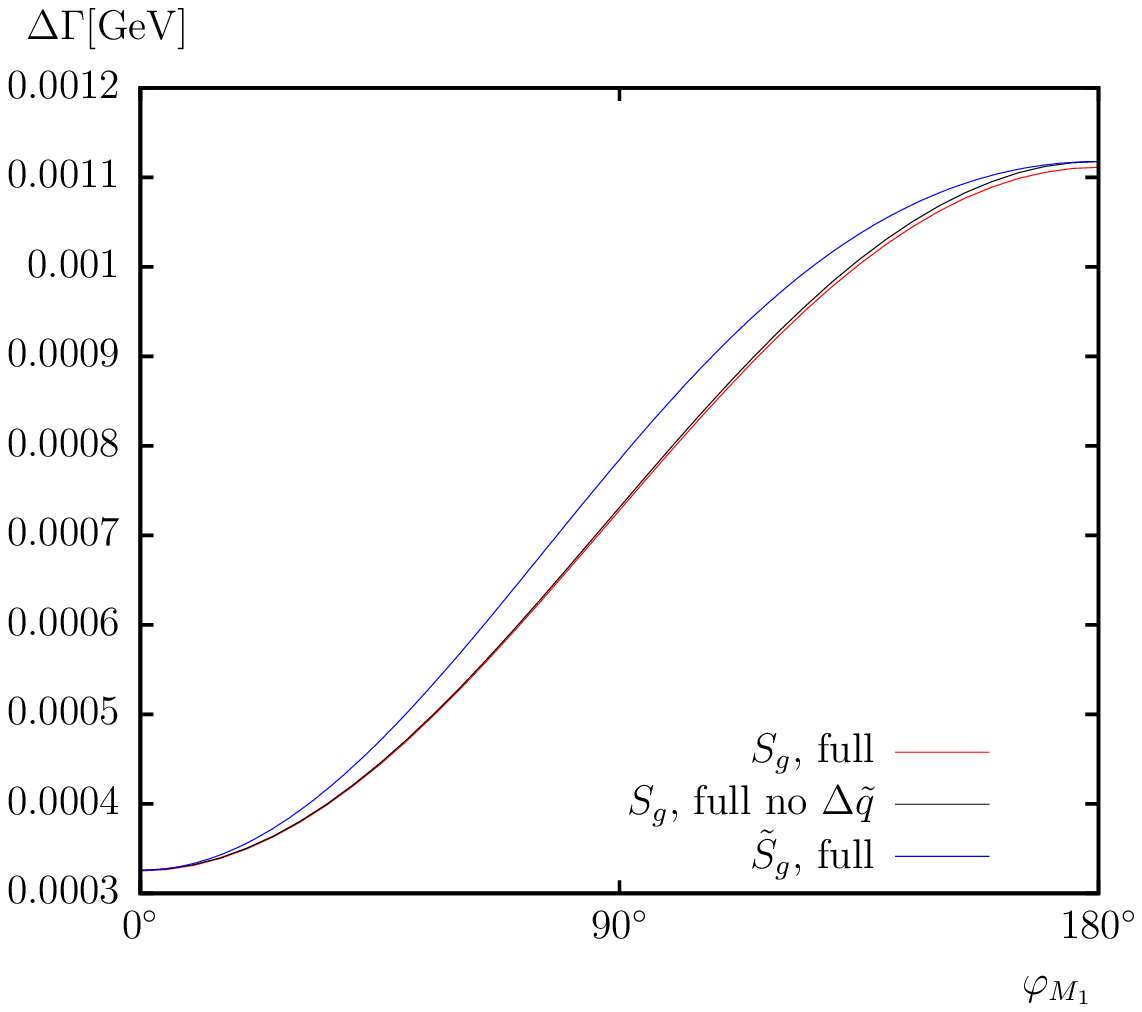}
\includegraphics[width=0.49\textwidth,height=7.5cm]{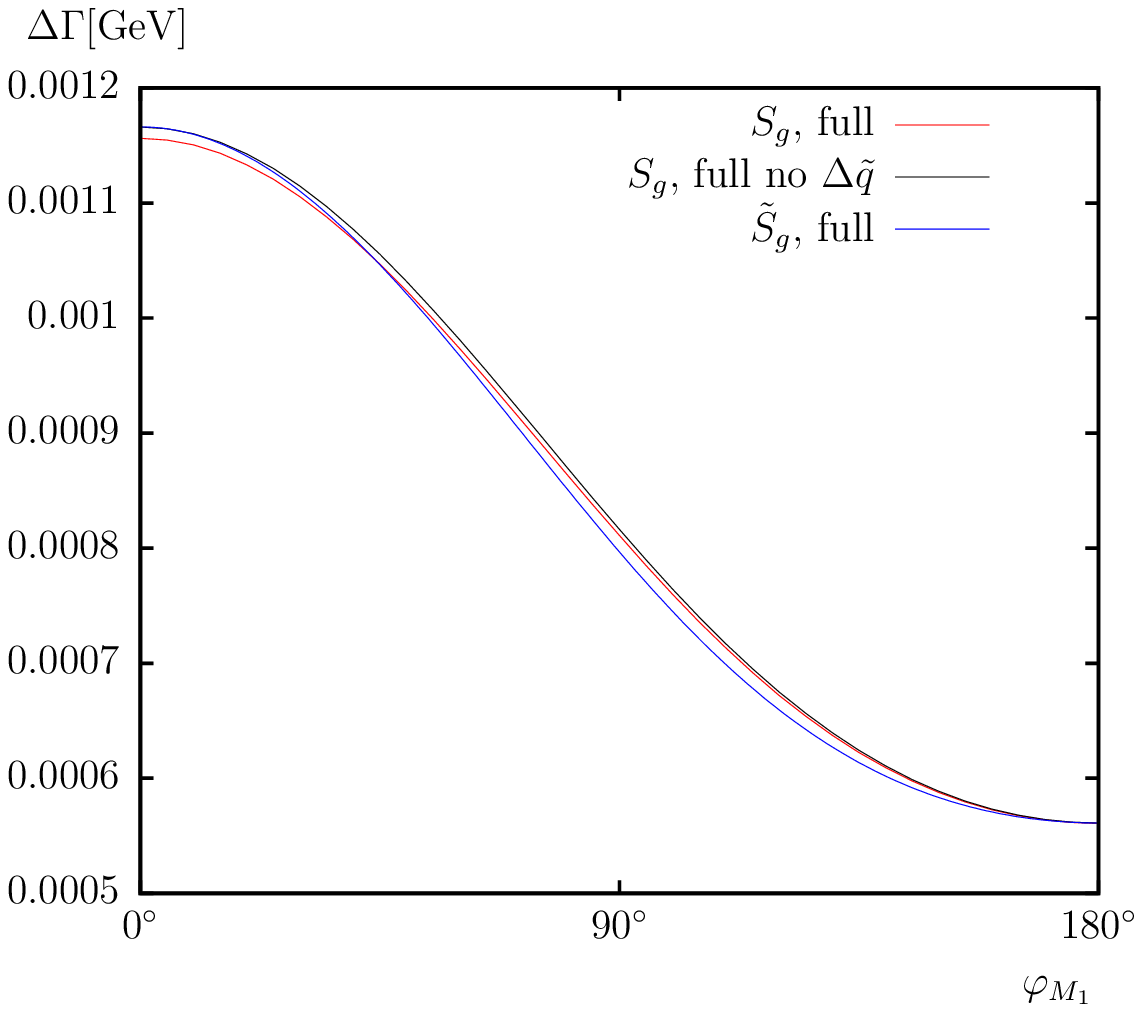} 
\\[2em]
\end{tabular}
\caption{$\Ga(\neu{4} \to\neu{1}Z)$ and $\Ga(\neu{4} \to\neu{1}h_1)$. 
  The one-loop (``full'') correction to the partial decay widths are shown
  as a function of $\phiMe$.  The parameters are chosen according to 
  \refta{tab:para}.
  Also shown is the   correction without including the shifts in the squark sector (``full no $\Delta \tilde q$'').
}
\label{fig:PhiM1.rendiff}
\end{center}
\end{figure}
%%%%%%%%%%%%%%%%%%%%%%%%%%%%%%%%%%%%%%%%%%%%%%%%%%%%%%%%%%%%%%%%%%%%%%%%%%%%%%%

In our benchmark scenarios $\Sg$ and $\Sh$ we have found remarkably good agreement between the two schemes. 
For most decay channels, the difference between the relative corrections to the partial widths $\De\Ga/\Ga$ for scheme I and II is of the order of $10^{-5}$. 
In \refta{tab:rendiff} we show this difference for all the decay channels, in both $\Sg$ and $\Sh$, 
and find that the largest differences are observed in $\DecayNNh{4}{1}{1,2,3}$, $\DecayNNh{4}{2}{2}$, $\DecayNNh{4}{3}{1}$, $\DecayNNZ{4}{1,2}$. 
In \reffi{fig:PhiM1.rendiff} we highlight these differences in $\Sg$ 
for $\DecayNNZ{4}{1}$  and $\DecayNNh{4}{1}{1}$.
To be precise, in this figure we compare the full one-loop correction $\De\Ga$
calculated in scheme~I both with and without the squark shifts 
(see end of \refse{sec:slepton})
to that calculated in scheme~II where squark shifts are not included. 
One can see that at $\phiMe=0^\circ$ and $180^\circ$ the difference between the results without squark shifts vanishes, 
confirming that the schemes differ only in the treatment of
the phases. 
One can also clearly see the impact of the squark shifts on the size of the one-loop correction.
Earlier in this section, it was noted that larger differences between the schemes are observed when the tree-level decay width depends strongly on the phase $\phiMe$. 
This explains why decays to the lightest neutralino are the most strongly affected ones. 
Also, the differences are in general much more pronounced in $\Sg$~than in $\Sh$, 
due to the mixing between the bino and Higgsino components, except for the
decays into sleptons or a $W$-boson, where the Higgsino component has a negligible r\^{o}le.

%%%%%%%%%%%%%%%%%%%%%%%%%%%%%%%%%%%%%%%%%%%%%%%%%%%%%%%%%%%%%%%%%%%%%%%%%%%%%%%
\section{Conclusions}
\label{sec:conclusions}

We have evaluated all non-hadronic two-body decay widths of neutralinos in 
the Minimal Supersymmetric Standard Model with complex parameters
(cMSSM). Assuming heavy scalar quarks we take into account all decay channels
involving charginos, neutralinos, (scalar) leptons, Higgs bosons and SM
gauge bosons. 
The decay modes are given in \refeqs{NNh} -- (\ref{NSnn}). 
The evaluation of the decay widths is based on a full one-loop calculation 
including hard and soft QED radiation. 
Such a calculation is necessary to derive a reliable
prediction of any two-body branching ratio.
Three-body decay modes can become sizable only if all the two-body channels
are kinematically (nearly) closed and have thus been neglected
throughout the paper. The same applies to two-body decay modes that
appear only at the one-loop level. 

We first reviewed the one-loop renormalization of the cMSSM, 
concentrating on the most relevant
aspects for our calculation, except for the details
for the Higgs boson sector which can be found in \citere{Stop2decay}. 
More importantly, we have given details for the char\-gi\-no/neu\-tra\-li\-no sector
in the two on-shell renormalization schemes which we have compared in this work.
The two schemes differ in the treatment of complex contributions in the chargino and neutralino sector.
The different renormalization of the 
$\cp$-violating phases leads to small differences in the cMSSM, which are however 
of higher order in the electroweak coupling
and vanish in the limit of real couplings.
Differences indicate the
  size of unknown higher-order corrections involving complex phases
  beyond the one-loop level.
We have also discussed the calculation of the one-loop diagrams, and the
treatment of UV- and IR-divergences that are canceled by the inclusion
of soft QED radiation. 
Our calculation set-up can easily be extended to other two-body decays
involving (scalar) quarks. 
We have taken into account all absorptive contributions,
explicitly inlcuding those of
self-energy type on external legs.
This ensures that all $\cp$-violating effects are correctly accounted for.
\medskip

In the numerical analysis we mainly concentrated on the decays of the 
  heaviest neutralino, $\neu{4}$.
For this analysis we 
have chosen a parameter set that allows
simultaneously {\em all} two-body decay modes under investigation, and
respects the current experimental bounds on Higgs boson and SUSY searches
(where the combination of low $\MHp$ and relatively
  large $\tb$ is in a potential conflict with the most recent LHC searches
  for the heavy MSSM Higgs bosons).
The masses of the charginos, and thus roughly those of the second and fourth neutralino, are in this scenario, respectively, $350$ and $600 \gev$. 
This leads to two representative scenarios for the char\-gi\-no/neu\-tra\-li\-no sector, 
for $\mu>M_2$ or $\mu<M_2$.
These benchmark scenarios allow copious 
production of the neutralinos in SUSY cascades at the LHC.
Furthermore, 
the production of  $\neu{4}\neu{j}$ at the ILC(1000), i.e.\ with $\sqrt{s} = 1000 \gev$, 
via $e^+e^- \to \neu{4}\neu{j}$ will be possible,
with all the subsequent decay modes (\ref{NNh}) -- (\ref{NSnn})
being (in principle) open. 
The clean environment of the ILC would
then permit a detailed, statistically dominated study of the neutralino
decays. Depending on the channel and the polarization, a precision at
the percent level seems to be achievable.
Special attention is paid to neutralino decays involving the 
Lightest Supersymmetric Particle (LSP), i.e.\ the lightest 
neutralino, or a neutral or charged Higgs boson.

We have shown results for 
varying $\phiMe$,
 the phase of the soft SUSY-breaking parameter~$\MOne$, which leads to
 $\cp$-violation in the chargino and neutralino sectors.
We have analyzed the tree-level and full one-loop results
for all kinematically open decay channels of the heaviest neutralino.
For the decays of the second and third neutralinos we have only shown the total widths.

We found sizable corrections in many of the decay channels. 
The higher-order corrections of the neutralino decay widths
involving the LSP are generically up to 
a level of about  $10\%$, and decay modes involving Higgs bosons can easily have corrections up
to $20-30\%$. 
The size of the full one-loop corrections to the decay
widths and the branching ratios also depends strongly on $\phiMe$,
especially for those decays in which an external neutralino is a mixed Higgsino-bino state.
We conclude that the largest effect of the radiative corrections is due to its effect on the mixing of the neutralinos.
All results on partial decay widths of $\neu{4}$
as well as the total decay widths of all neutralinos
are given in detail in
\refse{sec:numeval}.

\medskip

For the two on-shell renormalization schemes considered, we have found very
good  agreement:
where the difference between the relative size of the corrections is found to be $\lsim 0.1\%$.
The largestest differences have been found for decays with a strong dependence on the parameter $\phiMe$.
The good agreement between the two schemes is not unexpected, 
as they are found to be equivalent up to a higher order effect, as discussed in \refse{sec:rendiff}. 

The numerical results we have shown are of course dependent on choice of
the MSSM parameters. Nevertheless, they give an idea of the relevance
of the full one-loop corrections. 
For other choices of SUSY masses the 
corrections to the decay widths would stay the same, 
but the branching ratios would look very different. 
Channels for which the decay width (and its respective one-loop corrections) may look 
unobservable due to the smallness of the BR in our numerical examples
could become important if other channels are kinematically forbidden.

Following our analysis it is evident that the full one-loop corrections
are mandatory for a precise prediction of the various branching ratios.
This applies to LHC analyses,
but even more to analyses at the ILC or CLIC,
where a precision at the percent level is anticipated for the
determination of neutralino branching ratios (depending on the neutralino
masses, the center-of-mass energy and the integrated luminosity).
The results for the neutralino decays will be implemented into the
Fortran code {\fh}.

%%%%%%%%%%%%%%%%%%%%%%%%%%%%%%%%%%%%%%%%%%%%%%%%%%%%%%%%%%%%%%%%%%%%%%%%%%%%%%
\subsection*{Acknowledgments}
We thank  
A.~Fowler,
T.~Hahn, 
G.~Moortgat-Pick,
H.~Rzehak,
and
G.~Weiglein
for helpful discussions. 
A.B. gratefully acknowledges support of the DFG through the grant SFB 676, ``Particles, Strings, and the Early Universe''.
The work of S.H.\ was partially supported by CICYT (grant FPA
2007--66387 and FPA 2010--22163-C02-01).
F.v.d.P.\ was supported by 
the Spanish MICINN's Consolider-Ingenio 2010 Programme under grant MultiDark CSD2009-00064.

%%%%%%%%%%%%%%%%%%%%%%%%%%%%%%%%%%%%%%%%%%%%%%%%%%%%%%%%%%%%%%%%%%%%%%%%%%%%%%%

%----------------------------------------------------------------------------------------
\vspace*{1cm}

\begin{appendix}
\noindent{\Large\bf Appendix}
\setcounter{equation}{0}
\renewcommand{\thesubsection}{\Alph{section}.\arabic{subsection}}
\renewcommand{\theequation}{\Alph{section}.\arabic{equation}}

\setcounter{equation}{0}

% %%%%%%%%%%%%%%%%%%%%%%%%%%%%%%%%%%%%%%%%%%%%%%%%%%%%%%%%%%%%%%%%%%%%%%%%%%%%%
\section{Tree-level results}
\label{sec:treeresults}
For completeness we include here the expressions for the tree-level decay widths:
% %%%%%%%%%%%%%%%%%%%%%%%%%%%%%%%%%%%%%%%%%%%%%%%%%%%%%%%%%%%%%%%%%%%%%%%%%%%%%
\begin{align}
\label{NCmHtree}
\Gtree(\decayNCmH) &= 
\KKL 
\KL |C(\neu{i},\chap{j},H^-)_L|^2 + |C(\neu{i},\chap{j},H^-)_R|^2 \KR
(\mneu{i}^2+\mcha{j}^2-\MHp^2) 
\right.
\non \\ & 
\left.
+ 4 \re \KKKL C(\neu{i},\chap{j},H^-)_L^* C(\neu{i},\chap{j},H^-)_R \KKKR
\mneu{i}\mcha{j}
\KKR
\non \\ &
\times \frac{\la^{1/2}(\mneu{i}^2,\mcha{j}^2,\MHp^2)}{32\pi \mneu{i}^3}
\quad (i = 2,3,4, \; j = 1,(2))~, \\[0.5em]
\label{NCmWtree}
\Gtree(\decayNCmW) &=
\KKL 
\KL |C(\neu{i},\chap{j},W^{-})_L|^2 + |C(\neu{i},\chap{j},W^{-})_R|^2 \KR
\right.
\non \\ &
\left.
\times 
\KL \mneu{i}^2+\mcha{j}^2-2\MW^2+\frac{(\mneu{i}^2-\mcha{j}^2)^2}{\MW^2} \KR
\right.
\non \\ & 
\left.
- 12\re \KKKL C(\neu{i},\chap{j},W^{-})_L^* C(\neu{i},\chap{j},W^{-})_R \KKKR
\mneu{i}\mcha{j}
\KKR
\non \\ &
\times \frac{\la^{1/2}(\mneu{i}^2,\mcha{j}^2,\MW^2)}{32\pi \mneu{i}^3}
\quad (i = 2,3,4, \; j = 1,(2))~, \\[0.5em]
\label{NNHtree}
\Gtree(\decayNNh) &= 
\KKL 
\KL |C(\neu{i},\neu{j},h_k)_L|^2 + |C(\neu{i},\neu{j},h_k)_R|^2 \KR
(\mneu{i}^2+\mneu{j}^2-m_{h_k}^2) 
\right.
\non \\ & 
\left.
+ 4\re \KKKL C(\neu{i},\neu{j},h_k)_L^* C(\neu{i},\neu{j},h_k)_R \KKKR
\mneu{i}\mneu{j}
\KKR
\non \\ &
\times \frac{\la^{1/2}(\mneu{i}^2,\mneu{j}^2,m_{h_k}^2)}{32\pi \mneu{i}^3}
\quad  (i = 2,3,4, \; j = 1,(2),\; k = 1,2,3)~, 
%\\
%
\non \\ &= 
\KKL 
\KL  |C(\neu{i},\neu{j},h_k)_R|^2 \KR
(\mneu{i}^2+\mneu{j}^2-m_{h_k}^2) 
\right.
\non \\ & 
\left.
+2\re \KKKL \KL C(\neu{i},\neu{j},h_k)_R \KR^2 \KKKR
\mneu{i}\mneu{j}
\KKR
\non \\ &
\times \frac{\la^{1/2}(\mneu{i}^2,\mneu{j}^2,m_{h_k}^2)}{16\pi \mneu{i}^3}
\quad  (i = 2,3,4, \; j = 1,(2),\; k = 1,2,3)~, 
\\[0.5em]
\label{NNZtree}
\Gtree(\decayNNZ) &= 
\KKL 
\KL |C(\neu{i},\neu{j},Z)_L|^2 + |C(\neu{i},\neu{j},Z)_R|^2 \KR
\right.
\non \\ &
\left.
\times 
\KL \mneu{i}^2+\mneu{j}^2-2\MZ^2+\frac{(\mneu{i}^2-\mneu{j}^2)^2}{\MZ^2} \KR
\right.
\non \\ & 
\left.
- 12\re \KKKL C(\neu{i},\neu{j},Z)_L^* C(\neu{i},\neu{j},Z)_R \KKKR
\mneu{i}\mneu{j}
\KKR
\non \\ &
\times \frac{\la^{1/2}(\mneu{i}^2,\mneu{j}^2,\MZ^2)}{32\pi \mneu{i}^3}
%~, \\
\non \\ &= 
\KKL 
|C(\neu{i},\neu{j},Z)_R|^2 
\right.
\left.
\KL \mneu{i}^2+\mneu{j}^2-2\MZ^2+\frac{(\mneu{i}^2-\mneu{j}^2)^2}{\MZ^2} \KR
\right.
\non \\ & 
\left.
+ 6\re \KKKL \KL C(\neu{i},\neu{j},Z)_R\KR^2 \KKKR
\mneu{i}\mneu{j}
\KKR
\non \\ &
\times 
\frac{\la^{1/2}(\mneu{i}^2,\mneu{j}^2,\MZ^2)}{16\pi \mneu{i}^3}
\quad (i = 2,3,4, \; j < i)~, \\[0.5em]
\Gtree(\decayNlSl) &=  
\KKL 
\KL |C(\neu{i},\bar \ell,\slk)_L|^2 + |C(\neu{i},\bar \ell,\slk)_R|^2 \KR
(\mneu{i}^2 + \ml^2 - \mslk^2)
\right.
\non \\ &
\left.
+ 4\re \KKKL C(\neu{i},\bar \ell,\slk)_L^* C(\neu{i},\bar \ell,\slk)_R \KKKR \mneu{i} \ml
\KKR
\non \\ & 
\times \frac{\la^{1/2}(\mneu{i}^2,\ml^2,\mslk^2)}{32\pi \mneu{i}^3}
\quad (i = 2,3,4, \; \ell = e, \mu, \tau,\; k=1,2)~,  
\label{NSlltree}
\\[0.5em]
\Gtree(\decayNnSn) &=  
 |C(\neu{i},\bar\nu_\ell,\Sn)_R|^2 
(\mneu{i}^2 - \msn^2)
\non \\ & 
\times \frac{\la^{1/2}(\mneu{i}^2,0,\msn^2)}{32\pi \mneu{i}^3}
\quad (i = 2,3,4, \; \ell = e, \mu, \tau)~,
\label{NSnntree}
\end{align}
where $\lambda(x,y,z) = (x - y - z)^2 - 4yz$ and the couplings 
$C(a, b, c)$ can be found in the \fa~model files~\cite{feynarts-mf}.
$C(a, b, c)_{L,R}$ denote the part of the coupling which
is proportional to $\om_\mp = \ed{2}(\id \mp \ga_5)$.

%%%%%%%%%%%%%%%%%%%%%%%%%%%%%%%%%%%%%%%%%%%%%%%%%%%%%%%%%%%%%%%%%%%%%%%%%%%%%%%
\end{appendix}
%\clearpage

%%%%%%%%%%%%%%%%%%%%%%%%%%%%%%%%%%%%%%%%%%%%%%%%%%%%%%%%%%%%%%%%%%%%%%%%%%%%%


\begin{thebibliography}{99}
\bibitem{feynarts-mf} The couplings can be found in the files
                      {\tt MSSM.ps.gz}, {\tt MSSMQCD.ps.gz} and 
                      {\tt HMix.ps.gz} as part of the
                      \fa~package~\cite{feynarts}.
\end{thebibliography}
\end{document}